\documentclass[useAMS,usenatbib]{mn2e}
\usepackage[dvipdfm]{graphicx}
\usepackage{aas_macros}
\usepackage{times}

\newcommand\ion[2]{#1\,{\sc\romannumeral #2}}
\newcommand{\msol}{M_\odot}
\def\amfp{22.2 \pm 2.3}
\def\bmfp{15.1 \pm 1.8}
\def\cmfp{10.3 \pm 1.6}
\def\aval{37 \pm 2}
\def\gval{-5.4 \pm 0.4}
\def\nqso{163}
\def\nstack{145}
\def\nboot{2000}
\def\mhmpc{h^{-1}_{70} \, \rm Mpc}
\def\hmpc{$\mhmpc$}

\def\mased{\delta\alpha_\mathrm{T}}
\def\ased{$\mased$}
\def\mglym{\gamma_\tau}
\def\glym{$\mglym$} 
\def\mfobs{f_\lambda^{\rm obs}}
\def\fobs{$\mfobs$}
\def\mnD{\left<n_\mathrm{c}D\right>}
\def\nD{$\mnD$}

\def\mzq{z_{\rm q}} 
\def\zq{$\mzq$}

\def\intl{\int\limits}

\def \zem {$z_{\rm em}$}
\def \mzem {z_{\rm em}}

\def \ozll {$z_{\rm 912}^{\tau=1}$}
\def\kll{$\kappa_{912}$}
\def\mkll{\kappa_{912}}
\def \mtll  {\tau^{\rm LL}_{\rm 912}}

\def \kms  {km~s$^{-1}$}
\def \mkms  {{\rm km~s^{-1}}}

\def \lya  {Ly$\,\alpha$}

\def \nhi  {$N_{\rm HI}$}
\def \mnhi  {N_{\rm HI}}

\def \fnz {$f(\mnhi,z)$}
\def \mlmfp {\lambda_{\rm mfp}^{912}}
\def \lmfp {$\lambda_{\rm mfp}^{912}$}
\def \teff {$\tau_{\rm eff}^{\rm LL}$}
\def \mteff {\tau_{\rm eff}^{\rm LL}}

\def \mtlyman {\tau_{\rm eff}^{\rm Lyman}}
\def \tlyman {$\tau_{\rm eff}^{\rm Lyman}$}
\def \tlya {$\tau_{\rm eff}^{\rm Ly\alpha}$}
\def \mtlya {\tau_{\rm eff}^{\rm Ly\alpha}}
\newcommand{\cm}[1]{\, {\rm cm^{#1}}}
\def \mrvir {r_{\rm vir}}
\def \rvir {$\mrvir$}
\title[The mean free path across cosmic time]{The Giant Gemini GMOS survey of $\bmath{z_\mathrm{em}>4.4}$ quasars --\\
I.\ Measuring the mean free path across cosmic time}
\author[G. Worseck et al.]{G\'{a}bor Worseck,$^{1,2}$\thanks{E-mail: gabor@mpia-hd.mpg.de}
J. Xavier Prochaska,$^{1,2}$
John M. O'Meara,$^3$
George D. Becker,$^4$\newauthor
Sara L. Ellison,$^5$
Sebastian Lopez,$^6$
Avery Meiksin,$^7$
Brice M\'enard,$^{8,9,10}$\newauthor
Michael T. Murphy$^{11}$ and
Michele Fumagalli$^{12,13}$\\
$^1$Max-Planck-Institut f\"{u}r Astronomie, K\"{o}nigstuhl 17, D-69117 Heidelberg, Germany\\
$^2$Department of Astronomy and Astrophysics, UCO/Lick Observatory, University of California, 1156 High Street, Santa Cruz, CA 95064, USA\\
$^3$Department of Chemistry and Physics, Saint Michael's College, One Winooski Park, Colchester, VT 05439, USA\\
$^4$Kavli Institute for Cosmology and Institute of Astronomy, Madingley Road, Cambridge CB3 0HA, UK\\
$^5$Department of Physics and Astronomy, University of Victoria, Victoria, BC V8P 1A1, Canada\\
$^6$Departamento de Astronom\'ia, Universidad de Chile, Casilla 36-D, Santiago, Chile\\
$^7$Scottish Universities Physics Alliance, Institute for Astronomy, University of Edinburgh, Blackford Hill, Edinburgh EH9 3HJ, UK\\
$^8$Department of Physics \& Astronomy, John Hopkins University, 3400 N. Charles Street, Baltimore, MD 21218, USA\\
$^9$Kavli IPMU (WPI), The University of Tokyo, Kashiwa 277-8583, Japan\\
$^{10}$Alfred P. Sloan Fellow\\
$^{11}$Centre for Astrophysics and Supercomputing, Swinburne University of Technology, Hawthorn, Victoria 3122, Australia\\
$^{12}$Department of Physics, Durham University, South Road, Durham DH1 3LE, UK\\
$^{13}$Carnegie Observatories, 813 Santa Barbara Street, Pasadena, CA 91101, USA
}

\date{Draft version \today}

\begin{document}

\label{firstpage}

\maketitle

\begin{abstract}
We have obtained spectra of \nqso\ quasars at $\mzem>4.4$ with the Gemini
Multi Object Spectrometers on the Gemini North and South telescopes, the largest
publicly available sample of high-quality, low-resolution spectra at these redshifts.
From this homogeneous data set, we generated stacked quasar spectra in three redshift
intervals at $z \sim 5$. We have modelled the flux below the rest-frame Lyman limit 
($\lambda_\mathrm{r} < 912$\,\AA) to assess the mean free path \lmfp\ of the
intergalactic medium to \ion{H}{1}-ionizing radiation. 
At mean redshifts $z_\mathrm{q} = 4.56, 4.86$ and $5.16$, we measure
$\mlmfp =\left(\amfp, \bmfp, \cmfp\right)h_{70}^{-1}$ proper Mpc with uncertainties 
dominated by sample variance. Combining our results with \lmfp\ measurements from
lower redshifts, the data are well modelled by a simple power-law
$\mlmfp = A\left[\left(1+z\right)/5\right]^\eta$ with $A=\left(\aval\right)h_{70}^{-1}$\,Mpc
and $\eta = \gval$ between $z=2.3$ and $z=5.5$.
This rapid evolution requires a physical mechanism -- beyond cosmological expansion --
which reduces the cosmic effective Lyman limit opacity. We speculate that the majority
of \ion{H}{1} Lyman limit opacity manifests in gas outside galactic dark matter
haloes, tracing large-scale structures (e.g.\ filaments) whose average density
(and consequently neutral fraction) decreases with cosmic time. 
Our measurements of the strongly redshift-dependent mean free path shortly after the
completion of \ion{H}{1} reionization serve as a valuable boundary condition for
numerical models thereof. Having measured $\mlmfp \approx 10$\,Mpc at $z=5.2$,
we confirm that the intergalactic medium is highly ionized by that epoch and that the
redshift evolution of the mean free path does not show a break that would indicate
a recent end to \ion{H}{1} reionization.
\end{abstract}


\section{Introduction}
\label{sec:intro}

The current cosmological paradigm posits that $\approx 1$\,Gyr after
the Big Bang compact sources -- stars, accreting black holes --
generate sufficient ionizing radiation to reionize the neutral
hydrogen throughout the bulk of the intergalactic medium (IGM). 
Indeed, the spectra of distant quasars and gamma-ray bursts reveal a forest of
\lya\ absorption lines which are characteristic of a highly ionized medium
at $z \la 6$ \citep[e.g.][]{gp65,cpb+05}. Resolving the epoch of reionization,
its timing and the nature of these ionizing sources stands as one of the outstanding
challenges of modern cosmology.

While the community eagerly awaits results of low-frequency radio
observations to probe the reionization epoch via the 21cm hyperfine
transition \citep[e.g.][]{zaroubi12,yatawatta13,pober13,beardsley13},
researchers have been studying effects of reionization on the IGM in
absorption through spectroscopy of background sources. 
These include the most distant quasars \citep{Fan06,mortlock11} and $z>8$
gamma-ray bursts `revealed' by their extremely faint optical fluxes
\citep{kka+06,cucchiara11a,chornock13}.
Analysis of these data suggest the transition to a predominantly ionized 
universe occurs at $z \ga 6$ \citep{wbf+03,bolton11}, where a sharp
increase in the effective \lya\ opacity may occur (but see \citet{brs07}).

Preferably, one would trace evolution in the ionization state of the
IGM in the Lyman continuum which has an effective opacity nearly four
orders of magnitude smaller than \lya\ and is therefore far more sensitive
to the neutral fraction of hydrogen. The effective Lyman Limit (LL) opacity
is frequently represented by the mean free path \lmfp, defined here to
be the physical distance a packet of ionizing photons can travel before
encountering an $\mathrm{e}^{-1}$ attenuation. By definition, \lmfp\
approaches zero as one transitions into the reionization epoch \citep[e.g.][]{gnedin00}. 
After reionization, the mean free path is set by the distribution and
evolution of residual neutral gas in the universe.  This will include
dense, collapsed structures (e.g.\ galaxies) but the opacity may be
dominated by more diffuse gas in the outskirts of dark matter haloes
\citep[e.g.][]{fumagalli11a} and even more distant and diffuse
structures in the IGM. By assessing the redshift evolution of \lmfp,
one constrains the nature of structures providing the universe's LL
opacity and, as importantly, its attenuation of the ionizing sources
which generate the extragalactic UV background \citep{flz+09,hm12}.

Many previous works have estimated \lmfp\ by first evaluating the
\ion{H}{1} frequency distribution $f(\mnhi)$ of IGM absorbers (and its
redshift evolution; \fnz) and then convolving this distribution with the
photoionization cross-section \citep[e.g.][]{mm93}. Varying \fnz\ within
the observational uncertainties, this indirect approach has yielded
estimates of 40 to 150\hmpc\ (proper) at $z \sim 3$ in a $\Lambda$CDM
cosmology \citep{mhr99,flh+08,mm93}. This approach is affected by uncertainty
in \fnz\ from line blending and clustering of absorbers contributing to the
LL opacity, at least at $z\sim 2.5$ \citep{rudie13,pmof14}.
Recently, the mean free path has been \textit{directly} evaluated through the
analysis of stacked rest-frame quasar spectra: at $z \approx 4$ using data
taken by the Sloan Digital Sky Survey \citep[SDSS;][]{sdssdr7} and at
$z \sim 2-3$ with space and ground-based programmmes \citep{pwo09,omeara13,fop+13}.
These results show a monotonic increase in \lmfp\ with decreasing redshift,
ranging from $\approx 30$\,\hmpc\ at $z=4$ to over 200\,\hmpc\ at $z=2.5$.

For $z>4$, the current constraints on \lmfp\ are much poorer owing to the small
sample of quasars observed at sufficient signal-to-noise (S/N) to assess the
Lyman continuum opacity. Our own analysis of the SDSS data set terminated at
$z=4.2$ \citep{pwo09}, and the SDSS-III survey offers few new bright sources at
these redshifts \citep{paris13}. 
\citet{songaila10} surveyed 25 quasars for Lyman limit systems (LLSs) with medium
resolution Keck spectroscopy and discovered 20 absorbers with $\mtll \ge 1$ at $z>4$,
but sampled only 10 systems at $z>4.5$. Combining their results with previous surveys
\citep{storrie94,peroux_dla03}, they measured the incidence of LLSs to $z \sim 5$
and inferred a mean free path of 50\hmpc\ at $z=3.5$ assuming a power-law frequency
distribution $f(\mnhi) \propto \mnhi^{-\beta}$ with a relatively flat $\beta = 1.3$.
Again, \fnz\ is poorly constrained at $z>3.5$, especially at column densities
$\mnhi \approx 10^{17} \cm{-2}$ implying a significant ($\sim 30$ per cent)
systematic uncertainty in \lmfp\ estimates.

In 2010, our group began a multi-semester campaign with the Gemini
Multi Object Spectrometers \citep[GMOSs;][]{gmos} on the twin Gemini 8\,m telescopes,
to survey approximately $150$ quasars at $\mzem>4.4$, discovered primarily by SDSS.
The primary goal of this Giant Gemini GMOS (GGG) survey is to precisely measure \lmfp\
at $z\sim 5$; this is the focus of this manuscript.  The data also enable estimates
of the average IGM Lyman series opacity, studies of high-$z$ quasar emission properties,
and a search for high-$z$ high-column density absorbers. Those topics will be considered
in future manuscripts.  In the following, we adopt a flat $\Lambda$CDM cosmology with
$H_0 = 70 h_{70}\,\mathrm{km}\,\mathrm{s}^{-1}\,\mathrm{Mpc}^{-1}$,
$\Omega_\mathrm{m} = 0.30$, and $\Omega_\Lambda = 0.70$. Unless noted otherwise, all
distances quoted in this paper are proper and corrected to the used cosmology.

\section{Sample Selection}
\label{sec:sample}

The primary goal of the GGG survey is to extend measurements of the mean free path
to ionizing radiation to $z>4.2$. Our methodology follows the techniques developed
in \citet{pwo09} and \citet{omeara13}, which we briefly summarize:
(i) generate stacked rest-frame spectra of a random sample of quasars with a narrow
range of emission redshift \zem;
(ii) model the average flux at rest wavelengths $\lambda_\mathrm{r} < 912$\,\AA\
with a standard quasar spectral energy distribution (SED), an evolving Lyman series
opacity from the \ion{H}{1} forest, and an opacity set by the cumulative LL
absorption of the IGM;
(iii) calculate the distance from \zem\ where the flux is attenuated by
$\mathrm{e}^{-1}$ from LL absorption alone.
Unlike previous approaches which relied on evaluations of the \ion{H}{1} frequency
distribution \fnz, our approach offers a nearly direct estimation of \lmfp. 
Generally, the uncertainty is driven by sample variance, and possible systematic
errors are estimated below. 

To perform the \lmfp\ measurement, we require a quasar data set with the following
characteristics:  
(i) well measured emission redshifts ($\sigma_v<1000\,\mkms$); 
(ii) a large sample of sources ($N\ga 50$) at nearly the same \zem; 
(iii) a homogeneous, spectral data set in resolution,
wavelength coverage and data reduction processing; 
(iv) coverage of the emission-line free $\lambda_\mathrm{r} \approx 1450$\,\AA\
spectral region to scale the spectra to one another;
(v) quasar spectra without strong broad absorption line features;
(vi) a relative spectral fluxing accurate to a few per cent;
(vii) a S/N per pixel in excess of $\approx 5 \, \rm \AA^{-1}$ to
minimize systematic error associated with continuum placement
(for other projects) and sky subtraction.

Upon publication of \citet{pwo09}, we recognized that no such data set existed
at $z>4$. While the SDSS had discovered and observed several hundreds of quasars
at those redshifts, the majority of these data had too low S/N for precise \zem\
measurements and may well suffer from systematic errors (e.g.\ poor sky subtraction)
that preclude the generation of robust stacked spectra. Therefore, we initiated
a programmes on the Gemini North and South telescopes with the Gemini Multi Object
Spectrometers \citep[GMOSs;][]{gmos} to observe over 150 quasars at $\mzem > 4.4$.
Details on the instrument configuration are presented in the following section.
At these high redshifts, the IGM strongly absorbs the quasar flux at wavelengths
$\lambda < 6500$\,\AA. This implies that the colour selection algorithms used by
SDSS to target these quasars are essentially free from any bias from the presence
of LL absorption \citep{wp11}. In this respect, we consider the sample to be unbiased.

\begin{figure}
\includegraphics[angle=90,width=1.0\linewidth]{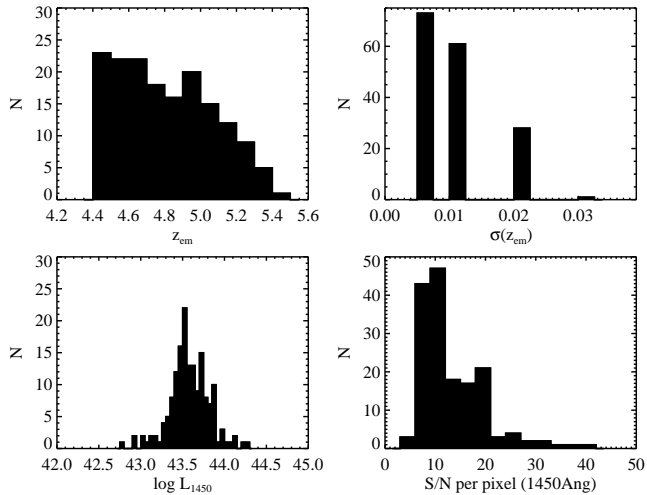}
\caption{\label{fig:demographics}
Histograms summarizing properties of the quasars and their spectra from the GGG survey
(\textit{top left:} quasar redshift; \textit{top right:} estimated redshift error;
\textit{bottom left:} specific luminosity at rest-frame wavelength $\lambda_\mathrm{r}=1450$\,\AA;
\textit{bottom right:} S/N per pixel at $\lambda_\mathrm{r}=1450$\,\AA).
The sample is restricted to quasars with $z_\mathrm{em} > 4.4$ and,
given the magnitude limit of SDSS, comprise a very luminous cohort
($L_{1450}\ga 10^{43}$\,erg\,s$^{-1}$\,\AA$^{-1}$).
}
\end{figure}

All of our targets were taken from the SDSS Data Release 7 \citep{sdssdr7}.
We began by retrieving the `best' 1D spectrum for every source flagged as a $z>4.4$
quasar by the SDSS automatic redshift assignment routine, and proceeded to vet each
of them through visual inspection of the SDSS spectra. We culled sources with
apparent broad absorption line (BAL) features and those where \zem\ had been
erroneously assigned. This provided a pool of $\approx 380$ sources.
From these we selected \nqso\ for observations with Gemini. The precise target list
represents a compromise between sampling the redshift interval
$\mzem = \left[4.4,5.5\right]$, selecting sources sufficiently bright for an
approximately one hour Gemini/GMOS observation, considerations on the number of
sources per observing semester, maximizing the number of sources for Gemini South, 
and the availability of suitable guide stars. We believe that none of these criteria
has an important impact on the LL absorption along the sight lines.
Table~\ref{tab:ggg_sample} lists all of the sources observed in our programme.
A visual summary of the main properties of our survey is provided in
Fig.~\ref{fig:demographics}. 

The only other survey of competitive size and quality at these redshifts is the
sample developed by \citet{songaila10}. Those authors observed 25 quasars at
$\mzem>4.17$ with Keck/ESI at high spectral resolution ($R\sim 5300$) and S/N,
and combined their sample with the literature \citep{peroux_dla03}. However, this
combined sample of 39 $\mzem > 4.4$ quasars yields a rough \lmfp\ estimate
at best, owing to sample variance and the broad emission redshift distribution.
As we will show below, $\ga 40$ sight lines per $\Delta z\simeq 0.3$ are needed to
track the redshift evolution of the mean free path at $z>4$.

\section{Observations and Data Reduction}
\label{sec:obs}

\begin{table}
\caption{\label{tab:ggg_obsprog}Summary of our programme allocations.}
\begin{center}
\begin{tabular*}{0.95\linewidth}{@{\extracolsep{\fill}}lccc}
\hline
Programme ID  & Allocation [h] & Band & Observed QSOs\\
\hline
GN-2010A-Q-33 & 40 & 2 & 40\\
GS-2010A-Q-1  & 18 & 1 & 18\\
GN-2010B-Q-71 & 24 & 2 & 21\\
GS-2010B-Q-28 & 20 & 1 & 20\\
GN-2011A-Q-1  & 33 & 1 & 35\\
GS-2011A-Q-1  & 27 & 1 & 29\\
\hline
\end{tabular*}
\end{center}
\end{table}

We proposed for Gemini observing time through all partners in Semesters
2010A, 2010B, and 2011A.  Altogether, we were allocated 162 hours for
observations in queue mode, with 60 per cent allocated on Gemini North to
efficiently cover the SDSS footprint. A summary of our programme is given in
Table~\ref{tab:ggg_obsprog}.

For every source, we obtained a spectrum through the 1\arcsec\ longslit with
two of the GMOS gratings: 
(i) the B600 grating blazed at 4610\,\AA\ with a FWHM resolution of
$\approx 320\,\mkms$ and an unbinned spectral dispersion of 
$0.46$\,\AA\,pixel$^{-1}$ and;
(ii) the R400 grating blazed at 7640\,\AA\ with FWHM $\approx 360\,\mkms$
and a dispersion of $0.69$\,\AA\,pixel$^{-1}$.
We binned the CCDs twice in the spatial dimension and 4 times spectrally,
resulting in a sampling of $\sim 3$ pixels per spectral resolution element.
During our observing campaign both GMOSs were equipped with their original
EEV 3 CCD mosaics.
For every target we acquired two 900\,s exposures with the B600 grating
and a single 480\,s exposure with the R400 grating. The exposures
were taken together in an approximately 1\,h continuous block including
overheads for target acquisition, readout and attached flatfield exposures.
We dithered by 50\,\AA\ between the B600 exposures to cover the spectral
gaps between the 3--CCD mosaic of GMOS.
The B600 grating tilt was set to cover down to a rest-wavelength of
$\approx 850$\,\AA\ and therefore depended on the quasar's emission
redshift. The R400 grating was tilted to cover the \ion{C}{4}
emission line of each quasar and to overlap the B600 spectrum at
approximately \lya\ emission.

To minimize slit losses from atmospheric dispersion, we designed
observations to be taken as close to parallactic as possible.
Almost all targets were placed near the centre of the slit to allow for
accurate sky subtraction (flexure) and to still approximately minimize
atmospheric dispersion when rotating the slit by 180\degr\ \citep{fil82}.
Typically, the two B600 exposures were taken with a 10\arcsec\ spatial
offset to mitigate CCD fringing at the reddest wavelengths after co-addition.
Occasionally, limitations on guide star availability meant that targets
could not be observed at parallactic angle or just with larger spatial
offsets. However, cross-checks with the available SDSS spectra revealed
that flux calibration was not compromised. We attribute part of this
success to target acquisition in a filter near the wavelength range of
interest (SDSS $r$).

Throughout our programme we obtained GMOS baseline calibrations.
One internal quartz halogen flat field exposure was attached to every
science exposure, while wavelength calibration spectra were taken during
daytime with a CuAr lamp. Night sky emission lines provided an approximate
flexure correction. As per standard Gemini operating procedure,
standard stars were observed at our various setups throughout each semester
under varying conditions that still allowed for relative flux calibration.
Bias frames were collected from the Gemini Science Archive.

More than half of the quasar spectra collected with GMOS-N in Semester 2010A
were affected by local persistence on the EEV CCDs due to the standard practice
to interleave GMOS science and flat field exposures in queue observations.
The persistence was stable and was corrected with dark exposures.
Spatial offsets ensured that the remaining 2010A spectra did not fall close to
affected pixels. In the following semesters GMOS-N science exposures were taken
before any flat field exposures. Our GMOS-S sample does not show persistence effects.

Our survey was carried out in varying observing conditions to maximize schedulability
and thus the final sample size. Our minimum acceptable conditions were:
(i) image quality 85 percentile (FWHM$\la 1.1\arcsec$),
(ii) sky background 50 percentile (dark--grey) and
(iii) sky transparency 70 percentile (patchy clouds or cirrus).
Eight of our targets received repeated observations due to clouds or bad seeing.
All of these exposures were usually co-added to increase the S/N.

All of the spectra were processed in identical fashion using two software packages
custom designed for Gemini/GMOS observations. Bias subtraction and flat fielding was
performed using the GMOS package (v1.9) distributed by Gemini within the
\textsc{iraf} software platform.
We found that the attached flat field frames showed constant fringing patterns
independent of the telescope pointing, so for any given setup we combined the
flat field frames to a high-quality master flat field and applied it to the data.

The remaining data reduction tasks were performed within the \textsc{lowredux} software
package\footnote{http://www.ucolick.org/\textasciitilde xavier/LowRedux/index.html}
developed by J. Hennawi, D. Schlegel, S. Burles, and J.~X. Prochaska.
Wavelength solutions were generated from the CuAr arc lamp spectra, yielding typical
RMS errors for the wavelength fits of $\sim 0.3$ pixels, corresponding to
$\sim 0.6$\,\AA\ and $\sim 0.8$\,\AA\ for the B600 and the R400 grating, respectively.
The accuracy of the wavelength solutions is limited by instrument flexure and the
almost critical sampling of the arc lines at the used spectral binning of four.
Objects were automatically identified in each of the three sub-slits of the GMOS
longslit and masked. Sky subtraction was performed on the remaining pixels.
A global solution was performed first for each sub-slit and then a local refinement
was made around each source in tandem with a spatial fit to the object profile. 
Each source was optimally extracted to produce a 1D spectrum, calibrated in wavelength.
A sky spectrum was also extracted and compared to an archived solution based on the
Paranal sky measurements \citep{hanuschik03} to estimate a rigid shift of the
wavelength solution due to instrument flexure. We then corrected these values to
vacuum and a heliocentric reference frame. The multiple exposures from the B600
grating were co-added, weighting by the S/N of the data
(nearly identical in most cases). For the few quasars that were observed on multiple
nights owing to variable observing conditions, we co-added all such data at this stage.

Standard stars observed throughout each semester were processed in identical
fashion except spectral extraction which was performed with a 100 pixel boxcar.
We compared these 1D extractions against catalogued
spectra\footnote{http://www.stsci.edu/hst/observatory/cdbs/calspec.html}$^,$\footnote{http://www.eso.org/sci/observing/tools/standards/spectra/stanlis.html}
to generate sensitivity functions that convert observed count rate to physical flux.
After re-scaling to correct for non-photometric conditions, sensitivity functions of
different standard stars taken throughout the survey showed an internal variation of
less than 5 per cent. Telluric absorption could not be corrected due to the varying
observing conditions and the lack of suitable standard stars.
We corrected for atmospheric absorption using the average extinction curve for
Mauna Kea \citep{beland88}. Slit losses were corrected by scaling the GMOS spectra
to the publicly available SDSS spectra of our targets in the regions where they
overlap. Assuming negligible quasar variability between the epochs of observation,
the resulting spectra have accurate absolute fluxes except at the ends of the covered
spectral range of GMOS where the fitting errors in the sensitivity curves exceed a
few per cent. The spectra collected at GMOS-S show fringing residuals at
$\lambda\ga 8300$\,\AA\ depending on the central wavelength.
Finally, we corrected the spectra for Galactic extinction using the \citet{ccm89}
extinction curve and the line-of-sight colour excess of \citet{sfd98}.

\begin{figure}
\includegraphics[bb=45 215 495 742,clip,width=1.0\linewidth]{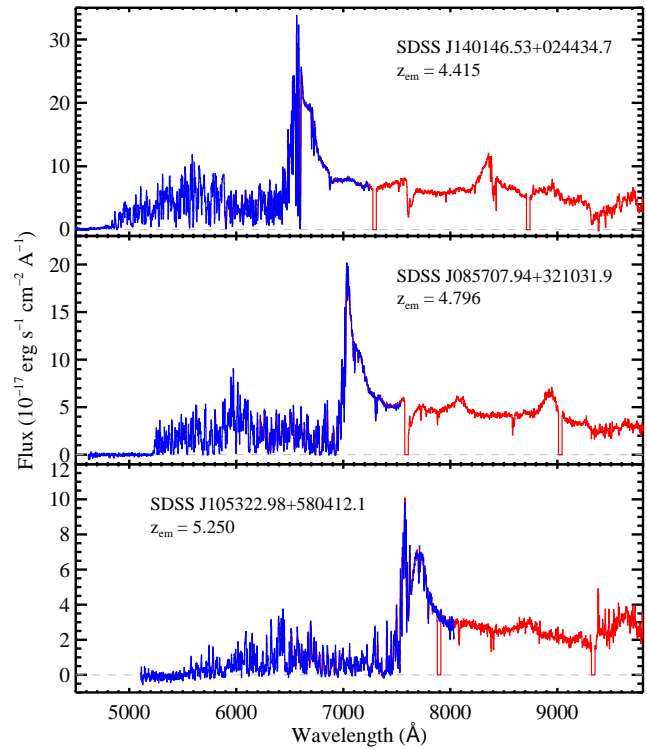}
\caption{\label{fig:ex_spec}
Gemini/GMOS spectra for three $z_\mathrm{em}>4.4$ quasars drawn from the
GGG survey. The blue lines indicate the data obtained with the B600
grating at a spectral resolution FWHM~$\approx 5.5$\,\AA. The red
lines, which overlap the blue data near \lya, trace the R400 grating
observations (FWHM~$\approx 8.0$\,\AA). Gaps between detectors give
the zero values in these red spectra. All spectra were fluxed using
several spectrophotometric standard stars and scaled to the available
SDSS spectra.
See the online edition of the Journal for a colour version of this figure.
}
\end{figure}

Figure~\ref{fig:ex_spec} shows the 1D spectra for three representative quasars
from our GGG survey. Plots of all \nqso\ quasar spectra are shown in
Appendix~\ref{app:gggspec}, available in the online edition of the Journal.
All spectra are publicly available in reduced
form\footnote{Link to VizieR will be provided upon acceptance of the manuscript.}.
The typical, absorbed S/N within the \lya\ forest is $\sim 20$ per
$1.85$\,\AA\ pixel, $\sim 7$ times higher than in the SDSS discovery spectra.
Due to the shorter exposure time with the R400 grating, the S/N at
$\lambda_\mathrm{r}=1450$\,\AA\ is generally lower (Fig.~\ref{fig:demographics}),
but still sufficient to accurately normalize the spectra.

\begin{figure}
\includegraphics[width=1.0\linewidth]{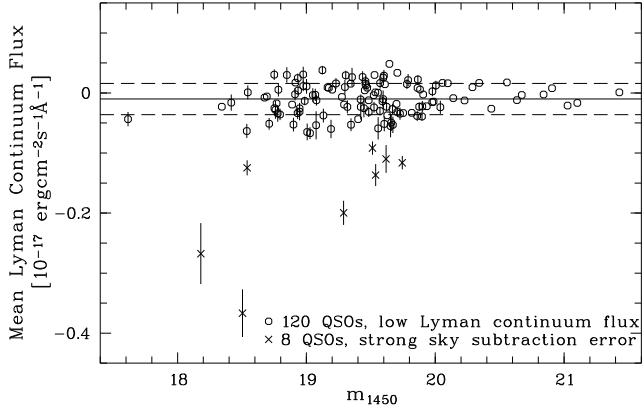}
\caption{\label{fig:skysub}
Mean Lyman continuum flux of the 128 quasars of our sample with a likely
optically thick LLS vs.\ their apparent AB magnitude at
$\lambda_\mathrm{r}=1450$\,\AA. Error bars are the standard error of the
mean flux. Error bars smaller than the symbol size are not plotted.
Crosses indicate the 8 quasars affected by systematic sky subtraction errors
(significantly negative flux). The solid and dashed lines mark the average
($-1\times 10^{-19}$\,erg\,cm$^{-2}$\,s$^{-1}$\,\AA$^{-1}$) and the standard
deviation ($2.6\times 10^{-19}$\,erg\,cm$^{-2}$\,s$^{-1}$\,\AA$^{-1}$),
of the 120 remaining measurements, respectively.
}
\end{figure}

The precise measurement of the mean free path at these high redshifts requires
very accurate sky subtraction at the bluest wavelengths, i.e.\ below the Lyman
limit of the quasar. Figure~\ref{fig:skysub} shows the mean Lyman continuum flux
of the 128 quasars in our sample with a likely optically thick LLS in the covered
wavelength range, as determined by visual inspection of the 1D and 2D spectra.
Eight quasars in our sample show significantly negative flux due to a systematic
overestimation of the sky background, the level of which increases with quasar
continuum flux. As the majority of these affected targets were taken in bad seeing,
we attribute this effect to an underestimation of the object spatial profile during
extraction, with a few per cent of quasar flux leaking into the sky subtraction windows.
After excluding these outliers, we still estimate a slightly negative average flux
($-1\times 10^{-19}$\,erg\,cm$^{-2}$\,s$^{-1}$\,\AA$^{-1}$) with considerable dispersion
($2.6\times 10^{-19}$\,erg\,cm$^{-2}$\,s$^{-1}$\,\AA$^{-1}$).
The residuals are generally less than 1 per cent of the quasar continuum flux
redward of \lya. We include the systematic sky subtraction error in our analysis,
but it does not significantly affect our results.

\section{Redshift Determination}
\label{sec:zem}

\begin{figure}
\centering
\includegraphics[bb=53 100 519 735,clip,width=0.75\linewidth]{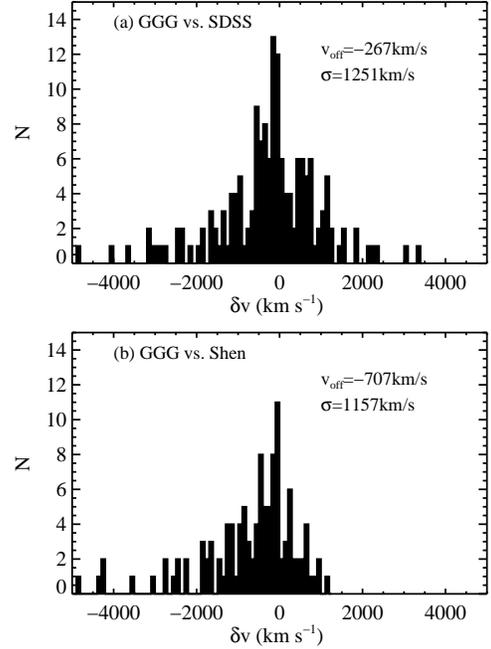}
\caption{\label{fig:zem}
Comparison of the GGG quasar redshifts to other redshift estimates of our targets
(see text). The tail to negative velocities in the lower panel arises because
these quasars have very weak emission lines (Fig.~\ref{fig:zoff}) and, therefore,
have redshifts best estimated from the onset of the \lya\ forest.
}
\end{figure}

\begin{figure}
\centering
\includegraphics[bb=47 180 488 742,clip,width=0.95\linewidth]{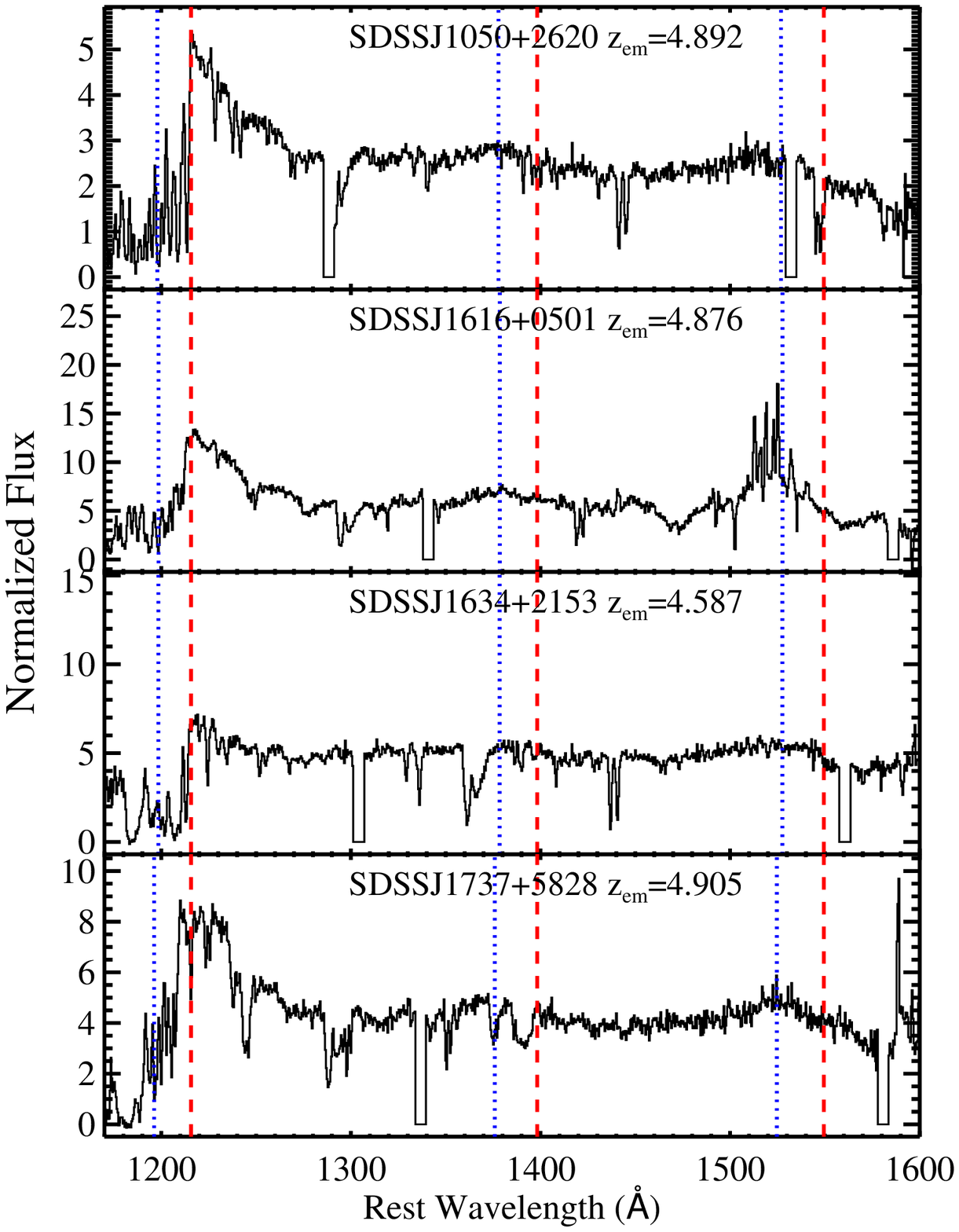}
\caption{\label{fig:zoff}
  Gemini/GMOS R400 spectra of four quasars which have large
  differences in $z_{\rm em}$ from our analysis (which sets the rest
  wavelength here) and from using the \citet{shen07} algorithm.  The
  vertical dashed lines indicate the rest wavelengths of
  \ion{H}{1} \lya, \ion{Si}{4}, and \ion{C}{4}.  The dotted lines show
  the positions for these emission lines when adopting the Shen
  estimate for $z_{\rm em}$.  The large offset occurs because
  these sources have very weak emission lines.  Indeed, our preferred
  values come from analysis of the onset of the \lya\ forest.  These
  values are adopted in the following mean free path analysis.
}
\end{figure}

Our science goals are dependent on precise estimations of the quasar emission
redshifts, especially analysis of the mean free path. It is now well
appreciated that the standard approach taken by the SDSS to automatically
estimate quasar redshifts gives values that are systematically in error
\citep{richards02,hw11}. For this reason, several groups have developed
algorithms to re-measure the values from resonant and fine-structure lines in
the rest-frame far-UV \citep{shen07,hw11}.

To fully explore the aspects of this challenging problem, we have analysed the
GGG spectra with several techniques and compare the results with estimates
from the literature. Our first method was to estimate \zem\ `by-eye', assuming
that the peaks of suitable UV emission lines are located at their laboratory
wavelengths. We focused especially on low-ionization emission lines
(\ion{O}{1}+\ion{Si}{2}\,$\lambda=1303.49$\,\AA\ and
\ion{C}{2}\,$\lambda=1335.30$\,\AA),
which are believed to have small offsets from systemic \citep{TF92,vanden01}.
We refer to this approach as `GGG'. Redshift errors were estimated based on the
emission line used, the presence of associated absorption (either BAL,
narrow associated or atmospheric), the onset of the \ion{H}{1} \lya\ forest, 
and the overall consistency between the suite of emission lines.
The estimated uncertainty for these redshift measurements ranges from $\sim 250$
to $\sim 1000$~\kms. Table~\ref{tab:ggg_zem} summarizes the results of this method.

Secondly, we applied the semi-automated algorithm of \citet{shen07} to estimate
\zem. This routine measures the centroids of each $5\sigma$-detected UV-emission
line from the following: \ion{H}{1} \lya, \ion{Si}{4}, \ion{C}{4}, and \ion{C}{3}].   
The algorithm then estimates \zem\ based on average velocity offsets measured
from systemic, as gauged from [\ion{O}{3}] nebular emission in lower redshift
quasars. Each fit was visually inspected and modifications were occasionally
made to the analysis (e.g.\ the elimination of a poorly-fit emission line).
The lines analysed and the \zem\ estimates and uncertainties are listed in
Table~\ref{tab:ggg_zem}. We refer to this technique as `Shen'. Lastly, we list
the best estimates from \citet{hw11} who applied an algorithm similar to that
of \citet{shen07} to SDSS Data Release 6. We refer to those measurements as `HW'.

Figure~\ref{fig:zem}a shows a histogram of the offsets in redshift between the
GGG--\zem\ measurements and the values reported in \citet{schneider+10}.
There is a considerable scatter between the two sets of measurements
($\sigma_v = 1250\,\mkms$ or $\sigma_z \approx 0.025$), but on average the GGG
redshifts are only slightly higher than the SDSS redshifts
(mean offset $v_\mathrm{off}\approx -270\,\mkms$). In contrast, a comparison of
the GGG--\zem\ measurements with the Shen--\zem\ values shows an
$\approx -700\,\mkms$ offset driven by an asymmetric tail to negative velocities
(Figure~\ref{fig:zem}b). A comparison to the \citet{hw11} redshifts reveals a
similar tail. In Figure~\ref{fig:zoff} we show the four spectra with
$\delta v < -4000\,\mkms$ between the GGG--\zem\ and Shen--\zem\ evaluations.
Each case is marked by very weak \ion{Si}{4} and \ion{C}{4} emission as well as
strong \ion{H}{1} absorption just blueward of rest-frame \lya. These cases have
had their GGG--\zem\ values estimated from the onset of  \lya\ forest absorption,
and we strongly prefer these values. After excluding outliers at
$\delta v < -2000\,\mkms$, the GGG vs.\ Shen--\zem\ velocity offset distribution
is approximately symmetric with a mean offset $v_\mathrm{off}\approx -380\,\mkms$
and standard deviation $\sigma_v\approx 700\,\mkms$.
As almost all Shen--\zem\ values are based on \ion{Si}{4} and \ion{C}{4}
(Table~\ref{tab:ggg_zem}), with the former being blended with \ion{O}{4}] and
the latter showing a blueshift that increases with luminosity \citep{richards11},
we consider the GGG--\zem\ values more reliable for our sample of very luminous
quasars (Fig.~\ref{fig:demographics}).

In the following analysis we adopt a Gaussian redshift error distribution with
zero mean and standard deviation $\sigma_z=0.01$
(corresponding to $\sigma_v\simeq 500\,\mkms$ at $z\simeq 4.8$) for the
entire sample. We show below that redshift errors do not significantly
affect our results.

\section{The Mean Free Path at $\bmath{z \sim 5}$}
\label{sec:results}

\subsection{Stacked rest-frame quasar spectra}
\label{sec:stacks}

Our technique for estimating the mean free path is to analyse the average flux
of a cohort of quasars at similar redshift and at rest wavelengths shortward of
the Lyman limit. The decrease in observed flux is attributed to the integrated,
average LL opacity of the universe and \lmfp\ is defined to be the average
distance from the source where a packet of photons suffers an e$^{-1}$ attenuation.
The next subsection describes the formalism and modelling in greater detail.

Central to the analysis is the generation of average (`stacked') rest-frame quasar
spectra which describe the mean flux of the background sources, as attenuated by
intergalactic hydrogen blueward of their \lya\ emission line. Because we are
interested foremost in the average properties of the \ion{H}{1} absorption, each
sight line is given equal weighting. Our parent sample is the \nqso\ quasars
observed in the GGG survey (Table~\ref{tab:ggg_sample}). Of these, 9 sources were
excluded from any further analysis due to strong BAL features, many of which were
not obviously apparent in the SDSS discovery spectra. The spectrum of one source,
SDSS~J120102.01$+$073648.1, is contaminated by flux from a neighbouring source,
presumably foreground to the quasar. An additional 8 sources were excluded due
to systematic sky subtraction errors (Section~\ref{sec:obs}). Therefore,
the final sample for generating stacked rest-frame spectra
totals \nstack\ quasars with $\mzem>4.4$.

Following the algorithm described in \citet{omeara13}, we have generated stacked
rest-frame quasar spectra in three redshift intervals designed to have roughly
equal numbers of quasars: $\mzem = \left[4.4, 4.7\right]$, $\left[4.7, 5.0\right]$,
and $\left[5.0, 5.5\right]$ with average redshifts $z_\mathrm{q} = 4.56$, $4.86$,
and $5.16$ (the median $z_\mathrm{q}$ values are similar).
The GMOS/B600 spectrum of each quasar was shifted to the rest-frame, normalized
to have unit flux at rest-wavelength $\lambda_\mathrm{r} = 1450$\,\AA, and then
binned on to a fixed wavelength grid with a constant dispersion of $0.45$\,\AA\
per pixel. This dispersion is sufficiently large to contain at least one pixel
from each of the original spectra; the mean is adopted when two or more of the
original pixels fall within a given pixel of the new grid. By taking a straight
average of all the processed quasar spectra in each redshift interval,
we generate a stacked spectrum that weights each sight line equally. Weighting by
S/N instead would introduce a bias towards sight lines without strong LL absorption.

\begin{figure}
\centering
\includegraphics[angle=90,width=0.98\linewidth]{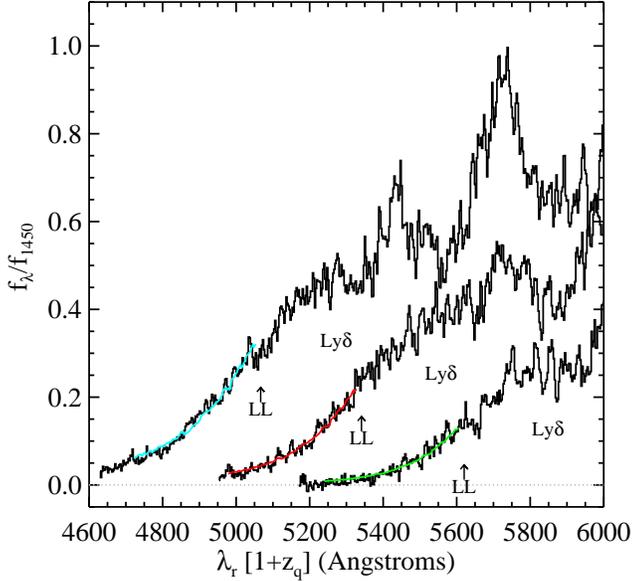}
\caption{\label{fig:stacks}
Stacked normalized rest-frame quasar spectra from the GGG survey generated for
three redshift intervals: $z_\mathrm{em} = \left[4.4, 4.7\right]$,
$\left[4.7, 5.0\right]$, and $\left[5.0, 5.5\right]$. These spectra are plotted
in a pseudo-observer frame defined as $\lambda_\mathrm{r}\left(1+\mzq\right)$
with, \zq\ the average redshift of the quasars in each interval. The Ly$\delta$
emission (strongly affected by IGM absorption for the two high-$z$ bins) and the
onset of the Lyman limit are marked for each stacked spectrum. Ly$\beta$ and
Ly$\gamma$ emission lines of the background quasars are clearly visible.
Overplotted on these stacked spectra are the best-fitting models which provide 
measurements for the mean free path \lmfp.
}
\end{figure}

Figure~\ref{fig:stacks} shows the stacked spectra, plotted in a pseudo-observer
frame $\lambda_\mathrm{pseudo}=(1+z_\mathrm{q})\lambda_\mathrm{r}$ for clarity
of presentation. Each spectrum shows the rest-frame quasar continuum with
readily visible Ly$\beta$ and Ly$\gamma$ emission after being absorbed by the
IGM with an effective Lyman series optical depth \tlyman\ and additional Lyman
continuum optical depth \teff\ at $\lambda_\mathrm{r} < 912$\,\AA\
\citep[e.g.][]{mad95,meiksin06,wp11}. At $\lambda_\mathrm{r} \approx 900$\,\AA,
we measure the scatter in the stack relative to a median smoothed version of
$\approx 7$ (30) per cent  in the lowest (highest) redshift interval.
This scatter arises from stochasticity in the IGM, not Poisson noise in the
individual spectra.

\begin{figure}
\includegraphics[bb=54 212 518 742,clip,width=1.0\linewidth]{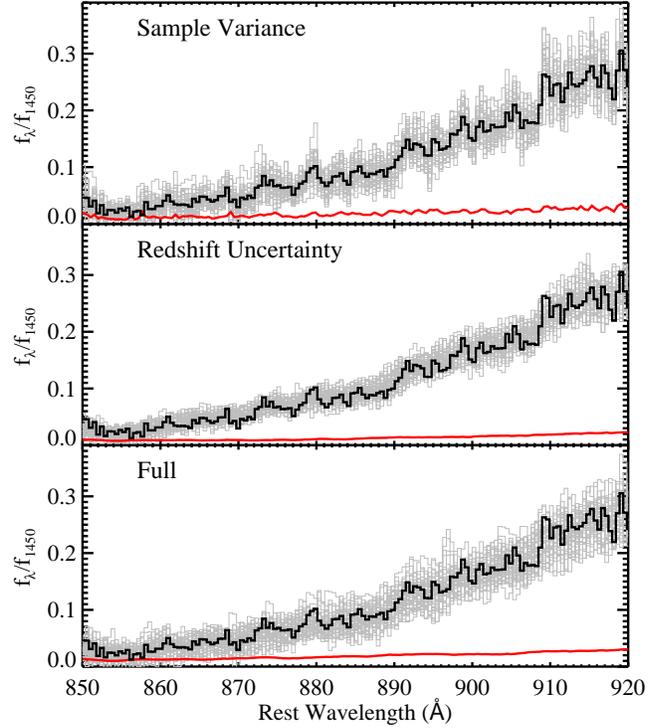}
\caption{\label{fig:RMS}
Bootstrap realizations of the rest-frame stacked spectrum for the
$z_\mathrm{em}=\left[4.7,5.0\right]$ cohort of quasars from the GGG survey.
The top panel shows the stacks when one allows for duplications, i.e.\ it explores
the effects of sample variance. The middle panel only allows for random offsets
in $z_\mathrm{em}$ for each quasar. The lowest panel shows the combined effect
of redshift error and sample variance. In all panels, the thick black curve
shows the actual stacked spectrum, the grey curves show individual bootstrap
realizations, and the lowest curve indicates the RMS of the bootstrap realizations
as a function of wavelength. We find that the redshift uncertainty has a small
effect, especially in comparison to that for sample variance.
}
\end{figure}

To assess uncertainties in the measurements that follow, we have generated three
sets of \nboot\ stacked spectra in each redshift interval. We start by
estimating sample variance by randomly sampling the quasars, allowing for
duplications. The upper panel of Fig.~\ref{fig:RMS} presents the full cohort of
spectra for the $\mzem=\left[4.4,4.7\right]$ interval. We measure an RMS per
pixel that ranges from $\approx 10$ per cent of the flux at 912\,\AA\ to $\ga 30$
per cent at 850\,\AA. These exceed the pixel-to-pixel scatter in the stacked
spectrum. To assess the effects of redshift uncertainty, we generated a set of
500 stacked spectra where \zem\ of each quasar was randomly offset from its
measured value by a Gaussian deviate with $\sigma_z = 0.01$. The middle panel of
Fig.~\ref{fig:RMS} reveals that redshift error is a relatively minor source of
uncertainty, especially compared to sample variance. Still, we have included the
combined uncertainty of sample variance and redshift error by generating a set of
\nboot\ stacked spectra by randomly sampling the quasars and varying \zem\ at the
same time (Fig.~\ref{fig:RMS}; lower panel).

%%%%%%%%%%%%%%%%%%%%%%%%%%%%%%%%%%%%
\subsection{Mean free path analysis and results}
\label{sec:analysis}

Using the stacked quasar spectra (Fig.~\ref{fig:stacks}), an estimation of the
mean free path to ionizing photons is made by modelling the flux below the Lyman
limit. This technique was developed and applied to $z<4$ quasars in previous
works \citep{pwo09,omeara13,fop+13}. Here, we refine the formalism to best match
the nature of the IGM at $z\sim 5$.

We model the intrinsic flux of the quasar stack using the radio-quiet quasar SED
of \citet{telfer02} modulated by a power-law and allowing for uncertainty in the
normalization at 912\,\AA. We find that apart from an obvious Baldwin effect in
the \ion{C}{4} emission line \citep[e.g.][]{baldwin77,richards11}, this is a
very good match to the data redward of the \lya\ forest.
At $\lambda_\mathrm{r}<1215$\,\AA, the flux is attenuated by the Lyman series
opacity beginning with \lya\ (i.e.\ an effective optical depth \tlya).
Below 912\,\AA, the flux is modulated by the full Lyman series effective optical
depth \tlyman\ and the Lyman limit effective optical depth \teff\
\citep[e.g.][]{mad95,meiksin06,wp11}. Explicitly, we may express the observed
flux at these wavelengths as
\begin{equation}
f^\mathrm{obs}_{\lambda} = f^\mathrm{SED}_{\lambda} \exp\left(-\mtlyman\right)\,\exp\left(-\mteff\right)\;\;\; ,
\end{equation}
where both \tlyman\ and \teff\ depend on redshift (and therefore wavelength; see below). 

In practice, we model the observed flux below the Lyman limit relative to the
observed flux at 912\,\AA\ (measured from the stacked spectrum) as
\begin{eqnarray}\label{eqn:model}
f^\mathrm{obs}_{\lambda<912} &=& f^\mathrm{obs}_{912}\left(\frac{C_{912} f^\mathrm{SED}_\lambda}{f^\mathrm{SED}_{912}}\right)\\\nonumber
& &\times\exp\left(-\left[\tau_{\mathrm{eff},\lambda}^\mathrm{Lyman}-\tau_{\mathrm{eff},912}^\mathrm{Lyman}\right]\right)
\exp\left(-\tau_\mathrm{eff}^\mathrm{LL}\right)\;\;\; ,
\end{eqnarray}
with each of these quantities defined below.

Altogether, the model described by Equation~\ref{eqn:model} has four model
parameters:
(i) a nuisance parameter $C_{912}$ for the overall normalization of the model.
This accounts for uncertainty in evaluating $f_{912}^\mathrm{obs}$ from the
stacked spectrum; 
(ii) a power-law tilt \ased\ applied to the assumed \citet{telfer02} SED,
normalized at 1450\,\AA
\begin{equation}
f_\lambda^\mathrm{SED} = \frac{f_\lambda^\mathrm{Telfer}}{f_{1450}^\mathrm{Telfer}}
\left(\frac{\lambda}{1450\,\rm \AA}\right)^{\mased}
 \;\;\; ;
\end{equation}
(iii) an exponent \glym\ which determines the redshift evolution of the effective
optical depth from Lyman series absorption, 
\begin{equation}
\tau_{\mathrm{eff},\lambda}^\mathrm{Lyman} = \tau_{\mathrm{eff},912}^\mathrm{Lyman}
\left(\frac{1+z}{1+z_{912}}\right)^{\mglym} \;\;\; ,
\end{equation}
where $z_{912} \equiv \lambda_\mathrm{r}\left(1+z_\mathrm{q}\right)/\left(911.7621\mathrm{\AA}\right) - 1$.
In practice, $\tau_{\mathrm{eff},912}^\mathrm{Lyman}$ is set to match the
observed flux at 912\,\AA\ given the SED, i.e.\  
$\tau_{\mathrm{eff},912}^\mathrm{Lyman} = \ln
\left(f_{912}^\mathrm{SED}/f_{912}^\mathrm{obs}\right)$;
(iv) an opacity \kll\ describing the effective LL optical depth 
\begin{eqnarray}\label{eqn:teff}
\mteff(z_{912},z_\mathrm{q}) & = &\mkll\ \frac{c}{H_0}\left(1+z_{912}\right)^{2.75}\\\nonumber
& &\times\intl_{z_{912}}^{z_\mathrm{q}} \left(1+z'\right)^{-5.25} \, \mathrm{d}z' \;\;\; .
\end{eqnarray}
The exponents in Equation~\ref{eqn:teff} are set by cosmology and an adopted
$\nu^{-2.75}$ dependence for the photoionization cross-section
\citep[see][]{pwo09,omeara13}. Unlike previous works, we do not fit for
redshift evolution in \kll\ because we find it to be highly degenerate with the
normalization when $\mlmfp \ll 100 \mhmpc$.

It is evident from Equation~\ref{eqn:model} that \fobs\ at wavelengths just
below 912\,\AA\ can change rapidly only if \teff\ is significant. Furthermore,
unless one adopts an extreme tilt for the SED that is ruled out by observations
($\mased > 0.5$), the LL opacity is the only factor which actually lowers \fobs.
Therefore, the sharp decline in flux observed in the stacked spectra at
$\lambda_\mathrm{r} < 912$\,\AA\ (Fig.~\ref{fig:stacks}) must be driven by \teff,
such that the model is most sensitive to \kll. In fact, we find substantial
degeneracy between the models if we attempt to constrain anything other then the
normalization term $C_{912}$ and \kll. Therefore, we solved for these two model
parameters and explore the dependency of the results on the other factors.
For our fiducial models, we set $\mased=0$ and $\mglym = 2.5$, where the latter
is motivated by observed redshift evolution in the Lyman series effective
optical depth of the IGM \citep{bhw+13,pmof14}. The model comparison to the data
is performed at wavelengths $\lambda_\mathrm{r} = 850 - 910$\AA\ where $\chi^2$
is minimized assuming $\sigma_\lambda = 0.02$ which is characteristic of the
scatter in the stacked spectrum. We caution that the stacked spectral fluxes are
not truly independent and therefore values of $\chi^2$ should not be interpreted
in the standard fashion.

\begin{figure}
\centering
\includegraphics[bb=48 217 438 742,clip,width=0.95\linewidth]{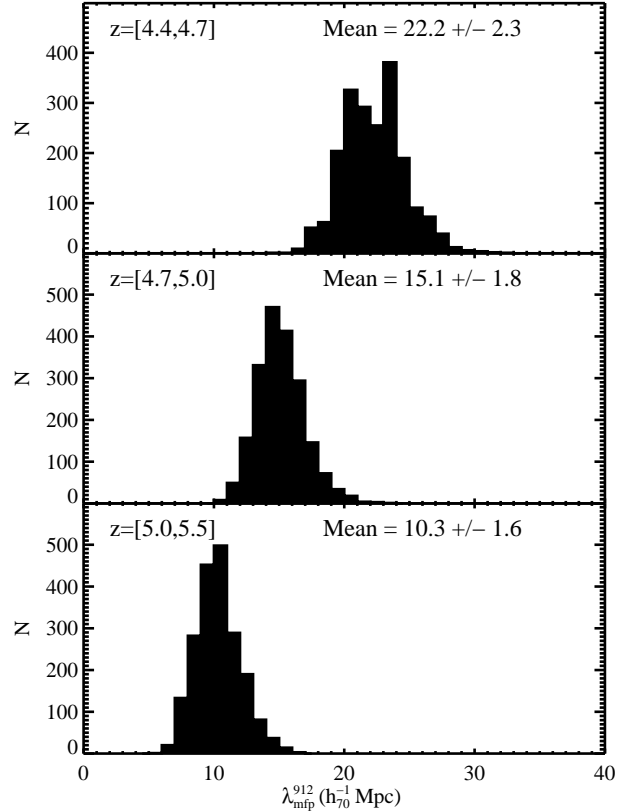}
\caption{\label{fig:boot}
Distribution of \lmfp\ values measured from \nboot\ bootstrap realizations of
the stacked spectra. The scatter is dominated by sample variance, e.g.\
fluctuations in the number of strong \ion{H}{1} absorbers in the stacks. Each
distribution is well-modelled by a Gaussian although we note the presence of a
small tail to larger \lmfp\ values.
}
\end{figure}

Figure~\ref{fig:stacks} presents the best-fitting models for each stacked
spectrum, which provide a good description of the observations
($\chi^2_\nu \la 1$). From these models, we assess the effective redshift \ozll\
where $\mteff = 1$ (Equation~\ref{eqn:teff}) and then measure \lmfp\ as the
proper separation between \ozll\ and \zq\ with our assumed cosmology
(Table~\ref{tab:mfp}). Uncertainties in the \lmfp\ values were estimated from the
\nboot\ stacked spectra generated with bootstrap techniques
($\S$~\ref{sec:stacks}; Fig.~\ref{fig:RMS}).  These include the combined effects
of redshift error and sample variance. Figure~\ref{fig:boot} presents the results
of this analysis. The distributions of \lmfp\ values are roughly Gaussian and we
adopt the measured RMS as the statistical error in the \lmfp\ values. These are
10--15 per cent of the central values.

\begin{figure}
\centering
\includegraphics[bb=45 104 463 737,clip,width=0.95\linewidth]{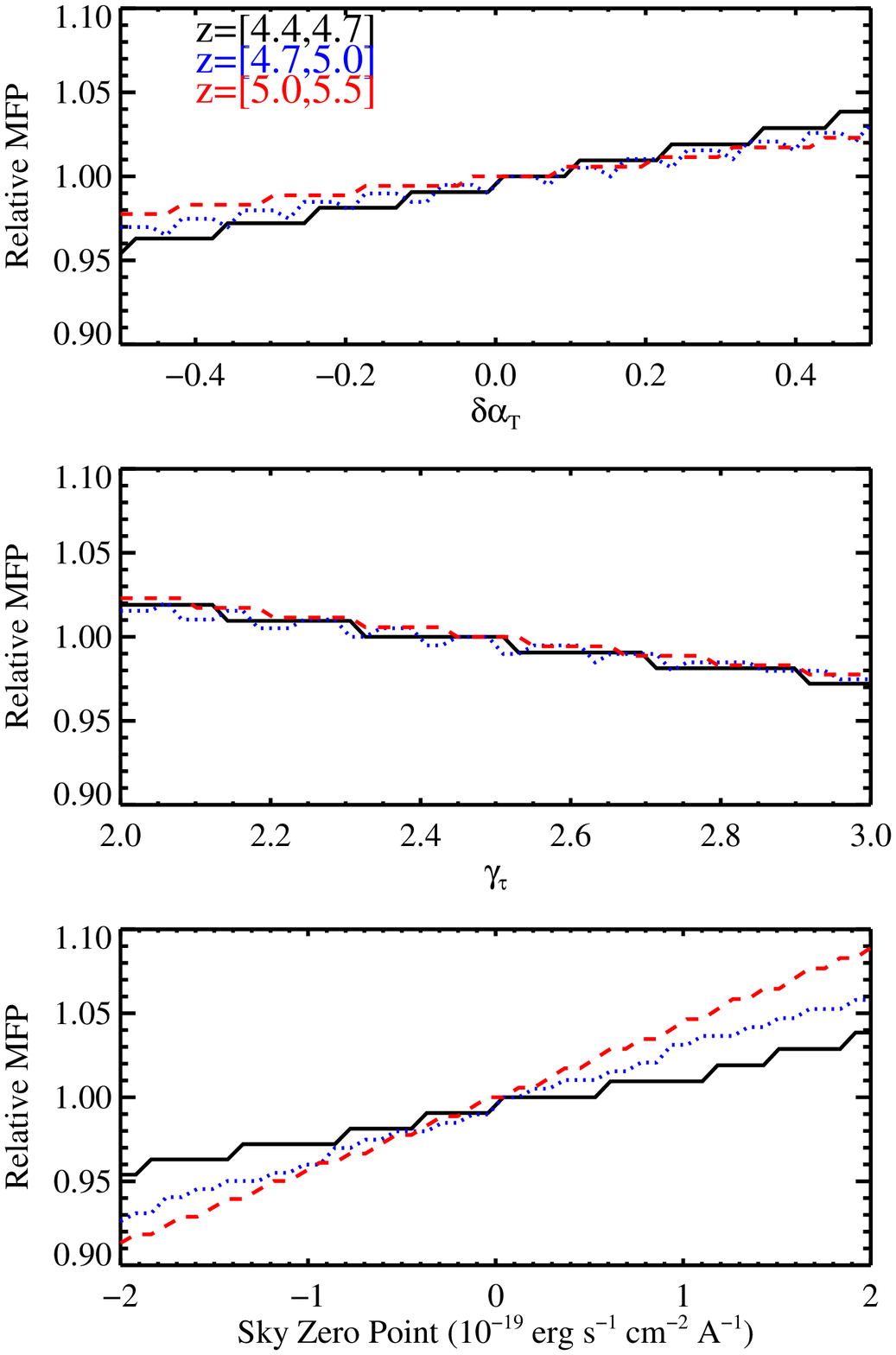}
\caption{\label{fig:explore}
Dependence of \lmfp\ on fixed model parameters. The panels plot the \lmfp\ value
measured relative to our fiducial estimate as one varies the tilt in the
intrinsic quasar SED (\ased; upper panel), the redshift evolution in the Lyman
series effective optical depth (\glym; middle panel), or the sky zero point
(lower panel). Our three subsamples are indicated by line style and colour
($z_\mathrm{em}=\left[4.4,4.7\right]$: black solid; 
$z_\mathrm{em}=\left[4.7,5.0\right]$: blue dotted;
$z_\mathrm{em}=\left[5.0,5.5\right]$: red dashed).
For a relatively broad range in \ased\ and \glym\ we find a $\la 5$ per cent
systematic dependence. Correction of the sky level zero point
(Fig.~\ref{fig:skysub}) increases the \lmfp\ values obtained from the
$z_\mathrm{em}=\left[4.7,5.0\right]$ and $z_\mathrm{em}=\left[5.0,5.5\right]$
cohorts by $\simeq 5$ per cent. See the online edition of the Journal for a
colour version of this figure.
}
\end{figure}

There are at least four sources of systematic uncertainty in our models which
affect the resultant \lmfp\ values. Two of these relate to using models with
assumed values for \ased\ and \glym. The first two panels of
Fig.~\ref{fig:explore} show the explicit dependence of the \lmfp\ values
(relative to the fiducial value) when \ased\ and \glym\ are varied.
The behaviour is as one expects, an increase (decrease) in \ased\ (\glym) implies
lower flux for the models prior to including LL attenuation resulting in smaller
\teff\ values (and larger \lmfp). The variations in \lmfp, however, are small;
for a plausible range of \ased\ and \glym\ values there is a less than 5 per
cent effect. We conclude that these two systematic errors are insignificant in
comparison with sample variance.

In addition, we estimated the systematic error incurred due to the
over-subtraction of the sky background level in our data set (Fig.~\ref{fig:skysub}).
The lowest panel in Fig.~\ref{fig:explore} shows that a typical upward correction
of $10^{-19}$\,erg\,cm$^{-2}$\,s$^{-1}$\,\AA$^{-1}$ results in a $\simeq 5$ per
cent larger \lmfp\ in the two higher redshift bins. This is simply due to the
low flux at the end of the spectral range used for the fit of
\lmfp\ (Fig.~\ref{fig:stacks}).

Another source of systematic uncertainty relates to detailed fluctuations in the
adopted SED, i.e. on $\sim 5$\,\AA\ scales. These presumably arise from
unresolved emission lines and also noise in the spectra analysed by
\citet{telfer02}. There is little reason to expect that this SED captures the
true flux modulations in $z_\mathrm{em} \sim 5$ quasars. To explore the effect
of small-scale variations in the SED, we repeated our analysis allowing for 10
per cent fluctuations in the Telfer SED on 5\,\AA\ scales using a Gaussian deviate.
From 500 trials in each composite we find a 10 per cent effect, comparable to
the uncertainty from sample variance.  In summation, we estimate that the
magnitude of systematic uncertainty is comparable to the $\approx 15$ per cent
statistical error associated with sample variance and redshift error.
This implies that future surveys would need to address these systematic effects
to substantially improve upon our measurements.

\section{Discussion}
\label{sec:discuss}

\subsection{Redshift evolution in the mean free path}
\label{sec:mfpzevol}

Studies of the IGM across cosmic time have revealed redshift evolution in many
properties of the \lya\ forest: temperature \citep[e.g.][]{lidz10,becker11},
line density \citep[e.g.][]{kim13}, and the flux PDF \citep[e.g.][]{brs07,kbv+07}.
Perhaps the best measured quantity has been the effective opacity of \ion{H}{1}
\lya\, \tlya, which numerous authors have found to decrease rapidly from $z=4$
to 2 \citep{kts+05,fpl+08,dww08,paris11}. A recent parametrization of the
redshift evolution finds
$\mtlya\left(z\right) = 0.751\left[\left(1+z\right)/4.5\right]^{2.9} - 0.132$
\citep{bhw+13}. The steep evolution in \tlya\ is attributed to the expansion of
the universe, an increase in the comoving number density of ionizing sources
(quasars), and the decrease in the mean density of the gas which implies a lower
hydrogen neutral fraction \citep[e.g.][]{bolton07,dave+10}. Several studies have
also traced redshift evolution in the incidence of strong \ion{H}{1} systems
$\ell(z)$, e.g.\ damped \lya\ systems (DLAs) and LLSs, which are expected to
trace non-linear and collapsed structures in the universe. Their incidences also
decline rapidly towards lower redshift with $\ell(z)$ scaling roughly as
$\left(1+z\right)^\eta$ for $1<\eta<3$
\citep{phw05,pow10,rtn06,songaila10,ribaudo11}. This exceeds the evolution
attributable to cosmic expansion alone and implies a reduction in the filling
factor of cool, dense gas in a given comoving volume. Including the results
presented here, we have now measured the mean free path with the stacked spectrum
technique from $z \approx 2.5$ to $\approx 5.2$ \citep{pwo09,omeara13,fop+13}.  
By exploring its redshift evolution, we may gain insight into the cosmological
distribution of gas dominating the \ion{H}{1} LL
opacity\footnote{Note that this likely includes gas that is both optically thick
at the Lyman limit (e.g.\ LLSs, DLAs) and gas that has $\mtll < 1$.},
its interplay with galaxies, and the formation/consumption of \ion{H}{1} gas on
cosmic scales. In turn, these results inform model predictions for evolution
in the extragalactic UV background \citep{flz+09,hm12,bb13}.

\begin{figure}
\includegraphics[angle=90,width=1.0\linewidth]{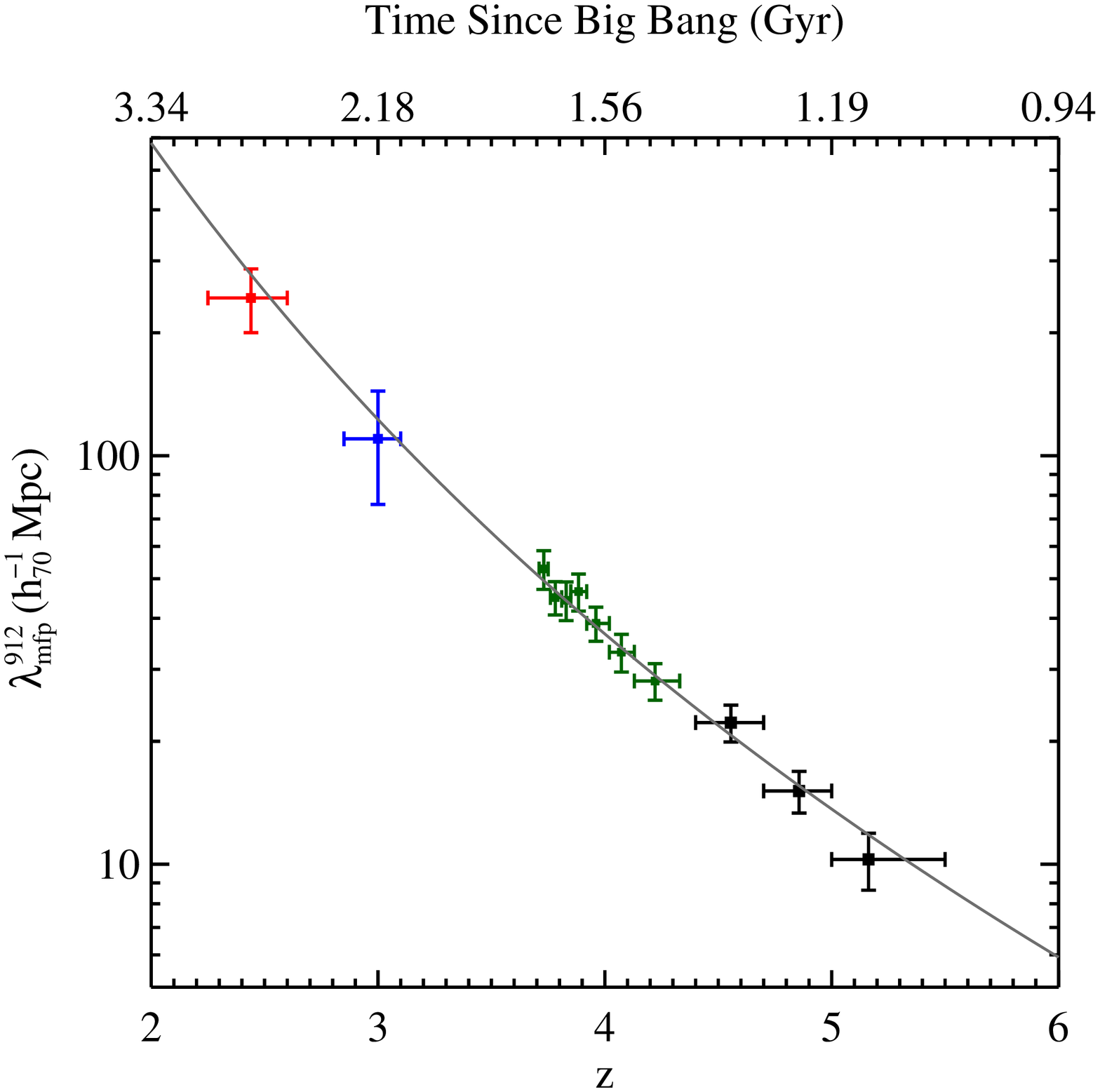}
\caption{\label{fig:mfpvsz}
The proper mean free path to Lyman limit photons in the intergalactic medium as
a function of redshift (and cosmic time). The data points show direct
measurements via the spectral stacking technique as estimated in this manuscript
(black), \citet[][green]{pwo09}, \citet[][blue]{fop+13}, and
\citet[][red]{omeara13}. One observes a monotonic decrease with increasing
redshift which is well modelled by a $\left(1+z\right)^\eta$ power-law with
$\eta = \gval$ (curve). See the online edition of the Journal for a colour
version of this figure.  
}
\end{figure}

\begin{table*}
\caption{\label{tab:mfp}\lmfp\ Measurements.}
\begin{tabular*}{0.80\linewidth}{@{\extracolsep{\fill}}ccccll}
\hline
$z_\mathrm{q}$ & $N_\mathrm{QSOs}$ & \lmfp & $\sigma\left(\mlmfp\right)^\star$ & Reference & Notes\\
               &  & $\left[h_{70}^{-1}\,\mathrm{Mpc}\right]$ & $\left[h_{70}^{-1}\,\mathrm{Mpc}\right]$ & &\\
\hline
\multicolumn{6}{c}{Direct Measurements}\\
\hline
2.44&53 &235.8&40.3&\cite{omeara13} &\\
3.00&61&110.0&34.0&\cite{fop+13} & Non-colour selected\\
3.73&150 &52.8&5.7&\cite{pwo09} &\\
3.78&150 &45.0&4.2&\cite{pwo09} &\\
3.83&150 &44.3&4.8&\cite{pwo09} &\\
3.88&150 &46.5&4.8&\cite{pwo09} &\\
3.96&150 &38.9&3.7&\cite{pwo09} &\\
4.07&150 &33.0&3.5&\cite{pwo09} &\\
4.22&150 &28.1&2.9&\cite{pwo09} &\\
4.56& 57&22.2&2.3&This paper &\\
4.86& 49&15.1&1.8&This paper &\\
5.16& 39&10.3&1.6&This paper &\\
\hline
\multicolumn{6}{c}{Indirect Estimates$^\dag$}\\
\hline
3.0 & $\sim 100$ & 85 & 65 &\cite{flh+08} & LLSs and $\beta=1.39, \gamma=1.5$ \\
3.5 & $\sim 100$ & 49 & 20 &\cite{songaila10} & LLSs and $\beta=1.3, \gamma=1.94$ \\
4.5 & $\sim 60$ & 20 &  8 &\cite{songaila10} & LLSs and $\beta=1.3, \gamma=1.94$ \\
\hline
\end{tabular*}
\begin{flushleft}
Notes: All of the estimates have been translated to a common cosmology
($\Omega_\Lambda = 0.7, \Omega_\mathrm{m} = 0.3$ with $H_0 = 70\,\mathrm{km}\,\mathrm{s}^{-1}\,\mathrm{Mpc}^{-1}$).\\
$^\star$For the direct measurements, these are estimated from the RMS of 2000
bootstrap realizations of the stacked spectra.\\
$^\dag$All of the indirect estimates are based on a measured incidence of LLSs
combined with assumptions on the \ion{H}{1} frequency distribution.  The values
listed are given at the redshift where the values were best estimated.
\end{flushleft}
\end{table*}

The complete set of \lmfp\ values measured with our technique are presented in
Fig.~\ref{fig:mfpvsz} and are listed in Table~\ref{tab:mfp}, each converted to
the cosmology used in this manuscript\footnote{Note that for the SDSS
measurements of \citet{pwo09} we have modified the analysis to conform to the
mean free path definition used here. This is a minor modification.}.   
It is evident that \lmfp\ increases by over one order of magnitude from $z=5$ to
2.5. This must be driven in large part by the expansion of the universe.
Therefore, one is motivated to model the redshift evolution in \lmfp\ as a
$(1+z)^\eta$ power-law. Adopting a two-parameter model,
$\mlmfp(z) = A [(1+z)/5]^\eta$, we minimize $\chi^2$ under the assumption of
Gaussian errors in the \lmfp\ measurements
(this assumption is not strictly true, but provides a good approximation,
e.g.\ \citealt{omeara13}). We find $A = \left(\aval\right)$\hmpc\ and
$\eta = \gval$ giving a reduced $\chi_\nu^2 = 0.8$. As is evident from
Table~\ref{tab:mfp}, the SDSS measurements have the smallest estimated errors
and therefore anchor the fit at $z \approx 4$. If we arbitrarily increase the
uncertainty in these measurements, then $\sigma(A)$ increases and $\chi_\nu^2$
decreases but there is very little effect on $\eta$ and its estimated uncertainty.
Therefore, we conclude at high confidence that \lmfp\ evolves more steeply than
$\left(1+z\right)^{-4}$ at $z<5.5$. We find a steeper redshift evolution than
recovered for \tlya. Clearly, the astrophysics governing gas absorbing
significantly at the LL differs from that of the canonical \lya\ forest.

Consider the physical significance of such strong evolution in \lmfp\ with cosmic
time. We assume first that the structures dominating the LL optical depth have a
characteristic physical size $D$ and comoving number density $n_\mathrm{c}$ at a
given redshift. Under this assumption, the redshift evolution of the mean free
path scales as
\begin{equation}
\mlmfp \propto \frac{\left(1+z\right)^{-3}}{\mnD}
\end{equation}
Therefore, in a universe where such structures do not evolve in comoving number
density or physical size, one roughly predicts
$\mlmfp \propto \left(1+z\right)^{-3}$ from cosmological
expansion\footnote{This scaling assumes that all opacity comes from highly
optically thick absorbers and the mean free path is small. If, as we will argue
below, absorbers with $\mtll\la 1$ significantly contribute to \lmfp\ and the
mean free path is large, cosmological expansion can lead to a redshift evolution
that is steeper than $\left(1+z\right)^{-3}$ due to redshifting of Lyman continuum
photons \citep[e.g.][]{bb13}. Redshift effects become significant at $z\la 3$
when $\mlmfp\ga 100$\,Mpc.}.
This is strictly ruled out by the observations. Instead, \nD\ must decrease with
time as approximately $\left(1+z\right)^{2}$. Whereas galaxies are assuredly
growing in radius and number with decreasing redshift, structures dominating the
LL opacity are \textit{reduced} in number and/or physical size. This implies
that the majority of such gas is not associated to the central regions of
gravitationally collapsed structures (e.g.\ \ion{H}{1} discs).

A possible scenario is that the LL opacity is dominated by gas in the haloes of
galaxies (aka the circumgalactic medium or CGM) which then evolves across cosmic
time. Numerical simulations of galaxy formation do predict a significant
reservoir of cool, dense gas accreting on to galaxies via `cold streams' that
span the dark matter haloes
\citep{birnboim03,db06,ocvirk08,dbe+08,keres09a,freeke11}. Portions of these
streams are predicted to have significant LL opacity
\citep{fumagalli11a,freeke12b} and should contribute to \teff\ at $z>2$.
These simulations also predict a declining covering fraction $f_\mathrm{c}$ of
optically thick gas from these structures within the virial radius \rvir\ in time
\citep[\rvir;][]{fg11,fumagalli13b}. On the other hand, \rvir\ is increasing and
the physical cross-section remains roughly constant or even increases in
galaxies of a given halo mass \citep{fumagalli13b}. Similarly, the central
galaxies and the dark matter haloes only grow with cosmic time. Therefore, simple
models for the evolution of optically thick gas in haloes could, in principle,
predict a \textit{decreasing} mean free path with decreasing $z$. Indeed,
\citet{fop+13} have argued that a significant fraction of LLSs with $\mtll>2$
must reside outside dark matter haloes at $z>3.5$. We draw a similar inference
for the gas dominating the \ion{H}{1} LL opacity, which may hold to $z<3$.
For dark matter haloes to dominate the integrated LL opacity at high-$z$,
one may need to invoke scenarios where low mass haloes contribute a majority of
the opacity at $z \sim 5$ and then evaporate shortly after
\citep[e.g.\ mini-haloes;][]{am98}. Presently, we consider this to be an
improbable scenario but we encourage the analysis of halo gas in lower mass
haloes and also the properties of gas with $\mtll < 1$ in all haloes.

We argue that the gas absorbing LL photons arises predominantly within
large-scale structures near the collapsed regions of dark matter haloes 
(e.g.\ filaments, the cosmic web), consistent with current numerical results
exploring the frequency distribution of \ion{H}{1} gas
\citep{fumagalli11a,mcquinn11,altay11,rahmati+13} and recent analysis of the
cross-correlation between LLSs and quasars \citep{qpq6}. But what then drives
the rapid evolution in \lmfp? There are three obvious possibilities: 
(i) the structures themselves decrease in physical size;
(ii) their mass decreases;
(iii) the gas becomes more highly ionized yielding lower LL opacity.  
We consider the first option to be very unlikely. If anything, structures
outside dark matter haloes are likely to increase in size via cosmological
expansion. There could be gravitational contraction along one dimension
(possibly two), but this would be balanced by expansion in at least one other.
The second effect, reduced mass, may follow from the funnelling of gas into
galaxies and their haloes. In turn, this reduces the surface and volume
densities of the gas. From $z=5$ to 2, the comoving mass density in dark matter
haloes with $M > 10^{10} \msol$ increases by a factor of 25. A significant
fraction of the mass must come from the surrounding environment, but this could
be replenished by new material from the even more distant IGM. We suspect that
mass evolution is a sub-dominant effect for \lmfp\ evolution but we encourage
exploration in cosmological simulations on scales of a few \rvir\ around
high-$z$ galaxies.

We posit that most structures giving rise to LL absorption are cosmologically
expanding, yielding a lower density  $n_\mathrm{H}$ that drives a substantial
decrease in the \ion{H}{1} fraction. Consider an idealized volume $d^3$ of
constant density $n_\mathrm{H}$ expanding with the universe. The average column
density $N_\mathrm{H} \sim n_\mathrm{H} d$ declines with time as
$\left(1+z\right)^{2}$ but the physical area of the structure increases by the
same factor. Therefore, the average number of hydrogen nuclei that an ensemble
of sight lines would intersect remains constant. The volume density evolves as
$\left(1+z\right)^{3}$, however, and an optically thin medium bathed in radiation
would see its neutral fraction lowered with time by the same scaling.  
Indeed, current estimates for the photoionization rate per hydrogen atom
$\Gamma_\mathrm{HI}$ at $z\sim 4$ based on measurements of the \lya\ opacity
yield a nearly constant value \citep[e.g.][]{flh+08,bb13}. We conclude that this
effect dominates the rapid evolution in \lmfp; cosmological expansion alone can
yield $\eta < -5$ by reducing the effective physical size with substantial LL opacity.

This scenario requires that a significant fraction of the LL opacity comes from
gas away from galaxies, i.e.\ with lower \ion{H}{1} column densities. Indeed,
\citet{pow10} and \citet{omeara13} have inferred that $\approx 50$ per cent of
the opacity arises from gas with $\mnhi < 10^{17.5} \cm{-2}$
\citep[see also][]{rudie13}. \citet{pow10} also found that the observed decline
in the incidence of LLSs is driven by gas with $\mnhi < 10^{19} \cm{-2}$.
If this gas is more subject to the effects of cosmological expansion, one may
predict significant evolution in the shape of the \nhi\ frequency distribution
at $\mnhi < 10^{19} \cm{-2}$ from $z=5$ to $2.5$, consistent with recent
numerical work \citep{rahmati+13}.

Before concluding this sub-section, we note that at present, the data do not
require a break in the power-law shown in Fig.~\ref{fig:mfpvsz}.
However, future studies at yet lower redshift (difficult to achieve) or improved
statistics at $2<z<3$ would test for such a break. This could indicate a change
in the origin of optically thick gas on cosmological scales. Likewise, we caution
against drawing firm conclusions from extrapolating our best-fitting power-law
outside the covered redshift range ($2.3\la z\la 5.5$).

\subsection{Comparison to models and implications for reionization}

\begin{figure}
\includegraphics[width=1.0\linewidth]{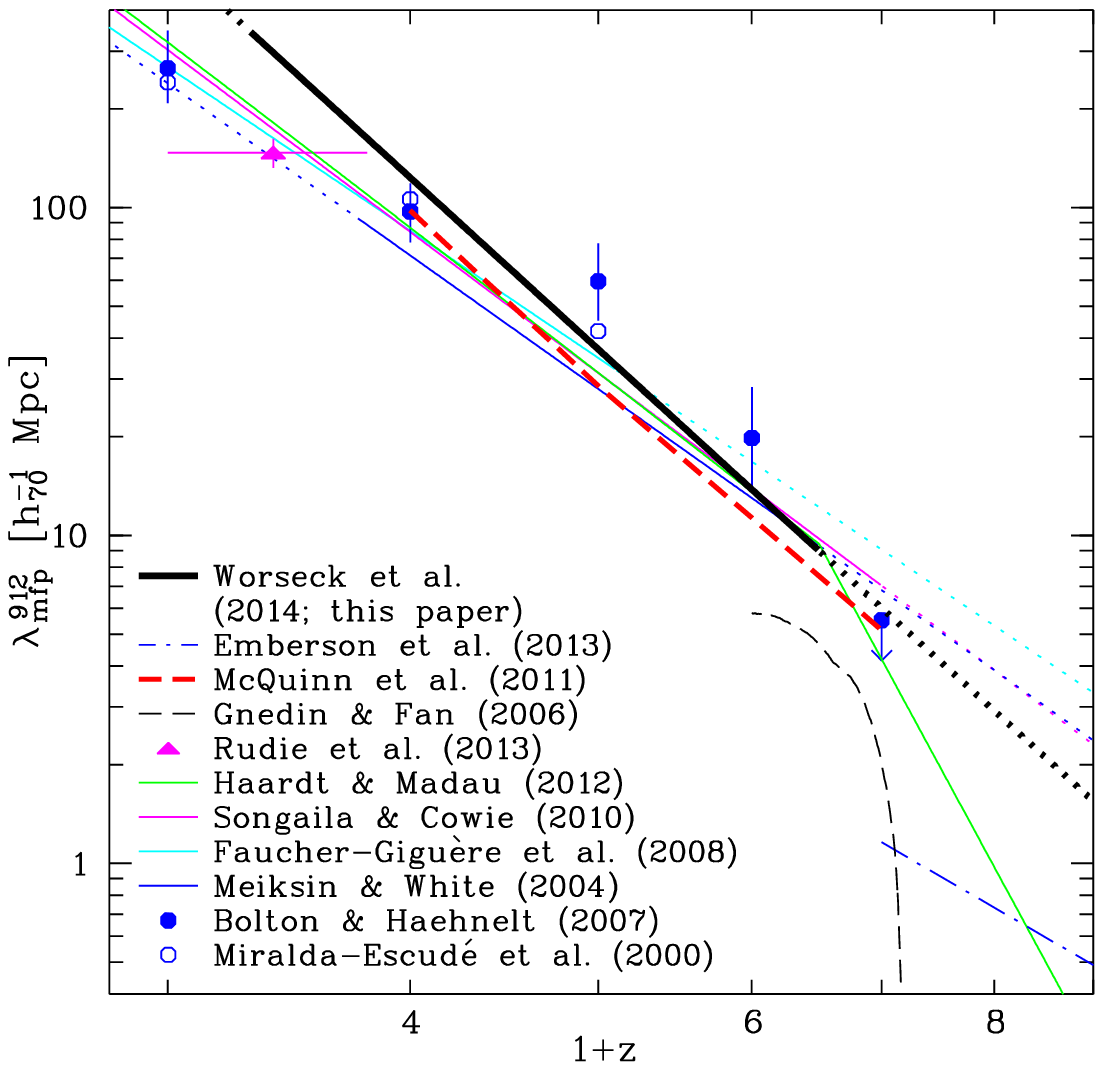}
\caption{\label{fig:mfpfits}
Comparison of various evaluations of the proper mean free path to Lyman limit
photons. Power-law fits $\mlmfp\propto\left(1+z\right)^\eta$ are shown as
straight solid (dotted) lines in (outside) the quoted redshift range of validity.
We show our power-law fit (thick black) and previous estimates based on
\ion{H}{1} absorber statistics \citep{flh+08,songaila10,hm12,rudie13},
semi-analytic additions to optically thin numerical simulations
\citep{mhr00,mw04,bolton07}, and estimates obtained in a fully numerical
framework \citep{gnedin06,mcquinn11,emberson13}. See the text for a discussion.
A colour version of this figure is shown in the online edition of the Journal.
}
\end{figure}

Despite recent progress it is still challenging to model LLSs in numerical
simulations due to their small abundance in small simulation volumes, the
involved high densities, necessary radiative transfer, and possible radiative
feedback from local sources. Therefore, three approaches have been developed to
model the IGM absorber population and the resulting \lmfp:
(i) empirical \ion{H}{1} absorber statistics,
(ii) semi-analytic additions to optically thin numerical simulations, and most recently
(iii) cosmological simulations post-processed with radiative transfer.
Figure~\ref{fig:mfpfits} presents a compilation of various estimates and
compares them to our power-law fit derived in Section~\ref{sec:mfpzevol}.
We discuss these in the following.

The first approach uses an empirical parametrization for the \ion{H}{1} absorber
redshift and column density distribution \fnz\ based on observations to calculate
$\mteff\left(z_{912},z_\mathrm{q}\right)$, with \lmfp\ defined as the distance at
which $\mteff\left(z_{912},z_\mathrm{q}\right)\equiv 1$
\citep[Table~\ref{tab:mfp};][]{mm93,mhr99,flh+08,songaila10,hm12,rudie13}.
Generally, these authors combined results on the observed incidence of LLSs with
estimations or assumptions on the frequency of absorbers with
$\mnhi\la 10^{17} \cm{-2}$. They also adopted differing approaches to evaluating
\lmfp\ \citep[e.g.\ whether to account for the redshifting of LL photons; see][]{bb13}. 
Recently, \citet{pmof14} have examined systematic uncertainties related t
evaluation \lmfp\ via \fnz. In addition to the difficulty in measuring \fnz\ at
$\mnhi \approx 10^{17} \cm{-2}$ \citep[e.g.][]{rudie13,kim13}, they identified
two systematic effects related to the clustering of strong absorbers: 
(i) the double counting of structures absorbing LL photons which yields an
underestimate of \lmfp; and 
(ii) a non-Poisson distribution of such absorbers which also increases \lmfp.
Given these issues and uncertainties, we consider it fortuitous that several of
the previous \lmfp\ estimations from \fnz\ are in good agreement with our results
(Fig.~\ref{fig:mfpfits}). Nevertheless, going forward we intend to combine our
constraints on \lmfp\ with measurements of \fnz\ to explore the clustering and
large-scale distributions of optically thick gas. We also note that our
power-law fit is steeper than any previous empirical estimate due to our
accounting for redshift effects at $z<3$ and the large redshift range now
covered by our direct measurements.

Besides the above empirical estimates, optically thick systems can be added to
the absorber frequency distribution obtained from optically thin numerical
simulations, either based on their observed frequency \citep{mw04} or by
semi-analytic approximations \citep{mhr00,bolton07}. Specifically, \citet{mw04}
added the observed number of LLSs from \citet{key_lls} (extrapolated to $z>4$)
to their particle mesh simulations and obtained
$\mlmfp=28\left[\left(1+z\right)/5\right]^{-4.2}$\,Mpc at $2.75<z<5.5$ for
our adopted cosmology. In the redshift range considered by \citet{mw04} we find
reasonable agreement with the direct \lmfp\ measurements from quasar stacks,
again probably limited by the uncertainty in the parametrization of LLSs.
Some physical insight may be gained by considering the \citet{mhr00} model of a
two-phase IGM composed of low-density fully ionized gas at
$\Delta=\rho/\bar{\rho}<\Delta_\mathrm{i}$ and optically thick neutral clumps at
$\Delta>\Delta_\mathrm{i}$. With this approximate treatment of self-shielding
and ignoring redshifting of the photons, the mean free path is the mean distance
between the neutral clumps
\begin{equation}\label{eqn:mhr00mfp}
\mlmfp=\lambda_0F_{\Delta_\mathrm{i}}^{-2/3} \;\;\; ,
\end{equation}
with the \ion{H}{1} volume filling fraction $F_{\Delta_\mathrm{i}}$ given by the
density distribution function $P\left(\Delta\right)$ as
\begin{equation}\label{eqn:mhr00fh1}
F_{\Delta_\mathrm{i}}= \int_{\Delta_\mathrm{i}}^{\infty}P\left(\Delta\right)\mathrm{d}\Delta \;\;\; .
\end{equation}
At low densities \citet{mhr00} parametrized $P\left(\Delta\right)$ using a
hydrodynamical simulation \citep{mco+96}, whereas at high densities below the
resolution limit of the simulation $P\left(\Delta\right)$ asymptotes to a
power-law density profile. \citet{mhr00} normalized the mean free path by noting
that $\lambda_0H\left(z\right)\simeq 60$\,km\,s$^{-1}$ reproduces the scales of
the \lya\ forest in their simulation. Calibrating $\Delta_\mathrm{i}$ with the
observed \lya\ forest effective optical depth and the density distribution from
their simulation, they obtained $\mlmfp\left(z\right)\simeq\left(241,106,42\right)\mhmpc$
at $z=\left(2,3,4\right)$. Despite the outdated cosmological parameters of the
\citet{mco+96} simulation, the latter two \lmfp\ values are in very good
agreement with the direct measurements \citep{pwo09,fop+13}. At $z=2$,
Equation~\ref{eqn:mhr00mfp} underpredicts \lmfp\ due to redshift effects in the
expanding universe \citep{mhr00}. \citet{bolton07} slightly varied this approach
by estimating the critical density for self-shielding analytically,
and extrapolating the \citet{mhr00} density distribution to $z=6$.
Their estimated \lmfp\ values at $z=4$ and $z=5$ are $\sim 50$ per cent larger
than ours, probably due to the strong assumptions of the \citet{mhr00} model
(extrapolated density distribution, two-phase IGM, fixed \lmfp\ normalization).

With these caveats in mind, we can combine the \lmfp\ parametrization of
\citet{mhr00} and our power-law fit to $\mlmfp\left(z\right)$ to estimate the
\ion{H}{1} volume filling fraction, yielding
$F_{\Delta_\mathrm{i}}\approx 2.3\times 10^{-4}\left[\left(1+z\right)/5\right]^{5.85}$.
Within the simplified framework of the \citet{mhr00} model, this
order-of-magnitude estimate confirms that the IGM is highly ionized at $z<5.5$.
Extrapolating the \citet{mhr00} parametrization and our fit of
$\mlmfp\left(z\right)$ to higher redshifts, we obtain
$F_{\Delta_\mathrm{i}}\approx 2\times 10^{-3}$ at $z=6$, indicating that
\ion{H}{1} reionization likely occurred at $z\gg 6$. Adopting the density
distribution of \citet{mhr00} we use Equation~\ref{eqn:mhr00fh1} to obtain an
approximate density threshold for self-shielding gas of
$\Delta_\mathrm{i}\approx 300$ ($\Delta_\mathrm{i}\approx 100$) at $z=3$ ($z=4$).
Assuming local hydrostatic equilibrium and typical values for the gas temperature
and the UV background, these thresholds correspond to column densities remarkably
similar to $\mtll\simeq 1$ LLSs \citep{schaye01_lya,fo05}. While this is a good
consistency check, we caution that $\mtll\simeq 1$ LLSs are translucent,
requiring radiative transfer models. Setting $\Delta_\mathrm{i}$ as the
characteristic density of $\mtll=1$ LLSs \citep{bolton07} neglects $\mtll<1$
absorbers and overestimates the number of optically thick ones.

Recent cosmological simulations post-processed with radiative transfer
calculations have significantly improved upon semi-analytic models of optically
thick gas \citep{mcquinn11,altay11,rahmati+13}. These studies predict an
\ion{H}{1} column density distribution shaped by radiative transfer, smoothly
transitioning from the optically thin IGM to fully neutral gas in the vicinity
of galaxy haloes. Self-shielding is significant at high overdensities, in good
agreement with our rough estimates obtained with the \citet{mhr00} formalism.
In particular, \citet{mcquinn11} self-consistently calculated the mean free path
of LL photons from their simulations. Figure~\ref{fig:mfpfits} shows
$\mlmfp\left(z\right)$ for their assumed power-law UV background spectrum
$J_\nu\propto\nu^{-1}$ yielding a UV background photoionization rate
$\Gamma_\mathrm{HI}=5\times 10^{-13}\,\mathrm{s}^{-1}$. A power-law fit to their
values (M.\ McQuinn, priv.\ comm.) yields
$\mlmfp=29.1\left[\left(1+z\right)/5\right]^{-5.26}$ at $3\le z\le 6$.
Interestingly, the power-law exponent is almost identical to the one we obtain
from the quasar stacks (Section~\ref{sec:mfpzevol}), but their lower
normalization results in a $\sim 20$ per cent smaller mean free path at all
redshifts (Fig.~\ref{fig:mfpfits}). For isothermal density profiles the
\citet{mhr00} model yields
$\Gamma_\mathrm{HI}\propto\left[\mlmfp\right]^{1.5}$ \citep{fo05} which
approximately holds in the numerical treatment by \citet{mcquinn11}. Thus, our
independent mean free path measurements imply a UV background photoionization rate
$\Gamma_\mathrm{HI}\approx 7\times 10^{-13}\,\mathrm{s}^{-1}$ at $z=4$, which
agrees with current estimates at the $1\sigma$ level \citep[e.g.][]{bb13}.
More importantly, as $\Gamma_\mathrm{HI}=$const.\ in the simulation by
\citet{mcquinn11}, the similarly steep redshift evolution of the mean free path
implies that $\Gamma_\mathrm{HI}$ should only weakly depend on redshift,
in agreement with independent estimates from the \lya\ forest \citep{flh+08,bb13}.
Remaining tension in the shape of \fnz\ between observations and the
\citet{mcquinn11} simulation can be alleviated by a softer UV background spectrum
\citep{altay11,rahmati+13}. We encourage further numerical work on \fnz\ with
varying SEDs of the UV background calibrated to $\Gamma_\mathrm{HI}$ and \lmfp.

At the highest redshifts $z>5$, i.e.\ in the immediate post-reionization IGM,
most previous inferences on the mean free path are brazen extrapolations from
lower redshifts, leaving very few constraints that are based on actual
measurements. At $z=6$ \citet{bolton07} give an upper limit $\mlmfp<5.5$ proper
Mpc (corrected to our cosmology) based on \ion{H}{1} Gunn-Peterson optical depth
measurements \citep{songaila04,Fan06} and the \citet{mhr00} model extrapolated
to $z=6$. At the same redshift the power-law fit to the LLS incidence by
\citet{songaila10} yields $\mlmfp\simeq 7$\,Mpc, but with an estimated
uncertainty of $\sim 30$ per cent due to the poorly constrained shape of \fnz\
at high redshift. Therefore it remains unclear whether their somewhat higher LLS
incidence at $5<z<6$ might correspond to a drop in \lmfp. Likewise, the accuracy
of our mean free path measurement from the GGG $z_\mathrm{em}>5$ quasar stack is
likely limited by sky subtraction errors and uncertainties in the adopted quasar
SED (Section~\ref{sec:analysis}). Current quasar samples are too small to track
rapid evolution in the mean free path at $z>5$, so that the power-law
parametrization from Section~\ref{sec:mfpzevol} is adequate. Extrapolating our
fit by $\Delta z=0.5$ beyond the range of validity, we obtain
$\mlmfp=\left(6.0\pm 0.9\right)$\,Mpc at $z=6$, in very good agreement with
\citet{songaila10}. 

The mean free path in the post-reionization IGM is an important boundary
condition for \ion{H}{1} reionization models. While in the early stages of
\ion{H}{1} reionization the mean free path strongly depends on the source
properties as the ionizing photons are absorbed within individual \ion{H}{2}
regions, it is predicted to rise rapidly by several orders of magnitude as
\ion{H}{2} regions merge and reionization is completed \citep{gnedin00}.
After overlap the remaining high-density regions are gradually ionized,
corresponding to a smoothly increasing mean free path. Hydrodynamical
simulations of reionization with approximate radiative transfer and sufficiently
large dynamic range predict the post-reionization mean free path to within a
factor of $\sim 2$, limited either by box size \citep{gnedin00,gnedin06} or
sub-cell physics \citep{kgh07}. As an example, Fig.~\ref{fig:mfpfits} shows the
mean free path evolution of the L8N256 run from \citet{gnedin06}, with an
overlap epoch (and hence a large jump in \lmfp) at $z\simeq 6.2$ by construction.
Recently, \citet{emberson13} presented adiabatic hydrodynamical simulations
post-processed with radiative transfer to predict the redshift evolution of the
mean free path for different amplitudes of the UV background. For characteristic
values of the UV background at $z\sim 6$
\citep[$\Gamma_\mathrm{HI}\simeq 3\times 10^{-13}$\,s$^{-1}$; e.g.][]{calverley11}
they underestimate the mean free path by a factor of $\simeq 5$
(Fig.~\ref{fig:mfpfits}), at least in part due to their neglect of photoheating
that would boost the mean free path by suppressing structure formation.
In semi-numerical approaches to study the large-scale morphology of reionization
the mean free path is an input parameter to impede the growth of \ion{H}{2}
regions, resulting in an extended reionization epoch and a spatially
inhomogeneous UV background \citep{crociani11,alvarez12}. Our measurements rule
out very large and redshift-independent mean free paths for absorption systems
assumed by \citet{alvarez12}.

Our \lmfp\ measurements also constrain models for spatial fluctuations in the UV
background in the post-reionization IGM. Adopting our power-law model,
a spherical volume with radius $r=\mlmfp$ contains roughly
$(11000, 1300, 180)$ $M_\mathrm{UV}<0.9M_\mathrm{UV}^{\star}$ star-forming
galaxies at $z=(5,6,7)$ \citep[e.g.][]{bouwens12}. Due to the steep faint-end
slope of the galaxy luminosity function, even a slight extrapolation beyond this
currently observable magnitude limit would dramatically boost the average number
of galaxies in this attenuation volume, arguing for a fairly homogeneous UV
background at $z\la 6$ if such galaxies dominate the photon budget.
However, galaxy clustering likely increases small-scale fluctuations in the UV
radiation field, requiring numerical approaches \citep{mesinger09}.

\subsection{Impact of quasar proximity zones}
\label{sec:pezone}

In the highest redshift interval considered ($z_\mathrm{em}>5$), we measure a
central value for the mean free path of only $\mlmfp \simeq 10$\,Mpc.
This implies, at least crudely, that a significant fraction $f$ of sources
will be strongly attenuated by gas within only a few Mpc. For example, we
estimate that $f(<3\,{\rm Mpc}) = 1 - \rm exp(-3Mpc/\mlmfp)\approx 25$ per cent
of the sight lines will be strongly attenuated by gas within $r = 3$\,Mpc of the
source. Such gas occurs within the so-called proximity zones of the quasars.
On these scales, there are at least two effects which influence the
characteristics of the \ion{H}{1} gas that differ from random regions of the
cosmological volume. First, quasars reside in massive galaxies
\citep[e.g.][]{white12} which themselves lie within large dark matter
overdensities relative to the cosmic mean. Indeed, observations of projected
quasar pairs at $z \sim$2--3 reveal excess \ion{H}{1} absorption to beyond 
1\,\hmpc\ transverse to the foreground quasar \citep{font13,qpq6}.
Similarly, one measures a strong clustering signal between quasars and optically
thick gas \citep{qpq2,qpq5,qpq6}. Therefore, gas within the proximity zone
apparently has large `intrinsic' opacity to LL photons. On the other hand,
the ionizing radiation emitted by the quasar illuminates the gas along our sight
line and can photoionize the foreground \ion{H}{1} to great distances
\citep[e.g.][]{qpq2}. Such a proximity effect is observed in the optically thin
\lya\ forest \citep[e.g.][]{sbd+00,dww08,calverley11}. The interplay of these
two effects is obviously complex and depends at least upon the mass of the host
galaxy and luminosity of the quasar \citep[e.g.][]{fg08_prox}.
In fact, one measures a lower incidence of optically thick LLSs within the
proximity zones of $z_\mathrm{em}\sim 4$ quasars \citep{pow10}.  

At $z \sim 5$, the incidence of such `proximate' LLSs (PLLSs) has not yet been
measured; this will be considered in a future paper of our series.
We report here on preliminary results finding 13 cases of strong quasi-continuous
absorption in the Lyman continuum at redshifts consistent with $r<3$\,Mpc from
the quasar. This represents one third of the $z_\mathrm{em}>5$ quasar sample,
consistent with our expectation based on the \lmfp\ analysis, although we
emphasize that these sight lines have contributed to the \lmfp\ measurement.
A refined treatment estimating the luminosity-dependent size of the proximity zone
at which the photoionization rate of the quasar equals that of the UV background
\citep[e.g.][]{calverley11} yields very similar results:
The measured mean free path is just $\simeq 1.8$ times larger than the average
proximity zone size $r_\mathrm{prox}\simeq 6$\,Mpc of the targeted
$z_\mathrm{em}>5$ quasars. At present, any such analysis is limited by quasar
redshift uncertainty ($\sigma_r\simeq 1$\,Mpc), strong contaminating Lyman
series absorption at lower redshifts, and the uncertainty in the likely evolving
UV background at $z>5$ \citep{calverley11}. The high incidence of PLLSs at $z>5$
is consistent with an extrapolation from $z\la 4$ \citep{pow10}.
While \lmfp\ at $z<4$ is sufficiently large that gas in the proximity zone
provides $\la 10$ per cent of the effective optical depth
\citep{pow10,omeara13,fumagalli13b}, our measurements at $z>5$ are influenced by
gas local to the quasar. We can estimate this potential bias by adopting the
power-law parametrization for $\mlmfp\left(z\right)$ at $z<5$ and comparing the
extrapolated \lmfp\ at $z>5$ to the measured value. The fit to $z<5$ yields
$\mlmfp\left(z\right) = \left(37\pm 2\right)\left[\left(1+z\right)/5\right]^{-5.3\pm 0.5}$\,Mpc
at a reduced $\chi_\nu^2=0.3$, consistent with the fit to all data.
The extrapolated mean free path at the mean redshift of the $z_\mathrm{em}>5$
stack ($z_\mathrm{q}=5.16$) agrees with the measured value within $1\sigma$,
suggesting that proximity zones do not strongly bias our \lmfp\ measurement.

However, measuring a mean free path that approaches the quasar proximity zone
size has several important implications. First, our results on \lmfp\ at $z>5$
may not apply to the `random' IGM. If the proximity zones of these quasars have
a higher (or lower) incidence of optically thick gas, this will bias the \lmfp\
values accordingly relative to the cosmological average at that epoch.
On the other hand, ionizing sources do not inhabit a random distribution of the
universe, but must occur within collapsed, overdense structures. Furthermore,
the majority of optically thick gas is almost surely associated to galaxies and
their surrounding environments \citep[e.g.][]{fumagalli11a,mcquinn11,freeke12b}
and also biased relative to random regions. Therefore, while \lmfp\ measurements
from quasars may be biased evaluations they are obviously valid for assessing
the attenuation of Lyman continuum photons emitted by luminous
$z_\mathrm{em}\sim 5$ quasars. It is unknown, however, how optically thick gas
is clustered to fainter quasars and/or star-forming galaxies at that epoch and
approaching reionization. Thus, the mean free path at $z>5$ likely depends on
the source environment and cannot be regarded as the mean separation between
LLSs in the IGM (that interpretation holds only for diffuse Lyman continuum
photons from recombination). Furthermore, clustering of the sources and
absorbers requires a more sophisticated treatment than simple Poisson statistics
\citep{pmof14}.

Another implication is that $z \simeq 5.2$ currently marks the highest redshift
at which \lmfp\ measurements from stacked quasar spectra can be safely related
to the IGM. At $z>5.5$ the mean free path is smaller than the typical proximity
zone size of luminous quasars \citep[see also][]{calverley11}, such that any
stacking analysis should be restricted to fainter quasars or star-forming
galaxies with smaller proximity zones. Alternatively, \lmfp\ estimates may be
obtained from the incidence of intervening LLSs assuming the shape of the column
density distribution, but this remains a challenging task at $z>5$ \citep{songaila10}.

%%%%%%%%%%%%%%%%%%%%%%%%%%%%%%%%%%%%%%%%%%%%%%%%%

\section{Conclusions}
\label{sec:conclusions}

We have presented first results from the Giant Gemini GMOS (GGG) survey of the
intergalactic medium at $z\ga 4$. In a long-term multi-partner Gemini programme
we have obtained high-quality (S/N$ \sim 20$ per $1.85$\,\AA\ pixel)
low-resolution (FWHM$\approx 320\,\mkms$) spectra of \nqso\ $\mzem>4.4$ quasars,
the largest sample of its kind at these redshifts. The reduced data are publicly
available\footnote{Link to VizieR will be provided upon acceptance of the manuscript.}.
The primary goal of this survey has been a precise measurement of the mean free
path to \ion{H}{1} Lyman limit photons in the high-redshift universe via the
analysis of stacked rest-frame quasar spectra, a technique pioneered by our team
\citep{pwo09}. Future papers in this series will determine the incidence rate of
DLAs and LLSs at $z>4$ and, by using the total Lyman limit opacity determined
from the mean free path, constrain the column density distribution of $\mtll<1$
absorbers. Our primary results are as follows:
\begin{enumerate}
\item The mean free path \lmfp\ monotonically decreases with redshift. Subsampling
our data set we measure $\mlmfp=\left(\amfp, \bmfp, \cmfp\right)h_{70}^{-1}$
proper Mpc (68 per cent confidence level) at mean redshifts of the stacks
$z_\mathrm{q}=\left(4.56, 4.86, 5.16\right)$, respectively
(Figs.~\ref{fig:stacks} \& \ref{fig:boot}).
\item When combining these measurements with lower-redshift results obtained
with the stacking technique \citep{pwo09,omeara13,fop+13}, we find that the mean
free path smoothly evolves between $z=2.3$ and $z=5.5$ (Fig.~\ref{fig:mfpvsz}),
well described by a power-law:
$\mlmfp\left(z\right)=A\left[\left(1+z\right)/5\right]^\eta$ with
$A=\left(\aval\right)h_{70}^{-1}$\,Mpc and $\eta=\gval$
(68 per cent confidence level). 
\end{enumerate}

The redshift evolution of the mean free path exceeds that expected from
cosmological expansion, indicating a reduction in number and/or physical size
of the absorbing structures with cosmic time. We conclude that a significant
fraction of the structures giving rise to \ion{H}{1} Lyman limit absorption are
in the IGM. These structures are likely cosmologically expanding, leading to a
substantial decrease of their \ion{H}{1} fraction with time in the otherwise
highly reionized IGM. Although our measurements are consistent with previous,
more uncertain estimates based on the statistics of Lyman limit systems and
(semi)-numerical models (Fig.~\ref{fig:mfpfits}), our inferred redshift
evolution of the mean free path is very steep, partly due to our correct
accounting for cosmological expansion at low redshifts. The smoothly evolving
mean free path tracks the Lyman limit absorption in the highly ionized IGM at
$2.3<z<5.5$, without any obvious indication of a more rapid decrease at the
highest redshifts that would signal the approach of the \ion{H}{1} reionization
epoch. Viable numerical models of \ion{H}{1} reionization must nevertheless
match the measured post-reionization mean free path and its evolution with
redshift \citep[e.g.][]{gnedin06}. 

Future work on the mean free path will likely focus on the lowest and highest
redshifts. At $z\sim 1.6$ the mean free path is expected to exceed the horizon
of the universe, such that all ionizing sources are expected to contribute to
the UV radiation field at any given point \citep[e.g.][]{mhr99,omeara13}.
Currently, there are efforts to constrain this `breakthrough' epoch with \textit{HST}
UV spectroscopy of a sample of $\mzem\sim 1$ quasars (Howk et al.\ in prep.).
At the highest redshifts $z\ga 6$ the mean free path is expected to drop rapidly,
indicating the epoch of \ion{H}{1} reionization \citep[e.g.][]{gnedin00}.
However, measurements of the mean free path are very challenging at $z\ga 5.5$
due to highly uncertain IGM absorber statistics \citep{songaila10} and
limitations of the quasar stacking technique (high sky subtraction accuracy
required for $\mlmfp\la 10$\,Mpc, uncertainty in the UV SED of quasars,
bias due to quasar proximity zones). A direct measurement of the mean free path
at $z>5.5$ from spectral stacks will require a modest sample of high-quality
spectra of either fainter quasars with smaller proximity zones
\citep[e.g.][]{willott10} or the brightest $z\sim 6$ galaxies
\citep[e.g.][]{willott13} to be collected with future 30\,m telescopes.

%%%%%%%%%%%%%%%%%%%%%%%%%%%%%%

\section*{Acknowledgements}
We thank Francesco Haardt and Piero Madau for useful discussions.
Matt McQuinn and Nick Gnedin kindly supplied tabulated mean free path values from
their numerical simulations.
GW and JXP acknowledge the support from the National Science Foundation (NSF) grant
AST-1010004. JXP thanks the Alexander von Humboldt Foundation for a visitor
fellowship to the MPIA where part of this work was performed, as well as the MPIA
for hospitality during his visits. MTM and JXP thank the Australian Research Council
for \textsl{Discovery Project} grant DP130100568 which supported this work.
JMO thanks the VPAA's office at Saint Michael's College for travel support.
GDB acknowledges support from the Kavli Foundation and the support of a STFC Ernest
Rutherford Fellowship. SL has been supported by FONDECYT grant number 1100214.
BM has been supported by NSF Grant AST-1109665 and the Alfred P. Sloan Foundation.

Based on observations obtained at the Gemini Observatory, which is operated by the 
Association of Universities for Research in Astronomy, Inc., under a cooperative
agreement with the NSF on behalf of the Gemini partnership: the NSF (United States),
the National Research Council (Canada), CONICYT (Chile), the Australian Research
Council (Australia), Minist\'{e}rio da Ci\^{e}ncia, Tecnologia e Inova\c{c}\~{a}o 
(Brazil) and Ministerio de Ciencia, Tecnolog\'{i}a e Innovaci\'{o}n Productiva (Argentina).

\bibliographystyle{mn2e}
\bibliography{allrefs}
\bsp

\appendix
\section{Online Tables and Figures}
\label{app:gggspec}

\begin{table*}
\caption{\label{tab:ggg_sample}GGG survey.}
\begin{tabular*}{0.65\linewidth}{@{\extracolsep{\fill}}lcccccc}
\hline
Quasar &$i$ [mag]&$z_\mathrm{SDSS}$ &flag$^\star$ &B600 [\AA] &S/N$^\dag$ &R400 [\AA]\\ 
\hline
SDSS~J001115.23$+$144601.8&18.28&4.967&0&4910--7830&41&6500--10780\\
SDSS~J004054.65$-$091526.8&19.20&4.976&3&4950--7870&10&6470--10710\\
SDSS~J010619.24$+$004823.3&18.61&4.449&0&4340--7240&13&5890--10160\\
SDSS~J012509.42$-$104300.8&19.43&4.492&0&4340--7240&7&5890--10160\\
SDSS~J021043.16$-$001818.4&19.17&4.733&0&4630--7550&10&6180--10460\\
SDSS~J023137.65$-$072854.4&19.55&5.413&0&5130--8060&7&6470--10720\\
SDSS~J033119.66$-$074143.1&19.12&4.739&0&4630--7550&10&6180--10460\\
SDSS~J033829.30$+$002156.2&20.07&5.032&2&4950--7870&7&6480--10720\\
SDSS~J073103.12$+$445949.4&19.23&5.004&0&4910--7830&27&6490--10780\\
SDSS~J075907.57$+$180054.7&19.18&4.862&1&4620--7530&13&6490--10780\\
SDSS~J080023.01$+$305101.1&19.06&4.685&0&4620--7530&19&6190--10470\\
SDSS~J080715.11$+$132805.1&19.43&4.875&0&4620--7530&17&6490--10780\\
SDSS~J081806.87$+$071920.2&18.61&4.581&0&4630--7560&11&6180--10460\\
SDSS~J082212.34$+$160436.9&19.05&4.488&0&4340--7240&10&5890--10170\\
SDSS~J082454.02$+$130217.0&20.00&5.188&0&5110--8030&13&6490--10780\\
SDSS~J083655.80$+$064104.6&19.04&4.436&0&4340--7240&7&5890--10160\\
SDSS~J083920.53$+$352459.3&19.54&4.777&0&4620--7530&7&6190--10470\\
SDSS~J084627.84$+$080051.7&19.84&5.030&0&4950--7870&7&6470--10720\\
SDSS~J084631.52$+$241108.3&19.22&4.743&0&4620--7530&11&6190--10470\\
SDSS~J085430.37$+$205650.8&19.43&5.179&0&5110--8030&8&6490--10780\\
SDSS~J085707.94$+$321031.9&18.73&4.776&0&4620--7530&23&6190--10470\\
SDSS~J090100.61$+$472536.2&19.53&4.608&0&4620--7530&15&6200--10470\\
SDSS~J090245.76$+$085115.8&20.15&5.226&0&5140--8060&7&6480--10720\\
SDSS~J090634.84$+$023433.8&18.50&4.511&0&4340--7240&10&5890--10160\\
SDSS~J091316.55$+$591921.6&20.48&5.122&0&5110--8030&7&6490--10770\\
SDSS~J091543.63$+$492416.6&19.54&5.196&6&5110--8030&16&6500--10780\\
SDSS~J092216.81$+$265358.9&20.24&5.032&0&4910--7830&7&6490--10770\\
SDSS~J093523.31$+$411518.5&19.58&4.787&0&4620--7530&17&6190--10470\\
SDSS~J094056.01$+$584830.2&19.32&4.659&0&4620--7530&19&6190--10470\\
SDSS~J094108.36$+$594725.7&19.30&4.790&0&4620--7530&15&6190--10470\\
SDSS~J094409.52$+$100656.6&19.30&4.748&0&4630--7560&10&6180--10460\\
SDSS~J095632.03$+$321612.6&19.24&4.647&0&4620--7530&21&6190--10470\\
SDSS~J095707.67$+$061059.5&19.27&5.185&0&5140--8060&7&6470--10720\\
SDSS~J100251.20$+$223135.1&19.40&4.744&0&4620--7530&12&6190--10470\\
SDSS~J100416.12$+$434739.0&19.39&4.872&0&4620--7530&10&6490--10780\\
SDSS~J100444.30$+$202520.0&20.24&5.084&1&4910--7830&12&6490--10780\\
SDSS~J101549.00$+$002020.0&19.28&4.403&0&4340--7240&10&5890--10160\\
SDSS~J102332.07$+$633508.0&19.69&4.881&6&4910--7830&13&6490--10770\\
SDSS~J102622.87$+$471907.2&18.74&4.943&0&4910--7830&20&6490--10780\\
SDSS~J102623.61$+$254259.5&20.04&5.303&0&5110--8030&7&6500--10780\\
SDSS~J103418.65$+$203300.2&19.79&4.998&2&4910--7830&18&6490--10780\\
SDSS~J103601.03$+$500831.7&19.23&4.470&1&4320--7220&27&5890--10160\\
SDSS~J103711.04$+$313433.5&19.52&4.885&0&4910--7830&16&6490--10780\\
SDSS~J103919.28$+$344504.5&19.33&4.420&0&4320--7230&20&5890--10160\\
SDSS~J104041.09$+$162233.8&18.96&4.814&0&4910--7830&13&6490--10780\\
SDSS~J104325.55$+$404849.5&19.09&4.934&6&4910--7830&10&6490--10780\\
SDSS~J104351.19$+$650647.6&19.10&4.471&0&4320--7220&20&5890--10160\\
SDSS~J105020.40$+$262002.3&19.47&4.796&0&4620--7530&10&6190--10470\\
SDSS~J105036.46$+$580424.6&19.66&5.132&0&5110--8030&7&6490--10780\\
SDSS~J105322.98$+$580412.1&19.81&5.215&0&5110--8030&9&6490--10780\\
SDSS~J105445.43$+$163337.4&20.22&5.187&6&5110--8030&7&6490--10780\\
SDSS~J110045.23$+$112239.1&18.85&4.707&0&4630--7560&7&6180--10460\\
SDSS~J110134.36$+$053133.8&19.25&4.987&4&4950--7870&7&6470--10720\\
SDSS~J111523.24$+$082918.4&19.56&4.640&2&4620--7550&7&6180--10460\\
SDSS~J111741.26$+$261039.2&19.52&4.635&0&4620--7530&21&6200--10470\\
SDSS~J111920.64$+$345248.1&20.05&5.011&0&4910--7830&10&6490--10780\\
SDSS~J112253.50$+$005329.7&19.10&4.551&0&4630--7560&15&6190--10460\\
SDSS~J112534.93$+$380149.3&19.50&4.618&0&4620--7530&15&6200--10470\\
SDSS~J112857.84$+$575909.8&19.49&4.978&0&4910--7830&11&6490--10770\\
SDSS~J113246.50$+$120901.6&19.76&5.167&0&5130--8060&7&6470--10710\\
SDSS~J114008.67$+$620530.0&19.24&4.521&0&4320--7230&18&5890--10160\\
SDSS~J114225.30$+$110217.3&19.39&4.590&0&4630--7560&10&6180--10460\\
SDSS~J114657.79$+$403708.6&19.41&5.005&0&4910--7830&13&6490--10780\\
SDSS~J114826.16$+$302019.3&20.06&5.142&0&5110--8030&10&6490--10780\\
SDSS~J114914.88$+$281308.7&18.52&4.553&0&4620--7530&38&6200--10470\\
\hline
\end{tabular*}
\end{table*}

\begin{table*}
\contcaption{GGG survey.}
\begin{tabular*}{0.65\linewidth}{@{\extracolsep{\fill}}lcccccc}
\hline
Quasar &$i$ [mag]&$z_\mathrm{SDSS}$ &flag$^\star$ &B600 [\AA] &S/N$^\dag$ &R400 [\AA]\\ 
\hline
SDSS~J115424.73$+$134145.7&20.18&5.010&4&4950--7870&7&6470--10720\\
SDSS~J115809.39$+$634252.8&19.34&4.494&0&4320--7230&20&5890--10160\\
SDSS~J120036.72$+$461850.2&19.28&4.730&0&4620--7530&11&6190--10470\\
SDSS~J120055.61$+$181732.9&19.67&4.984&0&4950--7870&7&6470--10710\\
SDSS~J120102.01$+$073648.1&19.43&4.443&7&4340--7250&10&5890--10160\\
SDSS~J120110.31$+$211758.5&18.73&4.579&6&4620--7530&20&6200--10470\\
SDSS~J120131.56$+$053510.1&19.48&4.830&1&4630--7560&10&6470--10720\\
SDSS~J120207.78$+$323538.8&19.27&5.292&0&5110--8030&7&6490--10770\\
SDSS~J120441.73$-$002149.6&19.28&5.032&0&4950--7870&7&6470--10710\\
SDSS~J120725.27$+$321530.4&19.15&4.643&0&4620--7530&16&6200--10470\\
SDSS~J120730.84$+$153338.1&18.99&4.465&0&4330--7240&12&5890--10160\\
SDSS~J120952.72$+$183147.2&19.86&5.158&0&5140--8060&5&6470--10720\\
SDSS~J121134.04$+$484235.9&19.23&4.505&0&4320--7220&17&5890--10160\\
SDSS~J121422.02$+$665707.5&18.88&4.639&1&4620--7530&7&6200--10470\\
SDSS~J122016.05$+$315253.0&19.08&4.891&0&4910--7830&24&6490--10780\\
SDSS~J122146.42$+$444528.0&19.96&5.206&0&5110--8030&7&6490--10780\\
SDSS~J122237.96$+$195842.9&20.06&5.189&2&5140--8060&5&6470--10710\\
SDSS~J123333.47$+$062234.2&20.06&5.289&0&5140--8060&7&6470--10710\\
SDSS~J124247.91$+$521306.8&20.01&5.018&0&4910--7830&10&6490--10780\\
SDSS~J124400.04$+$553406.8&19.59&4.625&1&4620--7530&7&6200--10470\\
SDSS~J124515.46$+$382247.5&19.66&4.933&6&4910--7830&16&6490--10770\\
SDSS~J125025.40$+$183458.1&19.34&4.599&5&4630--7560&13&6180--10460\\
SDSS~J125353.35$+$104603.1&19.42&4.909&0&4950--7870&14&6470--10710\\
SDSS~J125718.02$+$374729.9&19.23&4.750&0&4620--7530&19&6190--10470\\
SDSS~J130002.16$+$011823.0&18.81&4.613&0&4630--7560&13&6190--10470\\
SDSS~J130110.95$+$252738.3&19.41&4.660&0&4620--7530&17&6190--10470\\
SDSS~J130152.55$+$221012.1&19.67&4.829&6&4620--7530&20&6490--10780\\
SDSS~J130215.71$+$550553.5&19.03&4.436&0&4320--7230&20&5890--10160\\
SDSS~J130619.38$+$023658.9&19.67&4.837&4&4630--7560&7&6470--10710\\
SDSS~J130917.12$+$165758.5&18.94&4.692&0&4630--7560&7&6180--10460\\
SDSS~J131234.08$+$230716.3&19.38&4.996&1&4910--7830&10&6490--10770\\
SDSS~J132512.49$+$112329.7&19.18&4.412&0&4330--7240&11&5890--10160\\
SDSS~J133203.86$+$553105.0&19.23&4.734&0&4620--7530&17&6190--10470\\
SDSS~J133250.08$+$465108.6&19.62&4.855&0&4620--7530&9&6490--10770\\
SDSS~J133412.56$+$122020.7&19.94&5.134&0&5140--8060&7&6480--10720\\
SDSS~J133728.81$+$415539.8&19.55&5.015&0&4910--7830&10&6490--10780\\
SDSS~J134015.03$+$392630.7&19.53&5.026&0&4910--7830&16&6490--10780\\
SDSS~J134040.24$+$281328.1&20.05&5.338&0&5110--8030&7&6490--10770\\
SDSS~J134134.19$+$014157.7&18.91&4.670&0&4630--7560&10&6190--10470\\
SDSS~J134141.45$+$461110.3&20.22&5.023&0&4910--7830&7&6490--10780\\
SDSS~J134154.01$+$351005.6&19.71&5.267&0&5110--8030&10&6490--10780\\
SDSS~J134743.29$+$495621.3&17.62&4.510&0&4320--7230&20&5890--10160\\
SDSS~J134819.87$+$181925.8&19.21&4.961&6&4950--7870&10&6470--10720\\
SDSS~J140146.53$+$024434.7&18.59&4.441&6&4340--7250&27&5890--10160\\
SDSS~J140404.63$+$031403.9&19.53&4.870&4&4630--7560&7&6470--10720\\
SDSS~J140503.29$+$334149.8&18.83&4.459&4&4320--7220&19&5890--10160\\
SDSS~J141209.96$+$062406.9&19.45&4.467&3&4330--7240&10&5890--10170\\
SDSS~J141914.18$-$015012.6&19.07&4.586&1&4630--7560&7&6190--10470\\
SDSS~J142025.75$+$615510.0&19.17&4.434&0&4320--7230&17&5890--10160\\
SDSS~J142144.98$+$351315.4&18.96&4.556&5&4620--7530&10&6190--10470\\
SDSS~J142325.92$+$130300.6&19.71&5.037&0&4950--7870&7&6470--10720\\
SDSS~J142526.09$+$082718.4&18.81&4.945&0&4950--7870&10&6470--10710\\
SDSS~J142705.86$+$330817.9&18.90&4.718&0&4620--7530&20&6200--10470\\
SDSS~J143352.21$+$022713.9&18.34&4.721&4&4630--7560&10&6180--10460\\
SDSS~J143605.00$+$213239.2&20.01&5.250&0&5110--8030&12&6500--10780\\
SDSS~J143629.94$+$063508.0&19.65&4.851&0&4630--7560&7&6470--10720\\
SDSS~J143751.82$+$232313.3&19.52&5.317&0&5110--8030&10&6490--10770\\
SDSS~J143835.95$+$431459.2&17.64&4.611&0&4620--7530&25&6200--10470\\
SDSS~J143850.48$+$055622.6&19.30&4.437&3&4330--7240&20&5890--10170\\
SDSS~J144331.17$+$272436.7&19.02&4.443&6&4320--7230&35&5890--10160\\
SDSS~J144407.63$-$010152.7&19.30&4.518&0&4330--7240&10&5890--10170\\
SDSS~J144717.97$+$040112.4&19.17&4.580&1&4630--7560&7&6190--10470\\
SDSS~J145107.93$+$025615.6&19.19&4.481&0&4330--7240&10&5890--10170\\
SDSS~J150027.89$+$434200.8&19.01&4.643&0&4620--7530&18&6200--10470\\
SDSS~J150802.28$+$430645.4&19.09&4.694&0&4620--7530&17&6190--10470\\
\hline
\end{tabular*}
\end{table*}

\begin{table*}
\contcaption{GGG survey.}
\begin{tabular*}{0.65\linewidth}{@{\extracolsep{\fill}}lcccccc}
\hline
Quasar &$i$ [mag]&$z_\mathrm{SDSS}$ &flag$^\star$ &B600 [\AA] &S/N$^\dag$ &R400 [\AA]\\ 
\hline
SDSS~J151155.98$+$040802.9&19.57&4.686&0&4630--7560&10&6190--10460\\
SDSS~J151320.89$+$105807.3&19.39&4.618&4&4620--7550&13&6210--10490\\
SDSS~J151719.09$+$490003.2&19.57&4.681&0&4620--7530&14&6200--10470\\
SDSS~J152005.93$+$233953.0&19.16&4.484&0&4320--7230&20&5890--10160\\
SDSS~J152404.23$+$134417.5&19.50&4.810&0&4630--7560&7&6470--10720\\
SDSS~J153247.41$+$223704.1&18.32&4.417&0&4320--7230&20&5890--10160\\
SDSS~J153459.75$+$132701.4&19.88&5.059&0&4910--7840&7&6470--10710\\
SDSS~J153650.25$+$500810.3&18.52&4.927&0&4910--7830&33&6490--10780\\
SDSS~J154352.92$+$333759.5&19.58&4.602&0&4610--7530&13&6190--10470\\
SDSS~J155243.04$+$255229.2&19.58&4.667&2&4620--7530&10&6200--10470\\
SDSS~J155426.16$+$193703.0&18.00&4.612&0&4630--7560&26&6180--10460\\
SDSS~J160336.64$+$350824.3&18.55&4.460&4&4320--7220&32&5890--10160\\
SDSS~J160516.16$+$210638.5&19.01&4.475&0&4320--7230&10&5890--10160\\
SDSS~J160734.22$+$160417.4&19.24&4.798&0&4630--7560&10&6180--10460\\
SDSS~J161105.64$+$084435.4&18.91&4.545&0&4330--7240&21&5890--10160\\
SDSS~J161425.13$+$464028.9&20.11&5.313&0&5110--8030&7&6490--10770\\
SDSS~J161447.03$+$205902.9&20.03&5.091&0&4910--7830&9&6490--10770\\
SDSS~J161616.26$+$513336.9&19.47&4.536&0&4320--7230&21&5890--10160\\
SDSS~J161622.10$+$050127.7&18.83&4.872&0&4630--7550&11&6470--10720\\
SDSS~J162445.03$+$271418.7&18.68&4.496&0&4320--7230&29&5890--10160\\
SDSS~J162626.50$+$275132.4&19.30&5.275&0&5110--8030&10&6490--10780\\
SDSS~J162629.19$+$285857.5&19.94&5.022&0&4910--7830&5&6490--10780\\
SDSS~J163411.82$+$215325.0&19.03&4.529&3&4320--7230&20&5890--10160\\
SDSS~J163636.93$+$315717.1&18.61&4.559&0&4620--7530&17&6200--10470\\
SDSS~J165902.12$+$270935.1&19.50&5.312&0&5110--8030&7&6490--10770\\
SDSS~J173744.87$+$582829.6&19.34&4.916&2&4910--7830&14&6490--10780\\
SDSS~J205724.14$-$003018.7&18.70&4.663&0&4630--7560&11&6180--10460\\
SDSS~J214725.71$-$083834.6&18.32&4.588&0&4630--7560&12&6190--10470\\
SDSS~J220008.66$+$001744.9&19.15&4.817&0&4630--7560&11&6470--10720\\
SDSS~J222509.19$-$001406.9&19.34&4.885&0&4950--7870&12&6470--10720\\
SDSS~J222845.14$-$075755.3&19.88&5.142&0&5130--8060&7&6470--10720\\
SDSS~J225246.43$+$142525.8&19.80&4.904&0&4620--7550&10&6480--10720\\
SDSS~J234003.51$+$140257.1&18.98&4.559&0&4620--7550&10&6220--10500\\
\hline
\end{tabular*}
\begin{flushleft}
$^\star$Flag describing quasar spectrum: (1) BAL; (2) associated absorption; (3) weak-lined QSO; (4) candidate weak-lined QSO; (5) PDLA; (6) PLLS; (7) contaminated by neighbour.\\
$^\dag$Signal-to-Noise ratio per 2.76\,\AA\ pixel at rest-frame wavelength 1450\,\AA\ (R400).
\end{flushleft}
\end{table*}

\begin{table*}
\caption{\label{tab:ggg_zem}GGG Quasar emission redshifts.}
\begin{tabular*}{0.70\linewidth}{@{\extracolsep{\fill}}lcccccc}
\hline
Quasar &$z_\mathrm{SDSS}$ &$z_\mathrm{GGG}$ &Shen Lines &$z_\mathrm{Shen}$ &$z_\mathrm{HW}$\\ 
\hline
SDSS~J001115.23$+$144601.8&4.967&$4.970\pm0.005$&\ion{Si}{4}, \ion{C}{4}&$4.959\pm0.014$&$4.974$\\
SDSS~J004054.65$-$091526.8&4.976&$4.980\pm0.010$&\ion{Si}{4}&$4.981\pm0.016$&$4.977$\\
SDSS~J010619.24$+$004823.3&4.449&$4.449\pm0.005$&\ion{Si}{4}, \ion{C}{4}&$4.434\pm0.013$&$4.450$\\
SDSS~J012509.42$-$104300.8&4.492&$4.498\pm0.005$&\ion{Si}{4}, \ion{C}{4}&$4.500\pm0.013$&$4.503$\\
SDSS~J021043.16$-$001818.4&4.733&$4.770\pm0.020$&\ion{Si}{4}, \ion{C}{4}&$4.746\pm0.014$&$4.708$\\
SDSS~J023137.65$-$072854.4&5.413&$5.420\pm0.005$&\ion{Si}{4}, \ion{C}{4}&$5.418\pm0.015$&$5.415$\\
SDSS~J033119.66$-$074143.1&4.739&$4.734\pm0.010$&\ion{Si}{4}, \ion{C}{4}&$4.734\pm0.014$&$4.733$\\
SDSS~J033829.30$+$002156.2&5.032&$5.040\pm0.020$&\ion{Si}{4}, \ion{C}{4}&$5.030\pm0.014$&$5.033$\\
SDSS~J073103.12$+$445949.4&5.004&$4.998\pm0.005$&\ion{Si}{4}, \ion{C}{4}&$4.988\pm0.014$&$5.009$\\
SDSS~J075907.57$+$180054.7&4.862&$4.820\pm0.010$&\ion{Si}{4}, \ion{C}{4}&$4.798\pm0.014$&$4.815$\\
SDSS~J080023.01$+$305101.1&4.685&$4.676\pm0.005$&\ion{Si}{4}, \ion{C}{4}&$4.690\pm0.014$&$4.689$\\
SDSS~J080715.11$+$132805.1&4.875&$4.880\pm0.010$&\ion{Si}{4}, \ion{C}{4}&$4.858\pm0.014$&$4.872$\\
SDSS~J081806.87$+$071920.2&4.581&$4.625\pm0.010$&\ion{Si}{4}&$4.580\pm0.015$&$4.616$\\
SDSS~J082212.34$+$160436.9&4.488&$4.510\pm0.020$&\ion{C}{4}&$4.502\pm0.015$&$4.481$\\
SDSS~J082454.02$+$130217.0&5.188&$5.207\pm0.005$&\ion{Si}{4}, \ion{C}{4}&$5.193\pm0.015$&...\\
SDSS~J083655.80$+$064104.6&4.436&$4.435\pm0.010$&\ion{Si}{4}, \ion{C}{4}&$4.401\pm0.013$&$4.416$\\
SDSS~J083920.53$+$352459.3&4.777&$4.784\pm0.010$&\ion{Si}{4}&$4.790\pm0.015$&$4.794$\\
SDSS~J084627.84$+$080051.7&5.030&$5.028\pm0.010$&\ion{Si}{4}&$5.047\pm0.016$&$5.031$\\
SDSS~J084631.52$+$241108.3&4.743&$4.742\pm0.005$&\ion{Si}{4}, \ion{C}{4}&$4.710\pm0.014$&$4.722$\\
SDSS~J085430.37$+$205650.8&5.179&$5.179\pm0.005$&\ion{Si}{4}, \ion{C}{4}&$5.177\pm0.015$&$5.191$\\
SDSS~J085707.94$+$321031.9&4.776&$4.796\pm0.010$&\ion{Si}{4}, \ion{C}{4}&$4.783\pm0.014$&$4.799$\\
SDSS~J090100.61$+$472536.2&4.608&$4.598\pm0.010$&\ion{Si}{4}, \ion{C}{4}&$4.593\pm0.013$&$4.606$\\
SDSS~J090245.76$+$085115.8&5.226&$5.226\pm0.010$&\ion{C}{4}&$5.267\pm0.017$&$5.227$\\
SDSS~J090634.84$+$023433.8&4.511&$4.516\pm0.005$&\ion{Si}{4}, \ion{C}{4}&$4.521\pm0.013$&$4.504$\\
SDSS~J091316.55$+$591921.6&5.122&$5.122\pm0.005$&\ion{Si}{4}, \ion{C}{4}&$5.121\pm0.015$&$5.122$\\
SDSS~J091543.63$+$492416.6&5.196&$5.199\pm0.005$&\ion{Si}{4}, \ion{C}{4}&$5.210\pm0.015$&$5.197$\\
SDSS~J092216.81$+$265358.9&5.032&$5.042\pm0.010$&\ion{Si}{4}, \ion{C}{4}&$5.041\pm0.014$&$5.033$\\
SDSS~J093523.31$+$411518.5&4.787&$4.806\pm0.010$&\ion{Si}{4}, \ion{C}{4}&$4.763\pm0.014$&$4.782$\\
SDSS~J094056.01$+$584830.2&4.659&$4.664\pm0.005$&\ion{Si}{4}, \ion{C}{4}&$4.656\pm0.013$&$4.658$\\
SDSS~J094108.36$+$594725.7&4.790&$4.852\pm0.005$&\ion{Si}{4}, \ion{C}{4}&$4.850\pm0.014$&$4.861$\\
SDSS~J094409.52$+$100656.6&4.748&$4.776\pm0.005$&\ion{Si}{4}, \ion{C}{4}&$4.767\pm0.014$&$4.773$\\
SDSS~J095632.03$+$321612.6&4.647&$4.632\pm0.005$&\ion{Si}{4}, \ion{C}{4}&$4.601\pm0.013$&$4.618$\\
SDSS~J095707.67$+$061059.5&5.185&$5.167\pm0.005$&\ion{Si}{4}, \ion{C}{4}&$5.145\pm0.015$&$5.180$\\
SDSS~J100251.20$+$223135.1&4.744&$4.761\pm0.005$&\ion{Si}{4}, \ion{C}{4}&$4.758\pm0.014$&$4.755$\\
SDSS~J100416.12$+$434739.0&4.872&$4.872\pm0.010$&\ion{Si}{4}, \ion{C}{4}&$4.853\pm0.014$&$4.879$\\
SDSS~J100444.30$+$202520.0&5.084&$5.084\pm0.020$&\ion{C}{4}&$5.089\pm0.016$&$5.085$\\
SDSS~J101549.00$+$002020.0&4.403&$4.406\pm0.005$&\ion{Si}{4}, \ion{C}{4}&$4.406\pm0.013$&$4.416$\\
SDSS~J102332.07$+$633508.0&4.881&$4.872\pm0.010$&\ion{Si}{4}, \ion{C}{4}&$4.864\pm0.014$&$4.882$\\
SDSS~J102622.87$+$471907.2&4.943&$4.932\pm0.010$&\ion{Si}{4}, \ion{C}{4}&$4.928\pm0.014$&$4.948$\\
SDSS~J102623.61$+$254259.5&5.303&$5.254\pm0.020$&\ion{Si}{4}, \ion{C}{4}&$5.263\pm0.015$&$5.285$\\
SDSS~J103418.65$+$203300.2&4.998&$4.998\pm0.005$&\ion{Si}{4}, \ion{C}{4}&$4.994\pm0.014$&$4.999$\\
SDSS~J103601.03$+$500831.7&4.470&$4.480\pm0.020$&\ion{Si}{4}&$4.528\pm0.015$&$4.529$\\
SDSS~J103711.04$+$313433.5&4.885&$4.916\pm0.010$&\ion{Si}{4}, \ion{C}{4}&$4.881\pm0.014$&$4.873$\\
SDSS~J103919.28$+$344504.5&4.420&$4.421\pm0.005$&\ion{Si}{4}, \ion{C}{4}&$4.409\pm0.013$&$4.410$\\
SDSS~J104041.09$+$162233.8&4.814&$4.809\pm0.010$&\ion{Si}{4}, \ion{C}{4}&$4.822\pm0.014$&...\\
SDSS~J104325.55$+$404849.5&4.934&$4.923\pm0.005$&\ion{Si}{4}, \ion{C}{4}&$4.921\pm0.014$&$4.930$\\
SDSS~J104351.19$+$650647.6&4.471&$4.516\pm0.010$&\ion{C}{4}&$4.481\pm0.015$&$4.482$\\
SDSS~J105020.40$+$262002.3&4.796&$4.892\pm0.010$&\ion{Si}{4}, \ion{C}{4}&$4.807\pm0.014$&$4.838$\\
SDSS~J105036.46$+$580424.6&5.132&$5.151\pm0.005$&\ion{Si}{4}, \ion{C}{4}&$5.174\pm0.015$&$5.133$\\
SDSS~J105322.98$+$580412.1&5.215&$5.250\pm0.020$&\ion{Si}{4}, \ion{C}{4}&$5.240\pm0.015$&$5.260$\\
SDSS~J105445.43$+$163337.4&5.187&$5.154\pm0.005$&\ion{Si}{4}, \ion{C}{4}&$5.129\pm0.015$&$5.148$\\
SDSS~J110045.23$+$112239.1&4.707&$4.728\pm0.010$&\ion{Si}{4}, \ion{C}{4}&$4.736\pm0.014$&$4.734$\\
SDSS~J110134.36$+$053133.8&4.987&$5.045\pm0.020$&\ion{Si}{4}, \ion{C}{4}&$4.930\pm0.014$&$4.998$\\
SDSS~J111523.24$+$082918.4&4.640&$4.710\pm0.030$&\ion{Si}{4}, \ion{C}{4}&$4.684\pm0.014$&$4.641$\\
SDSS~J111741.26$+$261039.2&4.635&$4.626\pm0.010$&\ion{Si}{4}, \ion{C}{4}&$4.626\pm0.013$&$4.641$\\
SDSS~J111920.64$+$345248.1&5.011&$4.992\pm0.005$&\ion{Si}{4}, \ion{C}{4}&$4.981\pm0.014$&$5.012$\\
SDSS~J112253.50$+$005329.7&4.551&$4.586\pm0.005$&\ion{Si}{4}, \ion{C}{4}&$4.578\pm0.013$&$4.584$\\
SDSS~J112534.93$+$380149.3&4.618&$4.606\pm0.010$&\ion{Si}{4}, \ion{C}{4}&$4.581\pm0.013$&$4.588$\\
SDSS~J112857.84$+$575909.8&4.978&$4.992\pm0.020$&\ion{Si}{4}, \ion{C}{4}&$4.973\pm0.014$&$5.000$\\
SDSS~J113246.50$+$120901.6&5.167&$5.180\pm0.005$&\ion{Si}{4}&$5.118\pm0.016$&$5.168$\\
SDSS~J114008.67$+$620530.0&4.521&$4.523\pm0.005$&\ion{Si}{4}, \ion{C}{4}&$4.522\pm0.013$&$4.530$\\
SDSS~J114225.30$+$110217.3&4.590&$4.596\pm0.005$&\ion{Si}{4}, \ion{C}{4}&$4.588\pm0.013$&$4.590$\\
SDSS~J114657.79$+$403708.6&5.005&$4.996\pm0.010$&\ion{Si}{4}, \ion{C}{4}&$4.988\pm0.014$&$5.006$\\
SDSS~J114826.16$+$302019.3&5.142&$5.128\pm0.005$&\ion{Si}{4}, \ion{C}{4}&$5.104\pm0.015$&$5.143$\\
SDSS~J114914.88$+$281308.7&4.553&$4.556\pm0.005$&\ion{Si}{4}, \ion{C}{4}&$4.548\pm0.013$&$4.554$\\
\hline
\end{tabular*}
\end{table*}

\begin{table*}
\contcaption{GGG Quasar emission redshifts.}
\begin{tabular*}{0.70\linewidth}{@{\extracolsep{\fill}}lcccccc}
\hline
Quasar &$z_\mathrm{SDSS}$ &$z_\mathrm{GGG}$ &Shen Lines &$z_\mathrm{Shen}$ &$z_\mathrm{HW}$\\ 
\hline
SDSS~J115424.73$+$134145.7&5.010&$5.060\pm0.020$&\ion{Si}{4}&$5.005\pm0.016$&$5.011$\\
SDSS~J115809.39$+$634252.8&4.494&$4.479\pm0.010$&\ion{Si}{4}, \ion{C}{4}&$4.439\pm0.013$&$4.444$\\
SDSS~J120036.72$+$461850.2&4.730&$4.741\pm0.010$&\ion{Si}{4}, \ion{C}{4}&$4.752\pm0.014$&$4.754$\\
SDSS~J120055.61$+$181732.9&4.984&$4.995\pm0.020$&\ion{Si}{4}, \ion{C}{4}&$4.998\pm0.014$&...\\
SDSS~J120102.01$+$073648.1&4.443&$4.472\pm0.005$&\ion{C}{4}&$4.454\pm0.014$&$4.460$\\
SDSS~J120110.31$+$211758.5&4.579&$4.579\pm0.005$&\ion{Si}{4}, \ion{C}{4}&$4.564\pm0.013$&...\\
SDSS~J120131.56$+$053510.1&4.830&$4.840\pm0.020$&\ion{Si}{4}, \ion{C}{4}&$4.744\pm0.014$&$4.831$\\
SDSS~J120207.78$+$323538.8&5.292&$5.298\pm0.005$&\ion{Si}{4}, \ion{C}{4}&$5.256\pm0.015$&$5.275$\\
SDSS~J120441.73$-$002149.6&5.032&$5.094\pm0.010$&\ion{Si}{4}&$5.108\pm0.016$&$5.033$\\
SDSS~J120725.27$+$321530.4&4.643&$4.621\pm0.005$&\ion{Si}{4}, \ion{C}{4}&$4.614\pm0.013$&$4.609$\\
SDSS~J120730.84$+$153338.1&4.465&$4.452\pm0.010$&\ion{C}{4}&$4.456\pm0.014$&$4.450$\\
SDSS~J120952.72$+$183147.2&5.158&$5.127\pm0.005$&\ion{C}{4}&$5.157\pm0.016$&...\\
SDSS~J121134.04$+$484235.9&4.505&$4.544\pm0.005$&\ion{Si}{4}, \ion{C}{4}&$4.518\pm0.013$&$4.508$\\
SDSS~J121422.02$+$665707.5&4.639&$4.650\pm0.020$&\ion{Si}{4}, \ion{C}{4}&$4.626\pm0.013$&$4.650$\\
SDSS~J122016.05$+$315253.0&4.891&$4.900\pm0.010$&\ion{Si}{4}, \ion{C}{4}&$4.847\pm0.014$&$4.861$\\
SDSS~J122146.42$+$444528.0&5.206&$5.203\pm0.005$&\ion{Si}{4}, \ion{C}{4}&$5.201\pm0.015$&$5.207$\\
SDSS~J122237.96$+$195842.9&5.189&$5.120\pm0.020$&\ion{Si}{4}, \ion{C}{4}&$5.119\pm0.015$&...\\
SDSS~J123333.47$+$062234.2&5.289&$5.300\pm0.020$&\ion{Si}{4}, \ion{C}{4}&$5.339\pm0.015$&$5.290$\\
SDSS~J124247.91$+$521306.8&5.018&$5.036\pm0.005$&\ion{Si}{4}, \ion{C}{4}&$5.054\pm0.014$&$5.018$\\
SDSS~J124400.04$+$553406.8&4.625&$4.660\pm0.020$&\ion{Si}{4}&$4.628\pm0.015$&$4.637$\\
SDSS~J124515.46$+$382247.5&4.933&$4.963\pm0.010$&\ion{Si}{4}, \ion{C}{4}&$4.962\pm0.014$&$4.934$\\
SDSS~J125025.40$+$183458.1&4.599&$4.557\pm0.005$&\ion{Si}{4}, \ion{C}{4}&$4.543\pm0.013$&...\\
SDSS~J125353.35$+$104603.1&4.909&$4.918\pm0.005$&\ion{Si}{4}&$4.915\pm0.016$&...\\
SDSS~J125718.02$+$374729.9&4.750&$4.733\pm0.010$&\ion{Si}{4}, \ion{C}{4}&$4.728\pm0.014$&$4.741$\\
SDSS~J130002.16$+$011823.0&4.613&$4.619\pm0.005$&\ion{Si}{4}, \ion{C}{4}&$4.603\pm0.013$&$4.601$\\
SDSS~J130110.95$+$252738.3&4.660&$4.666\pm0.005$&\ion{Si}{4}, \ion{C}{4}&$4.664\pm0.013$&...\\
SDSS~J130152.55$+$221012.1&4.829&$4.805\pm0.020$&\ion{Si}{4}, \ion{C}{4}&$4.758\pm0.014$&...\\
SDSS~J130215.71$+$550553.5&4.436&$4.461\pm0.010$&\ion{Si}{4}, \ion{C}{4}&$4.444\pm0.013$&$4.454$\\
SDSS~J130619.38$+$023658.9&4.837&$4.860\pm0.020$&\ion{Si}{4}, \ion{C}{4}&$4.760\pm0.014$&$4.802$\\
SDSS~J130917.12$+$165758.5&4.692&$4.714\pm0.010$&\ion{Si}{4}, \ion{C}{4}&$4.720\pm0.014$&...\\
SDSS~J131234.08$+$230716.3&4.996&$4.960\pm0.020$&\ion{Si}{4}, \ion{C}{4}&$4.952\pm0.014$&...\\
SDSS~J132512.49$+$112329.7&4.412&$4.412\pm0.005$&\ion{Si}{4}, \ion{C}{4}&$4.416\pm0.013$&$4.415$\\
SDSS~J133203.86$+$553105.0&4.734&$4.737\pm0.005$&\ion{Si}{4}, \ion{C}{4}&$4.739\pm0.014$&$4.748$\\
SDSS~J133250.08$+$465108.6&4.855&$4.844\pm0.010$&\ion{Si}{4}, \ion{C}{4}&$4.857\pm0.014$&$4.867$\\
SDSS~J133412.56$+$122020.7&5.134&$5.130\pm0.010$&\ion{Si}{4}&$5.093\pm0.016$&$5.135$\\
SDSS~J133728.81$+$415539.8&5.015&$5.018\pm0.010$&\ion{Si}{4}, \ion{C}{4}&$5.021\pm0.014$&$5.016$\\
SDSS~J134015.03$+$392630.7&5.026&$5.048\pm0.010$&\ion{Si}{4}, \ion{C}{4}&$5.041\pm0.014$&$5.052$\\
SDSS~J134040.24$+$281328.1&5.338&$5.349\pm0.005$&\ion{Si}{4}, \ion{C}{4}&$5.339\pm0.015$&$5.339$\\
SDSS~J134134.19$+$014157.7&4.670&$4.696\pm0.010$&\ion{Si}{4}, \ion{C}{4}&$4.701\pm0.014$&$4.677$\\
SDSS~J134141.45$+$461110.3&5.023&$5.003\pm0.005$&\ion{Si}{4}, \ion{C}{4}&$5.008\pm0.014$&$4.998$\\
SDSS~J134154.01$+$351005.6&5.267&$5.252\pm0.005$&\ion{Si}{4}, \ion{C}{4}&$5.245\pm0.015$&$5.253$\\
SDSS~J134743.29$+$495621.3&4.510&$4.563\pm0.005$&\ion{Si}{4}, \ion{C}{4}&$4.536\pm0.013$&$4.538$\\
SDSS~J134819.87$+$181925.8&4.961&$4.954\pm0.005$&\ion{Si}{4}, \ion{C}{4}&$4.941\pm0.014$&...\\
SDSS~J140146.53$+$024434.7&4.441&$4.415\pm0.020$&\ion{Si}{4}, \ion{C}{4}&$4.412\pm0.013$&$4.409$\\
SDSS~J140404.63$+$031403.9&4.870&$4.924\pm0.020$&\ion{Si}{4}, \ion{C}{4}&$4.783\pm0.014$&$4.871$\\
SDSS~J140503.29$+$334149.8&4.459&$4.467\pm0.010$&\ion{C}{4}&$4.485\pm0.015$&$4.460$\\
SDSS~J141209.96$+$062406.9&4.467&$4.411\pm0.010$&\ion{C}{4}&$4.513\pm0.015$&$4.467$\\
SDSS~J141914.18$-$015012.6&4.586&$4.590\pm0.020$&\ion{Si}{4}, \ion{C}{4}&$4.545\pm0.013$&$4.571$\\
SDSS~J142025.75$+$615510.0&4.434&$4.448\pm0.005$&\ion{C}{4}&$4.447\pm0.014$&$4.452$\\
SDSS~J142144.98$+$351315.4&4.556&$4.599\pm0.005$&\ion{Si}{4}, \ion{C}{4}&$4.564\pm0.013$&$4.561$\\
SDSS~J142325.92$+$130300.6&5.037&$5.048\pm0.010$&\ion{Si}{4}&$5.074\pm0.016$&$5.030$\\
SDSS~J142526.09$+$082718.4&4.945&$4.955\pm0.005$&\ion{Si}{4}, \ion{C}{4}&$4.968\pm0.014$&$4.970$\\
SDSS~J142705.86$+$330817.9&4.718&$4.703\pm0.005$&\ion{Si}{4}, \ion{C}{4}&$4.694\pm0.014$&$4.708$\\
SDSS~J143352.21$+$022713.9&4.721&$4.729\pm0.010$&\ion{Si}{4}, \ion{C}{4}&$4.730\pm0.014$&$4.685$\\
SDSS~J143605.00$+$213239.2&5.250&$5.227\pm0.005$&\ion{Si}{4}, \ion{C}{4}&$5.233\pm0.015$&...\\
SDSS~J143629.94$+$063508.0&4.851&$4.828\pm0.005$&\ion{Si}{4}&$5.094\pm0.016$&$4.816$\\
SDSS~J143751.82$+$232313.3&5.317&$5.320\pm0.005$&\ion{Si}{4}, \ion{C}{4}&$5.316\pm0.015$&$5.333$\\
SDSS~J143835.95$+$431459.2&4.611&$4.686\pm0.010$&\ion{Si}{4}, \ion{C}{4}&$4.653\pm0.013$&$4.674$\\
SDSS~J143850.48$+$055622.6&4.437&$4.437\pm0.010$&\ion{C}{4}&$4.506\pm0.015$&$4.428$\\
SDSS~J144331.17$+$272436.7&4.443&$4.424\pm0.010$&\ion{Si}{4}, \ion{C}{4}&$4.410\pm0.013$&...\\
SDSS~J144407.63$-$010152.7&4.518&$4.530\pm0.005$&\ion{Si}{4}, \ion{C}{4}&$4.514\pm0.013$&$4.521$\\
SDSS~J144717.97$+$040112.4&4.580&$4.580\pm0.020$&\ion{Si}{4}, \ion{C}{4}&$4.514\pm0.013$&$4.584$\\
SDSS~J145107.93$+$025615.6&4.481&$4.483\pm0.005$&\ion{C}{4}&$4.483\pm0.015$&$4.482$\\
SDSS~J150027.89$+$434200.8&4.643&$4.641\pm0.005$&\ion{Si}{4}, \ion{C}{4}&$4.619\pm0.013$&$4.639$\\
SDSS~J150802.28$+$430645.4&4.694&$4.681\pm0.010$&\ion{Si}{4}, \ion{C}{4}&$4.680\pm0.014$&$4.683$\\
\hline
\end{tabular*}
\end{table*}

\begin{table*}
\contcaption{GGG Quasar emission redshifts.}
\begin{tabular*}{0.70\linewidth}{@{\extracolsep{\fill}}lcccccc}
\hline
Quasar &$z_\mathrm{SDSS}$ &$z_\mathrm{GGG}$ &Shen Lines &$z_\mathrm{Shen}$ &$z_\mathrm{HW}$\\ 
\hline
SDSS~J151155.98$+$040802.9&4.686&$4.679\pm0.005$&\ion{Si}{4}, \ion{C}{4}&$4.689\pm0.014$&$4.674$\\
SDSS~J151320.89$+$105807.3&4.618&$4.625\pm0.020$&\ion{Si}{4}, \ion{C}{4}&$4.607\pm0.013$&$4.618$\\
SDSS~J151719.09$+$490003.2&4.681&$4.660\pm0.010$&\ion{Si}{4}, \ion{C}{4}&$4.649\pm0.013$&$4.652$\\
SDSS~J152005.93$+$233953.0&4.484&$4.487\pm0.005$&\lya&$4.467\pm0.027$&...\\
SDSS~J152404.23$+$134417.5&4.810&$4.788\pm0.010$&\ion{Si}{4}, \ion{C}{4}&$4.786\pm0.014$&...\\
SDSS~J153247.41$+$223704.1&4.417&$4.434\pm0.005$&\ion{Si}{4}, \ion{C}{4}&$4.412\pm0.013$&...\\
SDSS~J153459.75$+$132701.4&5.059&$5.043\pm0.010$&\ion{Si}{4}, \ion{C}{4}&$5.072\pm0.014$&...\\
SDSS~J153650.25$+$500810.3&4.927&$4.929\pm0.005$&\ion{Si}{4}&$4.926\pm0.016$&$4.941$\\
SDSS~J154352.92$+$333759.5&4.602&$4.604\pm0.005$&\ion{Si}{4}, \ion{C}{4}&$4.589\pm0.013$&$4.594$\\
SDSS~J155243.04$+$255229.2&4.667&$4.645\pm0.010$&\ion{Si}{4}, \ion{C}{4}&$4.652\pm0.013$&$4.667$\\
SDSS~J155426.16$+$193703.0&4.612&$4.632\pm0.005$&\ion{Si}{4}, \ion{C}{4}&$4.566\pm0.013$&...\\
SDSS~J160336.64$+$350824.3&4.460&$4.485\pm0.010$&\ion{Si}{4}, \ion{C}{4}&$4.429\pm0.013$&$4.405$\\
SDSS~J160516.16$+$210638.5&4.475&$4.495\pm0.010$&\ion{Si}{4}, \ion{C}{4}&$4.491\pm0.013$&$4.496$\\
SDSS~J160734.22$+$160417.4&4.798&$4.786\pm0.010$&\ion{Si}{4}, \ion{C}{4}&$4.772\pm0.014$&$4.783$\\
SDSS~J161105.64$+$084435.4&4.545&$4.545\pm0.005$&\ion{Si}{4}, \ion{C}{4}&$4.561\pm0.013$&$4.551$\\
SDSS~J161425.13$+$464028.9&5.313&$5.313\pm0.005$&\ion{Si}{4}, \ion{C}{4}&$5.320\pm0.015$&$5.316$\\
SDSS~J161447.03$+$205902.9&5.091&$5.081\pm0.020$&\ion{Si}{4}, \ion{C}{4}&$5.025\pm0.014$&$5.092$\\
SDSS~J161616.26$+$513336.9&4.536&$4.528\pm0.005$&\ion{Si}{4}&$4.523\pm0.015$&$4.530$\\
SDSS~J161622.10$+$050127.7&4.872&$4.876\pm0.010$&\ion{Si}{4}, \ion{C}{4}&$4.794\pm0.014$&$4.863$\\
SDSS~J162445.03$+$271418.7&4.496&$4.498\pm0.005$&\ion{C}{4}&$4.451\pm0.014$&$4.476$\\
SDSS~J162626.50$+$275132.4&5.275&$5.265\pm0.020$&\ion{Si}{4}, \ion{C}{4}&$5.143\pm0.015$&$5.214$\\
SDSS~J162629.19$+$285857.5&5.022&$5.035\pm0.010$&\ion{Si}{4}, \ion{C}{4}&$4.994\pm0.014$&$5.023$\\
SDSS~J163411.82$+$215325.0&4.529&$4.587\pm0.010$&\ion{Si}{4}, \ion{C}{4}&$4.508\pm0.013$&$4.501$\\
SDSS~J163636.93$+$315717.1&4.559&$4.590\pm0.020$&\ion{Si}{4}, \ion{C}{4}&$4.559\pm0.013$&$4.570$\\
SDSS~J165902.12$+$270935.1&5.312&$5.316\pm0.005$&\ion{Si}{4}, \ion{C}{4}&$5.289\pm0.015$&$5.306$\\
SDSS~J173744.87$+$582829.6&4.916&$4.905\pm0.020$&\ion{Si}{4}, \ion{C}{4}&$4.811\pm0.014$&$4.919$\\
SDSS~J205724.14$-$003018.7&4.663&$4.686\pm0.005$&\ion{Si}{4}, \ion{C}{4}&$4.697\pm0.014$&$4.685$\\
SDSS~J214725.71$-$083834.6&4.588&$4.597\pm0.005$&\ion{Si}{4}, \ion{C}{4}&$4.550\pm0.013$&$4.583$\\
SDSS~J220008.66$+$001744.9&4.817&$4.782\pm0.010$&\ion{Si}{4}, \ion{C}{4}&$4.781\pm0.014$&$4.799$\\
SDSS~J222509.19$-$001406.9&4.885&$4.882\pm0.010$&\ion{Si}{4}, \ion{C}{4}&$4.837\pm0.014$&$4.883$\\
SDSS~J222845.14$-$075755.3&5.142&$5.150\pm0.010$&\ion{Si}{4}, \ion{C}{4}&$5.132\pm0.015$&$5.143$\\
SDSS~J225246.43$+$142525.8&4.904&$4.881\pm0.010$&\ion{Si}{4}, \ion{C}{4}&$5.000\pm0.014$&$4.905$\\
SDSS~J234003.51$+$140257.1&4.559&$4.548\pm0.010$&\ion{Si}{4}, \ion{C}{4}&$4.531\pm0.013$&$4.551$\\
\hline
\end{tabular*}
\end{table*}

\clearpage

\begin{figure*}
\includegraphics[bb=43 152 567 743,clip,width=1.0\linewidth]{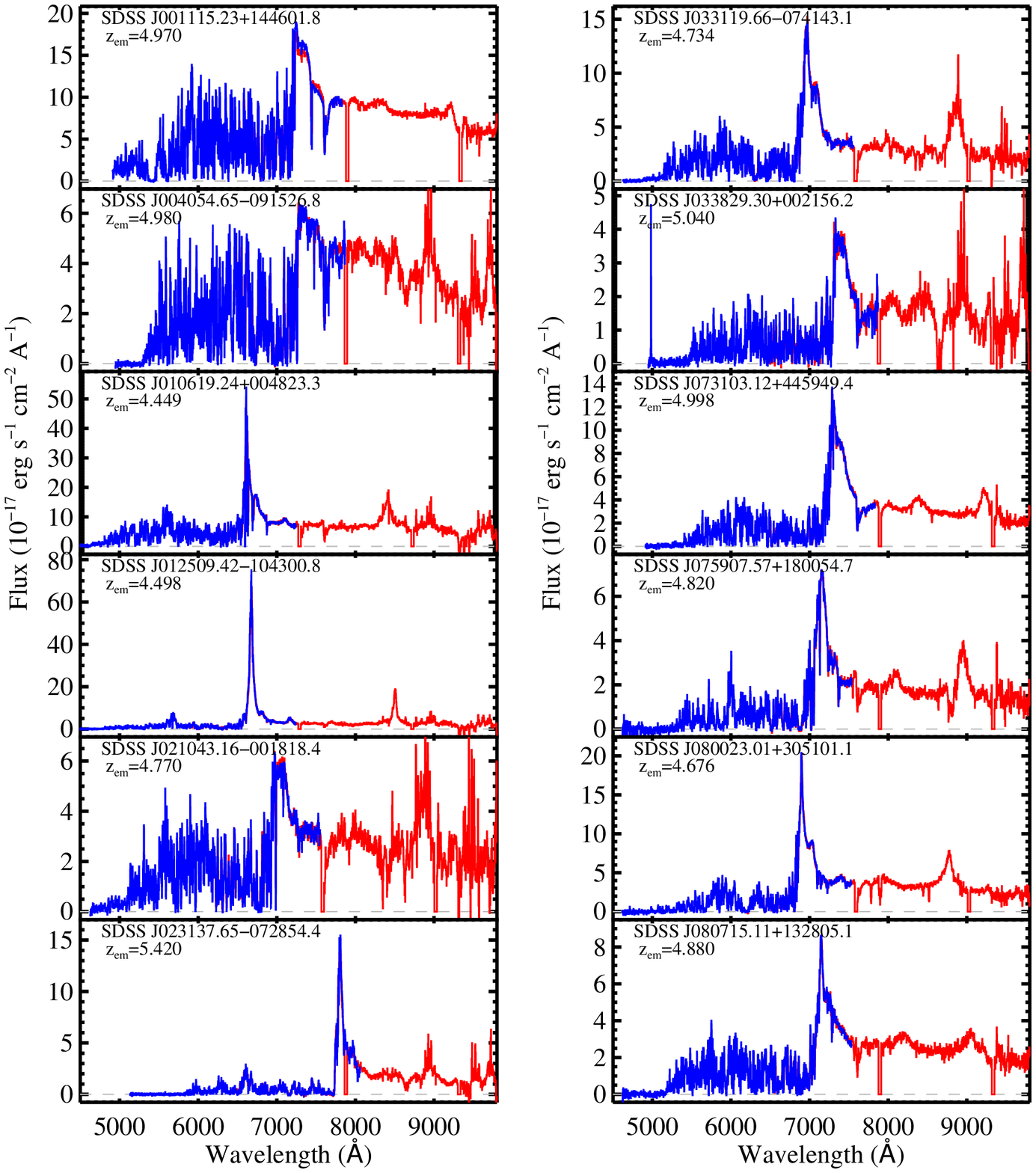}
\caption{\label{fig:all_spec_a}
Gemini/GMOS spectra of the 163 $z_\mathrm{em}>4.4$ quasars obtained in the
GGG survey.  The blue lines indicate the data obtained with the B600
grating at a spectral resolution FWHM~$\approx 5.5$\,\AA.  The red
lines, which overlap the blue data near \lya, trace the R400 grating
observations (FWHM~$\approx 8.0$\,\AA).  CCD gaps give
the zero values in these red spectra. All spectra were fluxed using
several spectrophotometric standard stars and scaled to the available
SDSS spectra. The R400 spectra taken with GMOS-S show fringing residuals
 at $\lambda\ga 8300$\,\AA, leading to enhanced sky line contamination.
}
\end{figure*}

\begin{figure*}
\includegraphics[bb=43 152 567 743,clip,width=1.0\linewidth]{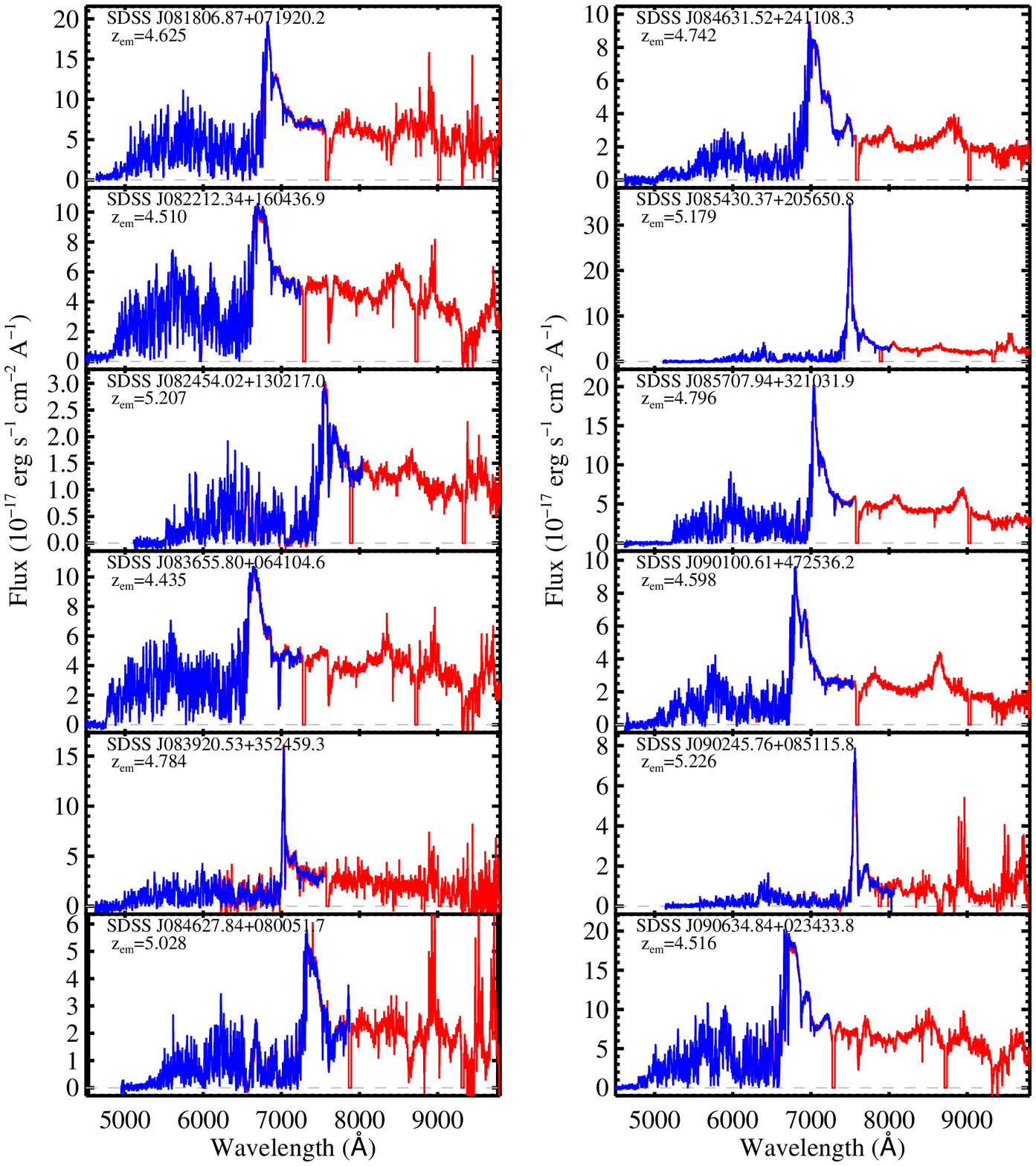}
\contcaption{}
\end{figure*}

\begin{figure*}
\includegraphics[bb=43 152 567 743,clip,width=1.0\linewidth]{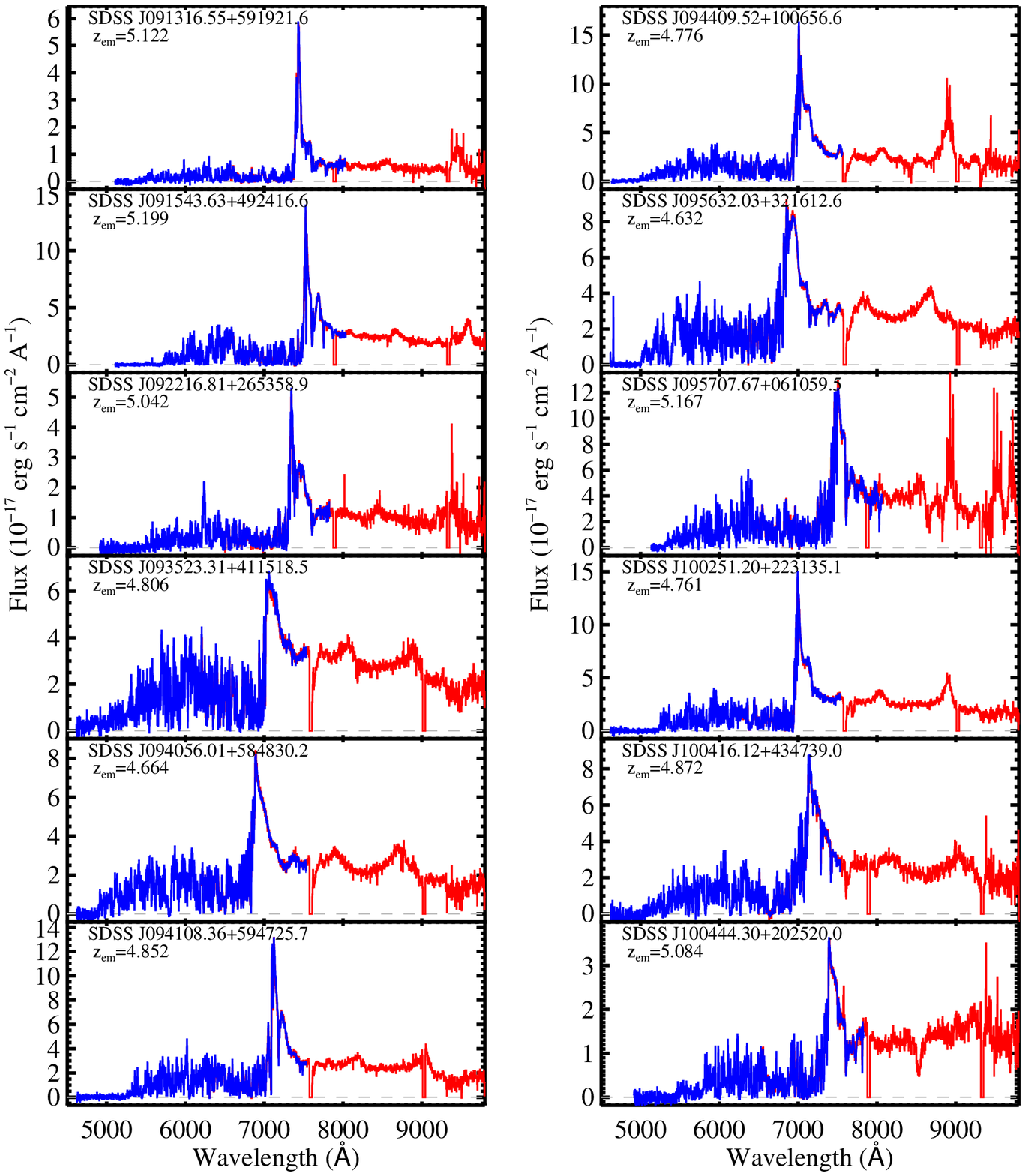}
\contcaption{}
\end{figure*}

\begin{figure*}
\includegraphics[bb=43 152 567 743,clip,width=1.0\linewidth]{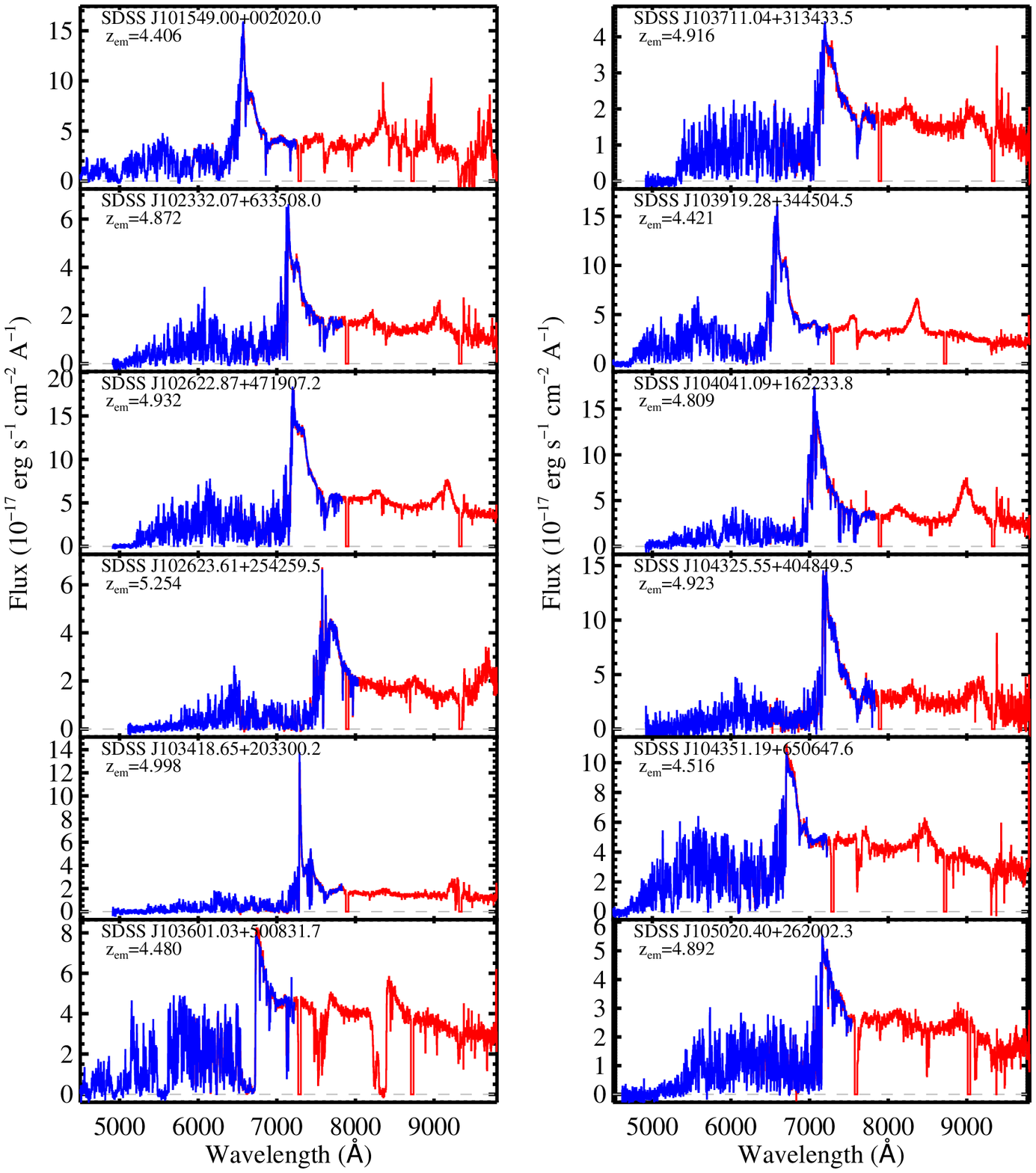}
\contcaption{}
\end{figure*}

\begin{figure*}
\includegraphics[bb=43 152 567 743,clip,width=1.0\linewidth]{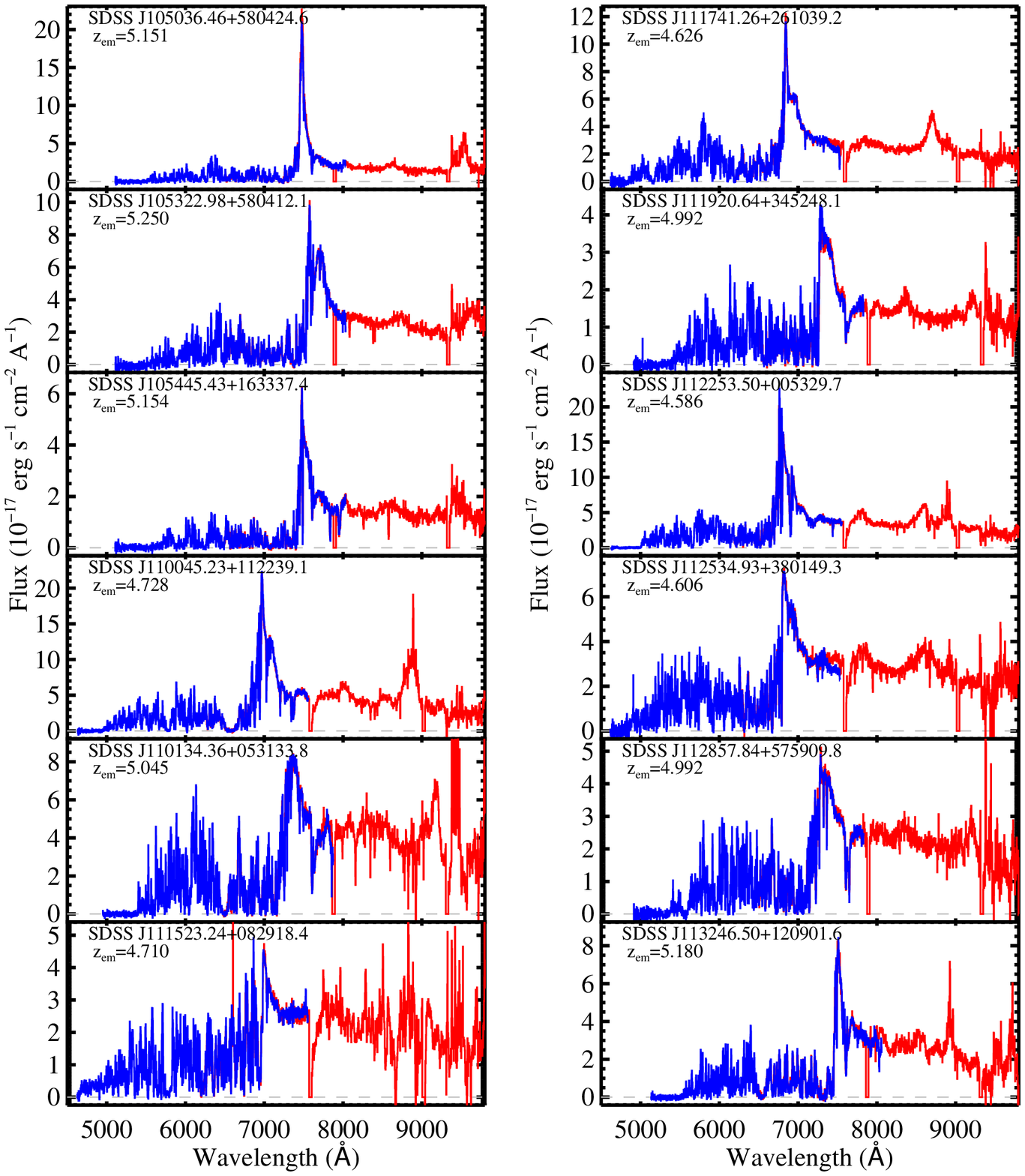}
\contcaption{}
\end{figure*}

\begin{figure*}
\includegraphics[bb=43 152 567 743,clip,width=1.0\linewidth]{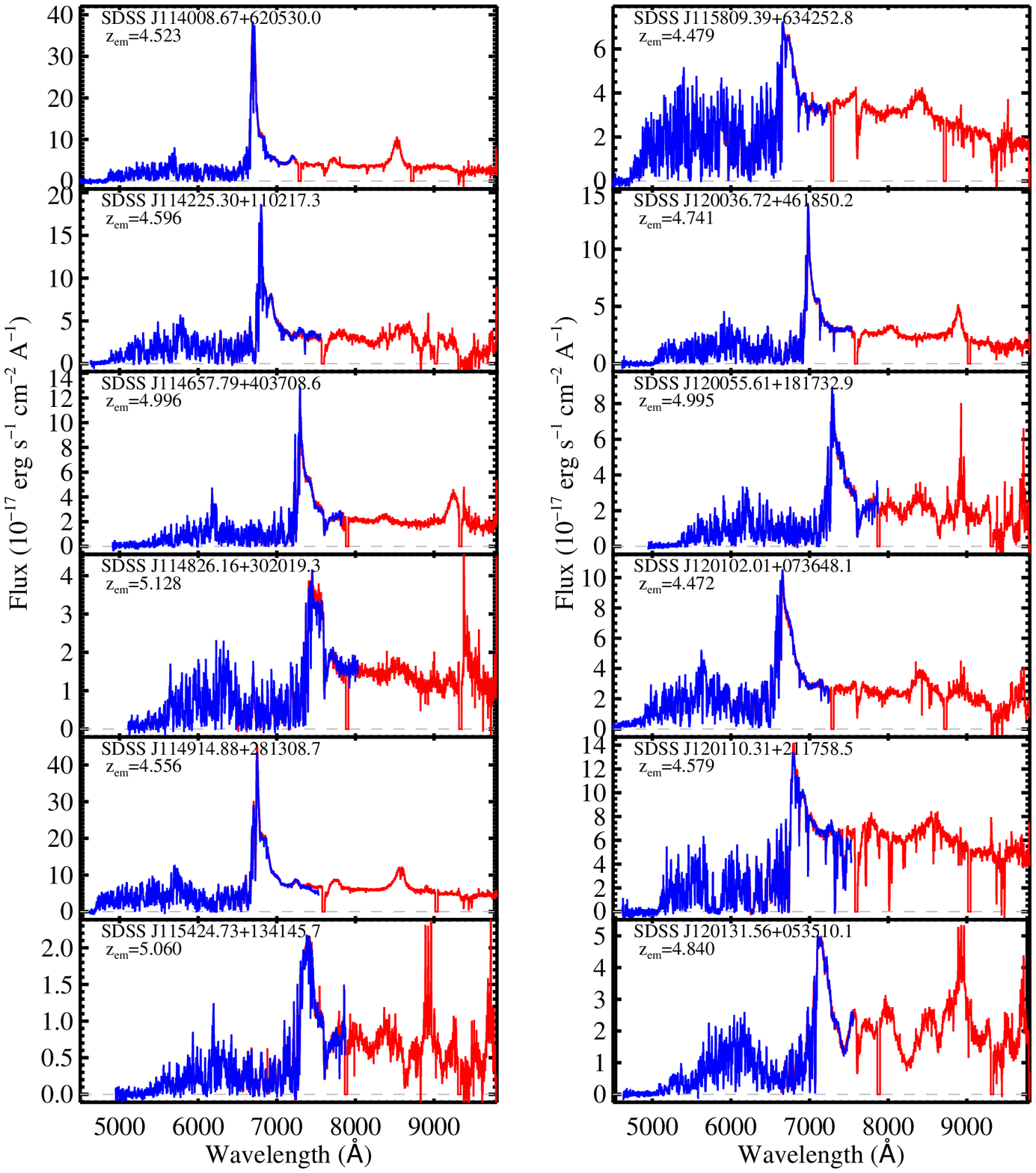}
\contcaption{}
\end{figure*}

\clearpage

\begin{figure*}
\includegraphics[bb=43 152 567 743,clip,width=1.0\linewidth]{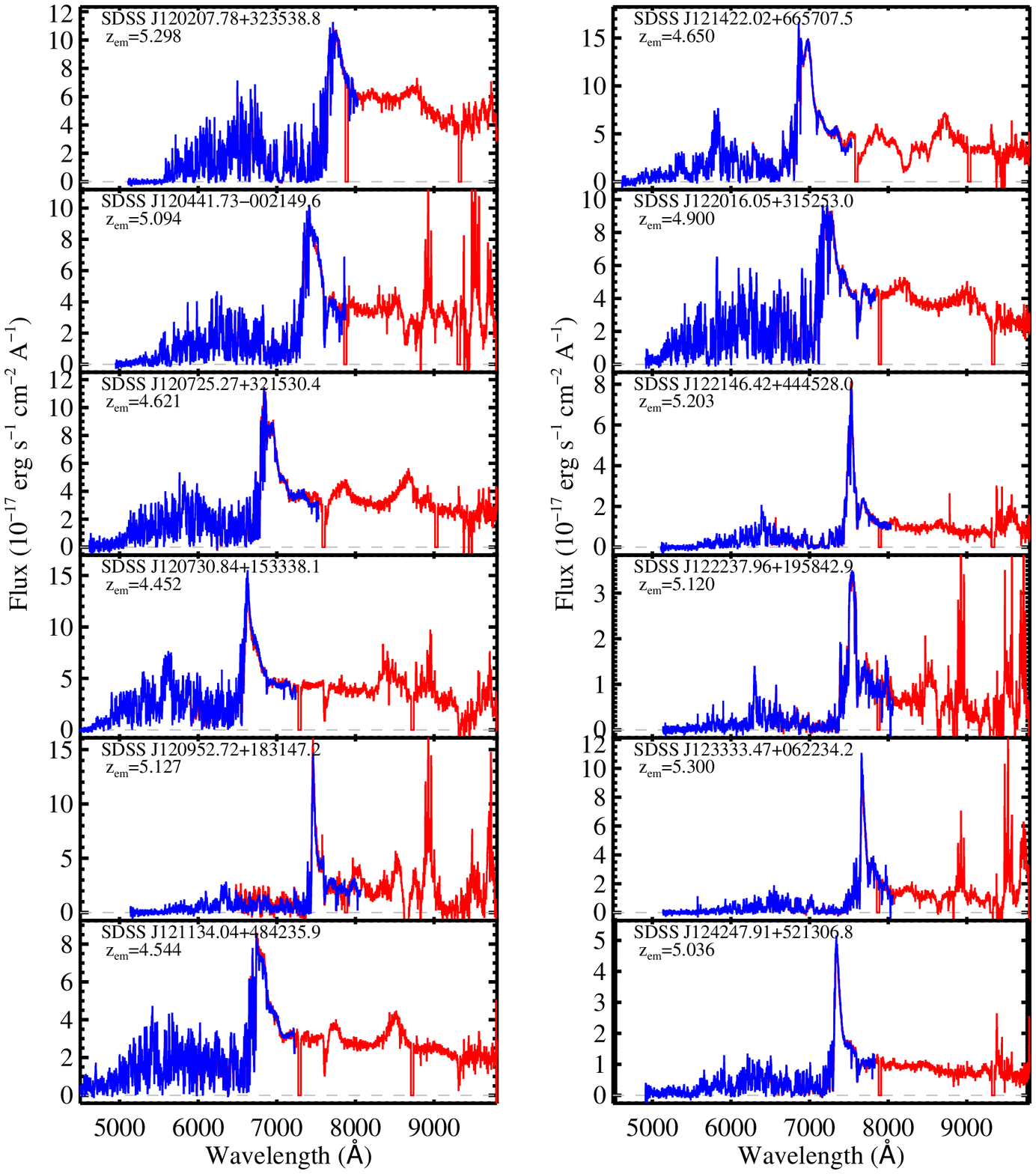}
\contcaption{}
\end{figure*}

\begin{figure*}
\includegraphics[bb=43 152 567 743,clip,width=1.0\linewidth]{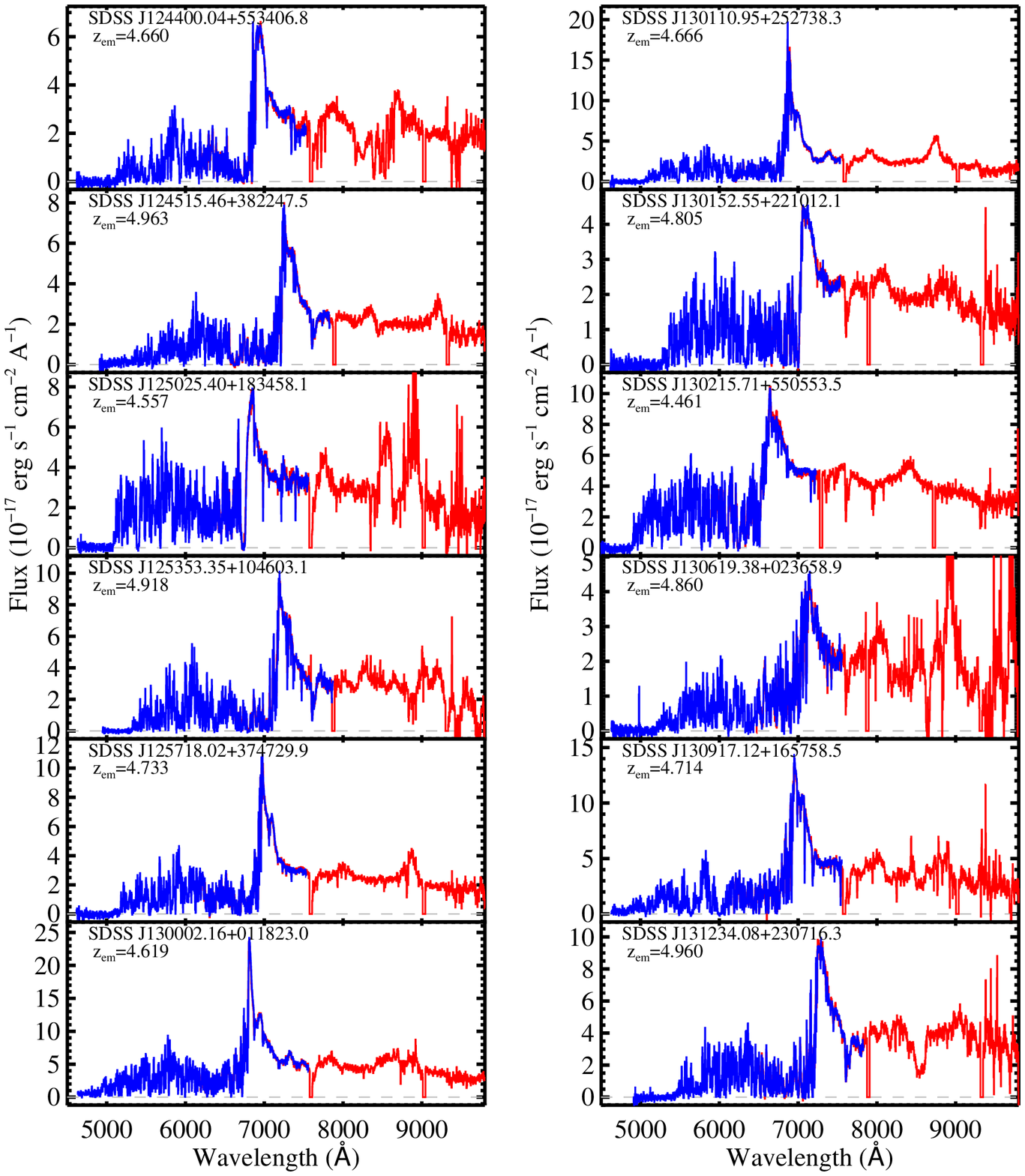}
\contcaption{}
\end{figure*}

\begin{figure*}
\includegraphics[bb=43 152 567 743,clip,width=1.0\linewidth]{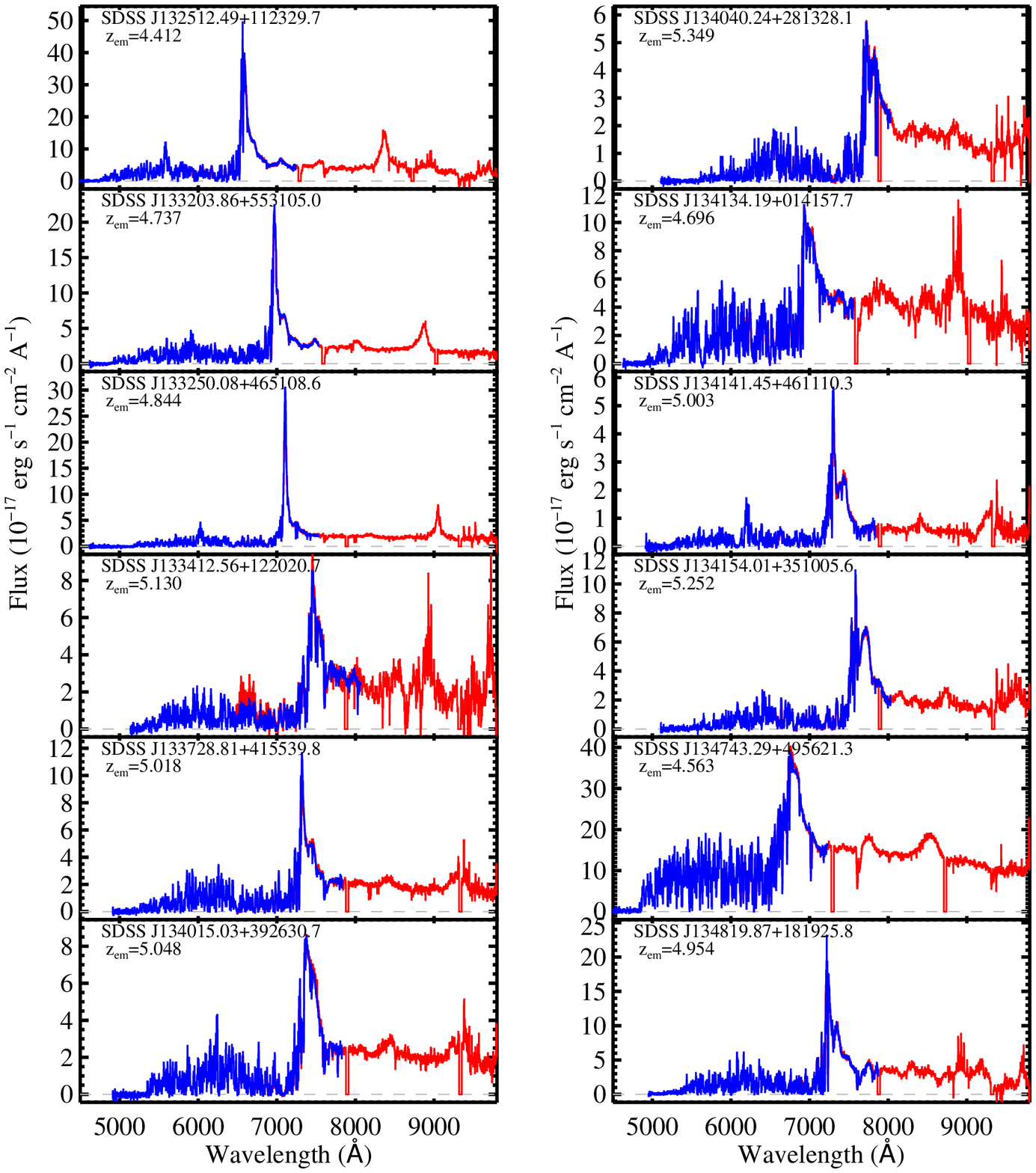}
\contcaption{}
\end{figure*}

\begin{figure*}
\includegraphics[bb=43 152 567 743,clip,width=1.0\linewidth]{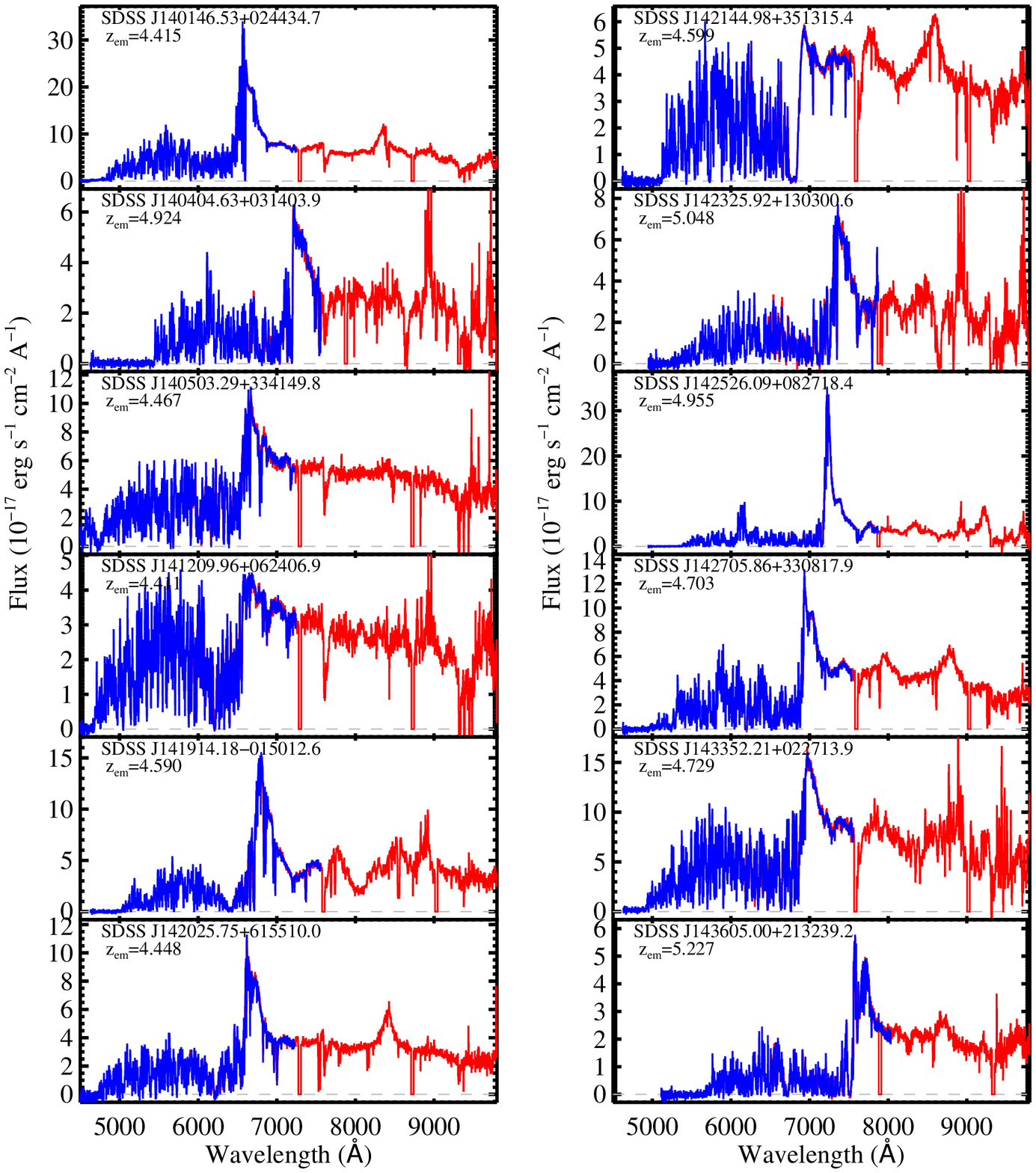}
\contcaption{}
\end{figure*}

\begin{figure*}
\includegraphics[bb=43 152 567 743,clip,width=1.0\linewidth]{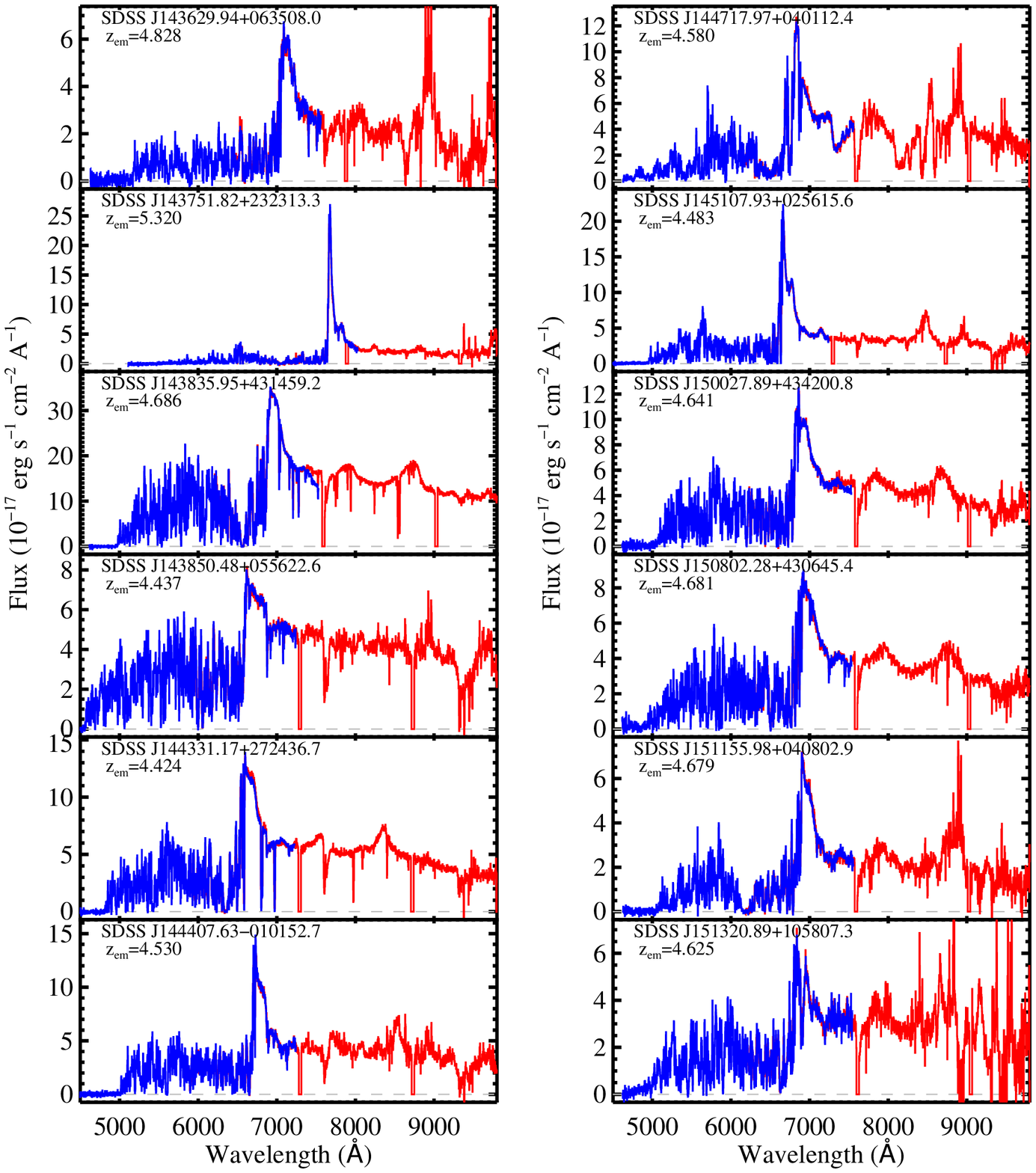}
\contcaption{}
\end{figure*}

\begin{figure*}
\includegraphics[bb=43 152 567 743,clip,width=1.0\linewidth]{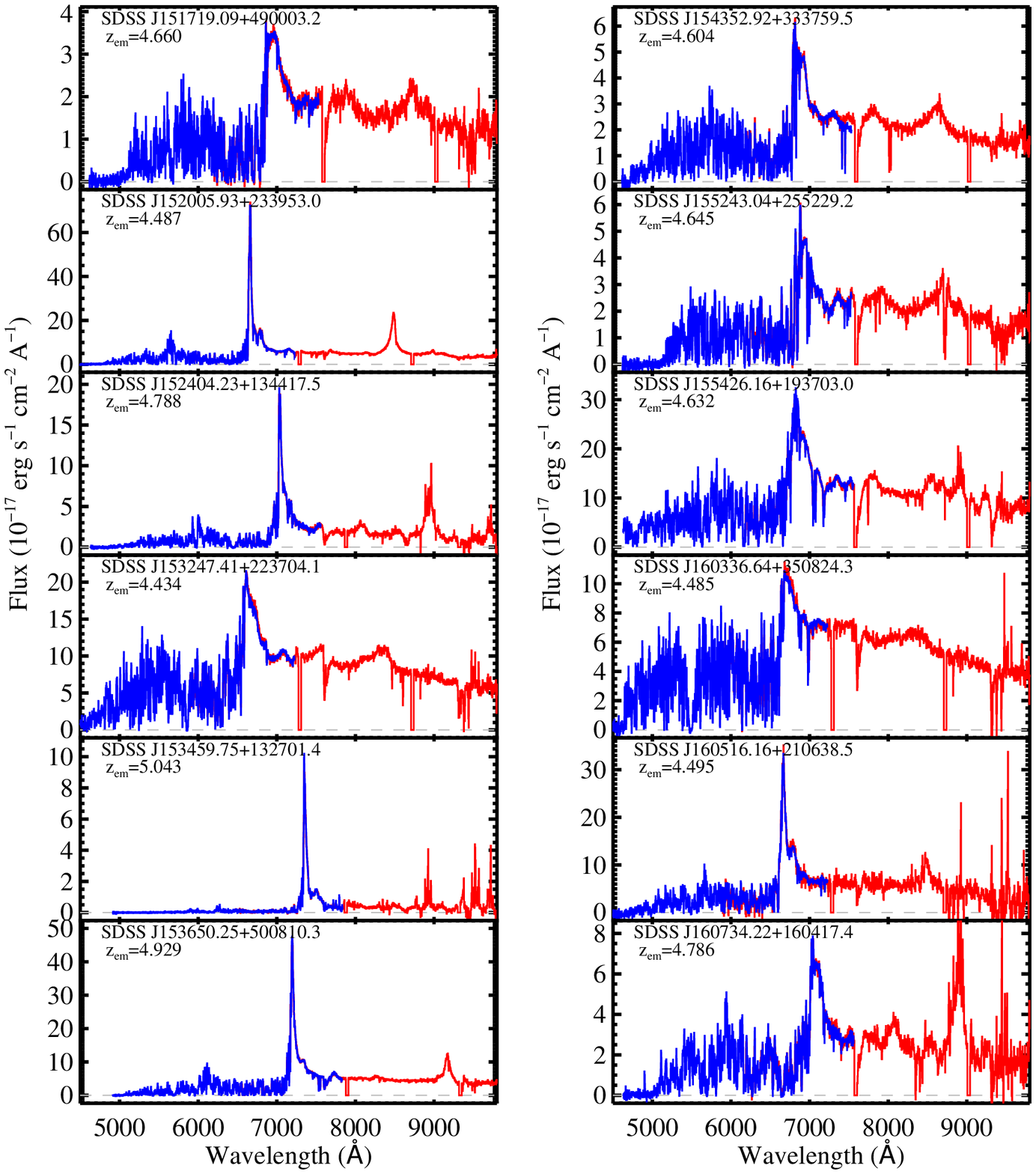}
\contcaption{}
\end{figure*}

\clearpage

\begin{figure*}
\includegraphics[bb=43 152 567 743,clip,width=1.0\linewidth]{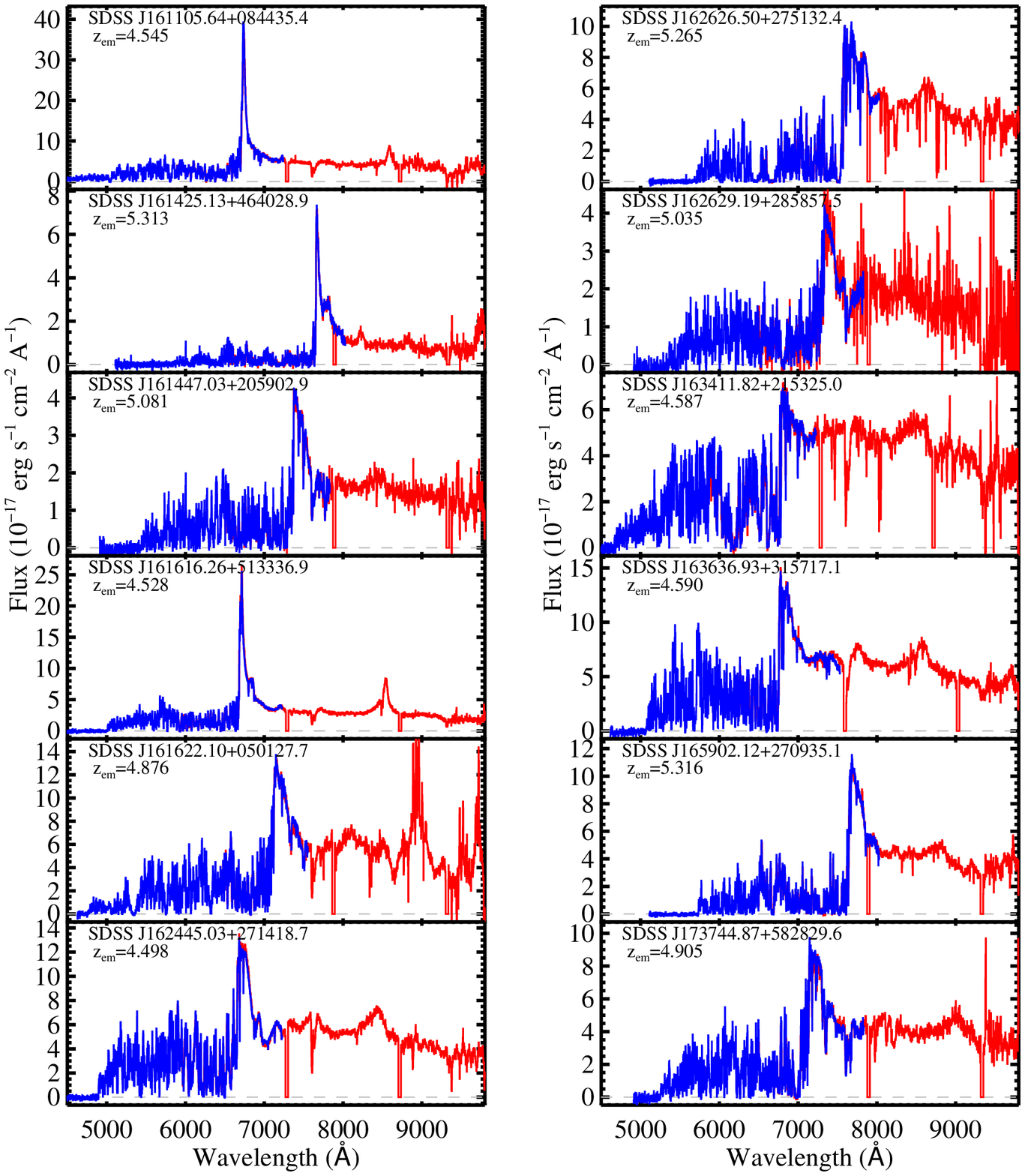}
\contcaption{}
\end{figure*}

\begin{figure*}
\includegraphics[bb=43 152 567 743,clip,width=1.0\linewidth]{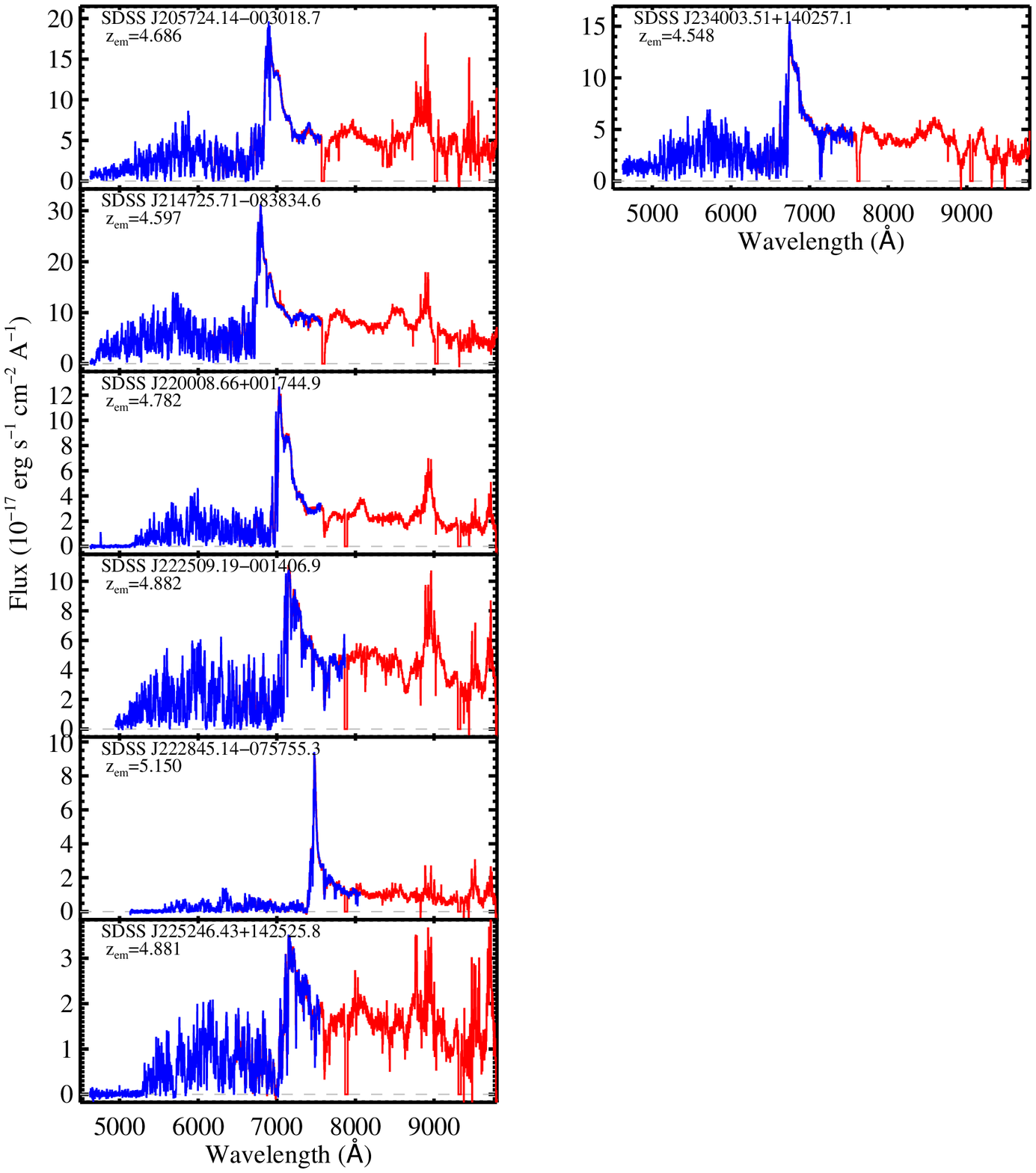}
\contcaption{}
\end{figure*}

\label{lastpage}

\end{document}